\documentclass[twocolumn,showpacs,superscriptaddress,preprintnumbers,amsmath,amssymb,floatfix]{revtex4}

\usepackage{amsmath}
\usepackage{graphicx}
\usepackage{dcolumn}
\usepackage{bm}

\begin{document}

\newcommand{\etal}{{\it et al.}}
\makeatother

\newcommand{\bra}[1]{\left\langle #1 \right|}
\newcommand{\brared}[1]{\langle #1 ||}
\newcommand{\ee}{\eta^{\ast},\eta}
\newcommand{\product}[2]{\left\langle #1 | #2 \right\rangle}

\newcommand{\kbar}{\bar{k}}
\newcommand{\ket}[1]{\left| #1 \right\rangle}
\newcommand{\ketred}[1]{|| #1 \rangle}
\newcommand{\ahat}{\hat{a}}
\newcommand{\adag}{a^{\dagger}}
\newcommand{\ahatdag}{\hat{a}^{\dagger}}
\newcommand{\Ahat}{\hat{A}}
\newcommand{\Adag}{A^{\dagger}}
\newcommand{\Ahatdag}{\hat{A}^{\dagger}}
\newcommand{\Atdag}{\hat{A}^{(\tau)\dagger}}
\newcommand{\Bdag}{\hat{B}^{\dagger}}
\newcommand{\Bhat}{\hat{B}}
\newcommand{\Bhatdag}{\hat{B}^{\dagger}}
\newcommand{\Btdag}{\hat{B}^{(\tau)\dagger}}
\newcommand{\Dhatp}{\hat{D}^{(+)}}
\newcommand{\Dp}{D^{(+)}}
\newcommand{\Ddotp}{\dot{D}^{(+)}}
\newcommand{\bhat}{\hat{b}}
\newcommand{\bdag}{b^{\dagger}}
\newcommand{\cdag}{c^{\dagger}}
\newcommand{\chat}{\hat{c}}
\newcommand{\chatdag}{\hat{c}^{\dagger}}
\newcommand{\degree}{^{\circ}}
\newcommand{\sprime}{s^{\prime}}
\newcommand{\Hhat}{\hat{H}}
\newcommand{\Hhatp}{\hat{H}^{\prime}}
\newcommand{\Ihat}{\hat{I}}
\newcommand{\Jhat}{\hat{J}}
\newcommand{\hhat}{\hat{h}}
\newcommand{\fp}{f^{(+)}}
\newcommand{\fpp}{f^{(+)\prime}}
\newcommand{\fm}{f^{(-)}}
\newcommand{\Fhat}{\hat{F}}
\newcommand{\Fhatdag}{\hat{F}^\dagger}
\newcommand{\Fhatp}{\hat{F}^{(+)}}
\newcommand{\Fhatm}{\hat{F}^{(-)}}
\newcommand{\Fhatpm}{\hat{F}^{(\pm)}}
\newcommand{\Fhatdagpm}{\hat{F}^{\dagger(\pm)}}
\newcommand{\Hc}{{\cal H}}
\newcommand{\Hcp}{{\cal H}^{\prime}}
\newcommand{\Ic}{{\cal I}}
\newcommand{\It}{\widetilde{I}}
\newcommand{\ITV}{{\cal I}_{\rm TV}}
\newcommand{\Jc}{{\cal J}}
\newcommand{\jp}{j^{\prime}}
\newcommand{\Qc}{{\cal Q}}
\newcommand{\Pc}{{\cal P}}
\newcommand{\Ec}{{\cal E}}
\newcommand{\Sc}{{\cal S}}
\newcommand{\Rc}{{\cal R}}

\newcommand{\bg}{\beta,\gamma}

\newcommand{\ddg}{d^{\dagger}}

\newcommand{\Nhat}{\hat{N}}
\newcommand{\Nt}{\widetilde{N}}
\newcommand{\Vt}{\widetilde{V}}
\newcommand{\nL}[1]{n_{L_{#1}}}
\newcommand{\nK}[1]{n_{K_{#1}}}
\newcommand{\nKb}{\mbox{\boldmath $n_K$}}
\newcommand{\nLb}{\mbox{\boldmath $n_L$}}

\newcommand{\mubar}{\bar{\mu}}

\newcommand{\Dc}{{\cal D}}
\newcommand{\Ddag}{\hat{D}^{\dagger}}
\newcommand{\dhat}{\hat{d}}
\newcommand{\Dhat}{\hat{D}}
\newcommand{\Ghat}{\hat{G}}
\newcommand{\Glambda}{G^{(\lambda)}}
\newcommand{\Gstarlambda}{G^{(\lambda)\ast}}
\newcommand{\Qhat}{\hat{Q}}
\newcommand{\Rhat}{\hat{R}}
\newcommand{\Phat}{\hat{P}}
\newcommand{\Pdag}{\hat{P}^{\dagger}}
\newcommand{\Psihat}{\hat{\Psi}}
\newcommand{\Qdag}{Q^{\dagger}}
\newcommand{\That}{\hat{\Theta}}
\newcommand{\Thatt}{\widetilde{\hat{\Theta}}}
\newcommand{\Tr}{{\rm Tr}}

\newcommand{\ktilde}{\tilde{k}}

\newcommand{\Pcirc}{\stackrel{\circ}{P}}
\newcommand{\Qcirc}{\stackrel{\circ}{Q}}
\newcommand{\Ncirc}{\stackrel{\circ}{N}}
\newcommand{\Tcirc}{\stackrel{\circ}{\Theta}}
\newcommand{\Pcircp}{\stackrel{\circ}{P^{\prime}}}
\newcommand{\Qcircp}{\stackrel{\circ}{Q^{\prime}}}
\newcommand{\Ncircp}{\stackrel{\circ}{N^{\prime}}}
\newcommand{\Tcircp}{\stackrel{\circ}{\Theta^{\prime}}}
\newcommand{\Fp}{F^{(+)}}
\newcommand{\Fm}{F^{(-)}}
\newcommand{\Fpm}{F^{(\pm)}}
\newcommand{\Rp}{R^{(+)}}
\newcommand{\Rm}{R^{(-)}}
\newcommand{\Bt}{\widetilde{B}}
\newcommand{\lambdat}{\widetilde{\lambda}}
\newcommand{\Phatt}{\widetilde{\hat{P}}}
\newcommand{\ab}{\bf a}

\newcommand{\Ab}{\mbox{\boldmath $A$}}
\newcommand{\Abdag}{\mbox{\boldmath $A$}^{\dagger}}
\newcommand{\Bb}{\mbox{\boldmath $B$}}
\newcommand{\cb}{\bf c}
\newcommand{\Db}{\mbox{\boldmath $D$}}
\newcommand{\Nb}{\mbox{\boldmath $N$}}
\newcommand{\Nbhat}{\hat{\mbox{\boldmath $N$}}}
\newcommand{\Qb}{\mbox{\boldmath $Q$}}
\newcommand{\Qhatt}{\widetilde{\hat{Q}}}
\newcommand{\Pb}{\mbox{\boldmath $P$}}
\newcommand{\phit}{\phi(t)}
\newcommand{\pdot}{\dot{p}}
\newcommand{\phix}[1]{\phi(#1)}
\newcommand{\qdot}{\dot{q}}
\newcommand{\phivib}{\phi(\eta^{\ast},\eta)}
\newcommand{\Ts}{{\cal T}}
\newcommand{\del}{\partial}
\newcommand{\eps}{\epsilon}
\newcommand{\beq}{\begin{equation}}
\newcommand{\beqa}{\begin{eqnarray}}
\newcommand{\eeq}{\end{equation}}
\newcommand{\eeqa}{\end{eqnarray}}
\newcommand{\Yb}{${}^{168}$Yb\ }
\newcommand{\Zhat}{\hat{Z}}
\newcommand{\rhodot}{\dot{\rho}}
\newcommand{\Khat}{\hat{K}}
\newcommand{\Kp}{K^{+}}
\newcommand{\Km}{K^{-}}
\newcommand{\Kz}{K^0}

\newcommand{\lb}{\bf l}
\newcommand{\sbold}{\bf s}

\newcommand{\Lp}{L^{+}}
\newcommand{\Lm}{L^{-}}
\newcommand{\Lz}{L^0}

\newcommand{\Mc}{{\cal M}}
\newcommand{\Mchat}{\hat{\cal M}}

\newcommand{\ddeta}{\frac{\partial}{\partial \eta}}
\newcommand{\ddetastar}{\frac{\partial}{\partial \eta^\ast}}
\newcommand{\etastar}{\eta^\ast}
\newcommand{\ketvib}{\ket{\phi (\etastar, \eta)}}
\newcommand{\bravib}{\bra{\phi (\etastar, \eta)}}
\newcommand{\zhateta}{\hat{z}(\eta)}
\newcommand{\zhat}{\hat{z}}
\newcommand{\oo}{\stackrel{\circ}{O}(\etastar,\eta)}
\newcommand{\oodag}{\stackrel{\circ}{O^{\dagger}}(\etastar,\eta)}
\newcommand{\oodagp}{\stackrel{\circ}{O^{\dagger\prime}}(\etastar,\eta)}
\newcommand{\oop}{\stackrel{\circ}{O^{\prime}}(\etastar,\eta)}
\newcommand{\Odag}{\hat{O}^{\dagger}}
\newcommand{\Ohat}{\hat{O}}
\newcommand{\Uinv}{U^{-1}(\etastar, \eta)}
\newcommand{\Uinvp}{U^{-1}(\etastar,\eta,\varphi,n)}
\newcommand{\U}{U(\etastar, \eta)}
\newcommand{\Up}{U(\etastar,\eta,\varphi,n)}
\newcommand{\etader}{\frac{\del}{\del \eta}}
\newcommand{\etastarder}{\frac{\del}{\del \etastar}}

\newcommand{\fb}{\mbox {\bfseries\itshape f}}
\newcommand{\SB}{\mbox {\bfseries\itshape S}}

\newcommand{\vbar}{\bar{v}}

\newcommand{\Udag}{U^{\dagger}}
\newcommand{\Vdag}{V^{\dagger}}

\newcommand{\Wc}{{\cal W}}
\newcommand{\Wcdag}{{\cal W}^{\dagger}}

\newcommand{\Xhat}{\hat{X}}
\newcommand{\Xdag}{\hat{X}^{\dagger}}

\renewcommand{\thanks}{\footnote}
\newcommand\tocite[1]{$^{\hbox{--}}$\cite{#1}}

\preprint{}

\title{
Microscopic description of large-amplitude shape-mixing dynamics with inertial functions derived in local quasiparticle random-phase approximation
}


\author{Nobuo Hinohara}
\affiliation{Theoretical Nuclear Physics Laboratory, RIKEN Nishina Center, Wako
351-0198, Japan}
\author{Koichi Sato}
\affiliation{Department of Physics, Graduate School of Science, Kyoto University,
606-8502 Kyoto, Japan}
\affiliation{Theoretical Nuclear Physics Laboratory, RIKEN Nishina Center,
Wako 351-0198, Japan}
\author{Takashi Nakatsukasa}
\affiliation{Theoretical Nuclear Physics Laboratory, RIKEN Nishina Center,
Wako 351-0198, Japan}
\author{Masayuki Matsuo}
\affiliation{Department of Physics, Faculty of Science, Niigata University,
Niigata 950-2181, Japan}
\author{Kenichi Matsuyanagi}
\affiliation{Theoretical Nuclear Physics Laboratory, RIKEN Nishina Center,
Wako 351-0198, Japan}
\affiliation{Yukawa Institute for Theoretical Physics, Kyoto University,
Kyoto 606-8502, Japan}

\date{\today}

\begin{abstract}
On the basis of the adiabatic self-consistent collective coordinate method, 
we develop an efficient microscopic method of deriving  the five-dimensional  
quadrupole collective Hamiltonian 
and illustrate its usefulness by applying it to the oblate-prolate shape 
coexistence/mixing phenomena in proton-rich $^{68,70,72}$Se. 
In this method, the vibrational and rotational collective masses (inertial functions) 
are determined by local normal modes built on constrained Hartree-Fock-Bogoliubov states.    
Numerical calculations are carried out using the pairing-plus-quadrupole 
Hamiltonian including the quadrupole-pairing interaction
within the two major-shell active model spaces both for neutrons and protons.  
It is shown that the time-odd components of the moving mean-field 
significantly increase the vibrational and rotational collective masses 
in comparison with the Inglis-Belyaev cranking masses. 
Solving the collective Schr\"odinger equation, we evaluate  
excitation spectra, quadrupole transitions and moments. 
Results of the numerical calculation are in excellent agreement with 
recent experimental data and indicate that 
the low-lying states of these nuclei are characterized 
as an intermediate situation between the oblate-prolate shape coexistence 
and the so-called $\gamma$ unstable situation where large-amplitude triaxial-shape 
fluctuations play a dominant role.  
  
\end{abstract}

\maketitle

\section{\label{sec:intro}Introduction}

The major purpose of this paper is to develop an efficient microscopic 
method of deriving  the five-dimensional (5D) quadrupole collective Hamiltonian 
\cite{BMvol2,Belyaev196517,Kumar1967608,0954-3899-36-12-123101}
and illustrate its usefulness by applying it to the oblate-prolate shape 
coexistence/mixing phenomena in proton-rich Se isotopes 
\cite{PhysRevC.67.064318,PhysRevLett.84.4064,obertelli:031304,ljungvall:102502}.
As is well known, the quadrupole collective Hamiltonian, also called 
the general Bohr-Mottelson Hamiltonian, contains six collective inertia masses 
(three vibrational masses and three rotational moments of inertia) as well as 
the collective potential. These seven quantities are functions of the quadrupole 
deformation variables $\beta$ and $\gamma$,  which represent the magnitude 
and triaxiality of the quadrupole deformation, respectively. 
Therefore, we also call the collective inertial masses `inertial functions.'
They are usually calculated by means of the adiabatic perturbation 
treatment of the moving mean field \cite{PhysRev.96.1059},
and the version taking into account nuclear superfluidity \cite{Beliaev1961322}
is called the Inglis-Belyaev (IB) cranking mass or the IB inertial function. 
Its insufficiency has been repeatedly emphasized, however
 (see e.g., Refs. \cite{Kumar1974189,Pomorski1977394,Rohozinski197766,Z.Phys.A294.341Dudek}). 
The most serious shortcoming is that the time-odd terms induced by the 
moving mean field are ignored, which breaks the self-consistency of the theory 
\cite{Baranger1978123,Dobaczewski1981123}.   
In fact, one of the most important motives of constructing 
microscopic theory of large-amplitude collective motion was 
to overcome such a shortcoming of the IB cranking mass 
\cite{Baranger1978123}. 

As fruits of long-term efforts, advanced microscopic theories of inertial  
functions are now available 
(see Refs. \cite{ Villars1977269,  Baranger1978123,Goeke1978328,
Can.J.Phys.54.1941Rowe,PTP.57.112,PhysRevC.60.054301,Dobaczewski1981123, 
ZPhysA272_387, PhysRevC.43.2254,Almehed2004163,PTP.64.1294,PhysRevC.21.2060} 
for original papers and Refs. \cite{Dang200093, 0954-3899-37-6-064018} for reviews). 
These theories of large-amplitude collective motion have been tested for schematic solvable models 
and applied to heavy-ion collisions and giant resonances \cite{Goeke1978328,PhysRevC.21.2060}.  
For nuclei with pairing correlations, 
Dobaczewski and Skalski studied the quadrupole vibrational 
mass with use of the adiabatic time-dependent Hartree-Fock-Bogoliubov 
(ATDHFB) theory and concluded that the contributions from 
the time-odd components of the moving mean-field 
significantly increase the vibrational mass compared to
the IB cranking mass \cite{Dobaczewski1981123}.        
Somewhat surprisingly, however, to the best of our knowledge, 
the ATDHFB vibrational masses have never been used 
in realistic calculations for low-lying quadrupole spectra of nuclei with superfluidity. 
For instance, in recent microscopic studies     
\cite{niksic:092502,niksic:034303,li:054301,PhysRevC.81.034316,Girod200939,PhysRevC.81.014303}  
by means of the 5D quadrupole Hamiltonian, 
the IB cranking formula are still used in actual numerical calculation  
for vibrational masses.  
This situation concerning the treatment of the collective kinetic energies is 
in marked contrast with the remarkable progress 
in microscopic calculation of the collective potential using 
modern effective interactions or energy density functionals 
(see Ref.~\cite{RevModPhys.75.121} for a review).

In this paper, on the basis of the adiabatic self-consistent collective coordinate 
(ASCC) method \cite{PTP.103.959}, 
we formulate a practical method of deriving the 5D quadrupole collective Hamiltonian.  
The central concept of this approach is local normal modes 
built on constrained Hartree-Fock-Bogoliubov (CHFB) states 
\cite{Ring-Schuck}
defined at every point of the ($\bg$) deformation space. 
These local normal modes are determined by the local QRPA (LQRPA) equation 
that is an extension of the well-known 
quasiparticle random-phase approximation (QRPA) to non-equilibrium 
HFB states determined by the CHFB equations.  
We therefore use an abbreviation `CHFB+LQRPA method' for this 
approach. This method may be used in conjunction with any 
effective interaction or energy density functional.  
In this paper, however,  we use, for simplicity, 
the pairing-plus-quadrupole (P+Q) force \cite{Bes-Sorensen,Baranger1968490}   
including the quadrupole-pairing force. 
Inclusion of the quadrupole-pairing force is essential because 
it produces the time-odd component of the moving field \cite{PTP.115.567}. 

To examine the feasibility of the CHFB+LQRPA method, 
we apply it to the oblate-prolate shape coexistence/mixing phenomena 
in proton-rich $^{68,70,72}$Se 
\cite{PhysRevC.67.064318,PhysRevLett.84.4064,obertelli:031304,
ljungvall:102502,0954-3899-28-10-307,PhysRevC.63.024313}. 
These phenomena are taken up because   
we obviously need to go beyond the traditional framework 
of describing small-amplitude vibrations around a single HFB equilibrium point 
to describe them; that is, they are very suitable targets for our purpose. 
We shall show in this paper that this approach successfully describes  
large-amplitude collective vibrations extending from the oblate to 
the prolate HFB equilibrium points (and vice versa). 
In particular, it will be demonstrated that we can describe very well  
the transitional region between the oblate-prolate shape coexistence 
and the $\gamma$ unstable situation where large-amplitude triaxial-shape 
fluctuations play a dominant role.  

This paper is organized as follows.
In Sec. \ref{sec:theory}, 
we formulate the CHFB+LQRPA as an approximation of the ASCC method 
and derive the 5D quadrupole collective Hamiltonian. 
In Sec. \ref{sec:collHresult}, we calculate the vibrational and rotational masses 
by solving the LQRPA equations,     
and discuss their properties in comparison with those calculated 
by using the IB cranking formula. 
In Sec. \ref{sec:reqresult}, 
we calculate excitation spectra, $B(E2)$, and spectroscopic quadrupole moments 
of low-lying states in $^{68,70,72}$Se and discuss 
properties of the oblate-prolate shape coexistence/mixing 
in these nuclei. 
Conclusions are given in Sec. \ref{sec:conclusion}.

\section{Microscopic derivation of the 5D quadrupole collective Hamiltonian 
\label{sec:theory}}

\subsection{The 5D quadrupole collective Hamiltonian}\label{sec:theory:5dcollH}

Our aim in this section is to formulate a practical method of microscopically deriving 
the 5D quadrupole collective Hamiltonian
\cite{BMvol2,Belyaev196517,Kumar1967608,0954-3899-36-12-123101}
\begin{align}
 \Hc_{\rm coll}&= T_{\rm vib} + T_{\rm rot} + V(\bg), \label{eq:collH_BM1}\\
 T_{\rm vib}   &= \frac{1}{2} D_{\beta\beta}(\bg)\dot{\beta}^2 +
 D_{\beta\gamma}(\bg)\dot{\beta}\dot{\gamma}
+ \frac{1}{2}D_{\gamma\gamma}(\bg)\dot{\gamma}^2, \label{eq:collH_BM2}\\
 T_{\rm rot} &= \frac{1}{2}\sum_{k=1}^3 \Jc_k(\bg) \omega^2_k, \label{eq:collH_BM3}
\end{align}
starting from an effective Hamiltonian for finite many-nucleon systems.  
Here, $T_{\rm vib}$ and $T_{\rm rot}$ denote the kinetic energies of vibrational 
and rotational motions, while $V(\beta,\gamma)$ represents the collective potential. 
The velocities of the vibrational motion are described in terms of 
the time-derivatives ($\dot{\beta}$, $\dot{\gamma}$) of 
the quadrupole deformation variables ($\beta$, $\gamma$)  
representing the magnitude and the triaxiality of the quadrupole deformation, 
respectively. The three components $\omega_k$ of the rotational angular velocity 
are defined with respect to the intrinsic axes associated with the rotating nucleus. 
The inertial functions for vibrational motions (vibrational masses),  
$D_{\beta\beta}$, $D_{\beta\gamma}$, and $D_{\gamma\gamma}$, 
and the rotational moments of inertia $\Jc_k$ are functions of $\beta$ and $\gamma$. 

As seen in the recent review by 
Pr\'{o}chniak and Rohozi\'{n}ski \cite{0954-3899-36-12-123101},  
there are numerous papers on microscopic approaches 
to the 5D quadrupole collective Hamiltonian;   
among them, we should quote at least early papers 
by Belyaev \cite{Belyaev196517}, 
Baranger-Kumar \cite{Baranger1968241,Kumar1968273}, 
Pomorski et~al. \cite{Pomorski1977394,Rohozinski197766},
and recent papers by Girod et~al. \cite{Girod200939}, 
Nik\v{s}i\'{c} et~al. \cite{niksic:092502,niksic:034303},
and Li et~al. \cite{li:054301,PhysRevC.81.034316}.  
In all these works, the IB cranking formula is used for 
the vibrational inertial functions.  
Below, we outline the procedure of deriving the vibrational and rotational inertial 
functions on the basis of the ASCC method.  

\subsection{Basic equations of the ASCC method}
\label{sec:theory:2DASCC}

To derive the 5D quadrupole collective Hamiltonian $\Hc_{\rm coll}$ 
starting from a microscopic Hamiltonian $\Hhat$, we use the ASCC method 
\cite{PTP.103.959,PTP.117.451}. 
This method enables us to determine a collective submanifold 
embedded in the large-dimensional TDHFB configuration space.  
We can use this method in conjunction with any 
effective interaction or energy density functional 
to microscopically derive the collective masses 
taking into account time-odd mean-field effects. 
For our present purpose, 
we here recapitulate a two-dimensional (2D) version of the ASCC method.  
We suppose existence of a set of two collective coordinates $(q^1, q^2)$ 
that has a one-to-one correspondence to the quadrupole deformation variable set $(\bg)$ 
and try to determine a 2D collective hypersurface 
associated with the large-amplitude quadrupole shape vibrations. 
We thus assume that the TDHFB states 
can be written on the hypersurface in the following form; 
\begin{align}
 \ket{\phi(\bm{q},\bm{p},\bm{\varphi},\bm{n})} &= 
e^{-i\sum_{\tau} \varphi^{(\tau)}\Nt^{(\tau)}}
 \ket{\phi(\bm{q},\bm{p},\bm{n})} \nonumber \\
 &= e^{-i\sum_{\tau} \varphi^{(\tau)}\Nt^{(\tau)}} e^{i
 \Ghat(\bm{q},\bm{p},\bm{n})} 
\ket{\phi(\bm{q})},
\end{align}
with
\begin{align}
 \Ghat(\bm{q},\bm{p},\bm{n}) =&  
\sum_{i=1,2} p_i \Qhat^i(\bm{q}) + 
\sum_{\tau=n,p} n^{(\tau)} \That^{(\tau)}(\bm{q}), \\
\Qhat^i(\bm{q}) &= \Qhat^{A}(\bm{q}) + \Qhat^{B}(\bm{q}) \nonumber \\
  =& \sum_{\alpha\beta} [ Q^A_{\alpha\beta}(\bm{q}) \adag_\alpha
 \adag_\beta + Q^{A \ast}_{\alpha\beta}(\bm{q}) a_\beta a_\alpha
 \nonumber \\
 & + Q^B_{\alpha\beta}(\bm{q})\adag_\alpha a_\beta], \\
 \That^{(\tau)}(\bm{q}) &= \sum_{\alpha\beta} 
 [ \Theta^{(\tau)A}_{\alpha\beta}(\bm{q}) \adag_\alpha \adag_\beta+
   \Theta^{(\tau)A\ast}_{\alpha\beta}(\bm{q}) a_\beta a_\alpha ]. 
\end{align}
For a gauge-invariant description of nuclei with superfluidity,  
we need to parametrize the TDHFB state vectors, as above, 
not only by the collective coordinates 
$\bm{q}=(q^1,q^2)$ and conjugate momenta $\bm{p}=(p_1,p_2)$ but also by 
the gauge angles $\bm{\varphi}=(\varphi^{(n)},\varphi^{(p)})$ 
conjugate to the number variables $\bm{n} =(n^{(n)},n^{(p)})$ representing 
the pairing-rotational degrees of freedom (for both neutrons and protons). 
In the above equations, $\Qhat^i(\bm{q})$ and $\That^{(\tau)}(\bm{q})$ are 
infinitesimal generators which are written in terms of the quasiparticle 
creation and annihilation operators  $(\adag_\alpha, a_\alpha)$ locally defined 
with respect to the moving-frame HFB states $\ket{\phi(\bm{q})}$. 
Note that the number operators are defined 
as $\Nt^{(\tau)}\equiv \Nhat^{(\tau)} - N_0^{(\tau)}$ subtracting 
the expectation values $(N_0^{(n)}, N_0^{(p)})$ of the neutron and proton numbers  
at $\ket{\phi(\bm{q})}$.  
In this paper, we use units with $\hbar=1$. 

The moving-frame HFB states $\ket{\phi(\bm{q})}$ 
and the infinitesimal generators $\Qhat^i(\bm{q})$
are determined as solutions of the moving-frame HFB equation, 
\begin{align}
 \delta \bra{\phi(\bm{q})} \Hhat_M(\bm{q})\ket{\phi(\bm{q})} = 0, \label{eq:ASCC0}
\end{align}
and the moving-frame QRPA equations, 
\begin{align}
 \delta & \bra{\phi(\bm{q})} [\Hhat_M(\bm{q}), \Qhat^i(\bm{q})]
- \frac{1}{i} \sum_{k} B^{ik}(\bm{q}) \Phat_k(\bm{q})
 \nonumber \\
&+ \frac{1}{2}\left[\sum_{k} \frac{\del V}{\del q^k}\Qhat^k(\bm{q}),
 \Qhat^i(\bm{q})\right]
\ket{\phi(\bm{q})} = 0, \label{eq:ASCC1}
\end{align}
\begin{align}
 \delta & \bra{\phi(\bm{q})} [\Hhat_M(\bm{q}),
 \frac{1}{i}\Phat_i(\bm{q})] - \sum_{j} C_{ij}(\bm{q}) \Qhat^j(\bm{q}) \nonumber \\
& - \frac{1}{2}\left[\left[\Hhat_M(\bm{q}), \sum_{k} \frac{\del V}{\del
 q^k}\Qhat^k(\bm{q})\right], \sum_{j} B_{ij}(\bm{q}) \Qhat^j(\bm{q})\right]  \nonumber \\
& - \sum_{\tau} \frac{\del\lambda^{(\tau)}}{\del q^i} \Nt^{(\tau)} \ket{\phi(\bm{q})}
 = 0, \label{eq:ASCC2}
\end{align}
which are derived from the time-dependent variational principle.
Here, $\Hhat_M(\bm{q})$ is the moving-frame Hamiltonian given by
\begin{align}
  \Hhat_M(\bm{q}) = 
 \Hhat - \sum_{\tau} \lambda^{(\tau)}(\bm{q})
 \Nt^{(\tau)} - 
 \sum_{i} \frac{\del V}{\del q^i}\Qhat^i(\bm{q}) 
\end{align}
and
\begin{align}
 C_{ij}(\bm{q}) = \frac{\del^2 V}{\del q^i \del q^j} - \sum_k \Gamma_{ij}^k\frac{\del V}{\del q^k} 
\end{align}
with
\begin{align}
 \Gamma_{ij}^k(\bm{q}) = \frac{1}{2} \sum_l B^{kl}( \frac{\del B_{li}}{\del q^j}
 + \frac{\del B_{lj}}{\del q^i} - \frac{\del B_{ij}}{\del q^l} ).  
\end{align}
The infinitesimal generators $\Phat_i(\bm{q})$ are defined by 
\begin{align}
 \Phat_i(\bm{q}) \ket{\phi(\bm{q})} = i \frac{\del}{\del q^i} \ket{\phi(\bm{q})},
\end{align}
with
\begin{align}
 \Phat_i(\bm{q}) = i \sum_{\alpha\beta}[ P_{i\alpha\beta}(\bm{q})
 \adag_\alpha \adag_\beta - P_{i\alpha\beta}^\ast(\bm{q}) a_\beta a_\alpha], 
\end{align}
and determined as solutions of the moving-frame QRPA equations.

The collective Hamiltonian is given as the expectation value of the microscopic 
Hamiltonian with respect to the TDHFB state:  
\begin{align}
 \Hc(\bm{q},\bm{p},\bm{n}) 
=& \bra{\phi(\bm{q},\bm{p},\bm{n})}\Hhat
 \ket{\phi(\bm{q},\bm{p},\bm{n})} \nonumber \\
=& V(\bm{q}) + \sum_{ij} \frac{1}{2} B^{ij}(\bm{q})p_i p_j + \sum_{\tau}
 \lambda^{(\tau)}(\bm{q}) n^{(\tau)}, \label{eq:collH_ASCC}
\end{align}
where 
\begin{align}
 V(\bm{q}) =&  \Hc(\bm{q},\bm{p},\bm{n})
\Big\arrowvert_{\bm{p}=\bm{0},\bm{n}={\textbf 0}}, \\
 B^{ij}(\bm{q}) =& \frac{\del^2 \Hc}{\del p_i \del p_j}
 \Big\arrowvert_{\bm{p}=\bm{0},\bm{n}={\textbf 0}}, \label{eq:defB} \\
  \lambda^{(\tau)}(\bm{q}) =& \frac{\del \Hc}{\del n^{(\tau)}}\Big\arrowvert_{\bm{p}=\bm{0},\bm{n}={\textbf 0}}, \label{eq:deflambda}
\end{align}
represent the collective potential, inverse of the collective mass,
and the chemical potential, respectively. 
Note that the last term in Eq.~(\ref{eq:ASCC2})
can be set to zero adopting the QRPA gauge-fixing condition,  
$d\lambda^{(\tau)}/dq^i = 0$ \cite{PTP.117.451}. 

The basic equations of the ASCC method are invariant against
point transformations of the collective coordinates ($q^1 , q^2$).
The $B^{ij}(\bm{q})$ and  $C_{ij}(\bm{q})$ can be diagonalized simultaneously
by a linear coordinate transformation at each point of $\bm{q}=(q^1, q^2)$. 
We assume that we can introduce the collective coordinate system in which
the diagonal form is kept globally. Then, we can choose, without
losing generality and for simplicity, the scale of the collective coordinates $\bm{q}=(q^1, q^2)$ 
such that the vibrational masses become unity. 
Consequently, the vibrational kinetic energy 
in the collective Hamiltonian (\ref{eq:collH_ASCC}) is written as
\begin{align}
 T_{\rm vib} = \frac{1}{2} \sum_{i=1,2} (p_i)^2 
 = \frac{1}{2} \sum_{i=1,2} (\dot{q^i})^2. 
\label{eq:Tvib}
\end{align}

\subsection{CHFB+LQRPA equations}

The basic equations of the ASCC method can be solved with an iterative 
procedure. This task was successfully carried out for extracting 
a one-dimensional (1D) collective path embedded in the TDHFB configuration space
\cite{PTP.119.59,hinohara:014305}.  To determine a 2D hypersurface, however, 
the numerical calculation becomes too demanding at the present time.   
We therefore introduce practical approximations as follows:  
First, we ignore the curvature terms 
(the third terms in Eqs. (\ref{eq:ASCC1}) and (\ref{eq:ASCC2})), 
which  vanish at the HFB equilibrium points where $dV/dq^i=0$, 
assuming that their effects are numerically small.  
Second, we replace the moving-frame HFB Hamiltonian $\Hhat_M(\bm{q})$ and  
the moving-frame HFB state $\ket{\phi(q^1,q^2)}$ with   
a CHFB Hamiltonian $\Hhat_{\rm CHFB}(\bg)$ and 
a CHFB state $\ket{\phi(\bg)}$, respectively, 
on the assumption that the latters are good approximations to the formers.  

The CHFB equations are given by  
\begin{align}
 \delta \bra{\phi(\bg)} \Hhat_{\rm CHFB}(\bg) \ket{\phi(\bg)} = 0,  \label{eq:CHFB} 
\end{align}
\begin{align}
 \Hhat_{\rm CHFB}(\bg) =& \Hhat -
 \sum_{\tau}\lambda^{(\tau)}(\bg)\Nt^{(\tau)} \nonumber \\
 &- \sum_{m = 0, 2} \mu_{m}(\bg) \Dhatp_{2m}  \label{eq:H_CHFB} 
\end{align}
with four constraints
\begin{align}
 \bra{\phi(\bg)} \Nhat^{(\tau)} \ket{\phi(\bg)} = N^{(\tau)}_0,  
\quad (\tau = n, p) \\
 \bra{\phi(\bg)} \Dhatp_{2m} \ket{\phi(\bg)} = \Dp_{2m}, \quad (m = 0, 2)
\end{align}
where $\Dhatp_{2m}$ denotes Hermitian quadrupole operators, 
$\Dhat_{20}$ and $(\Dhat_{22} + \Dhat_{2-2})/2$ 
for  $m = 0$ and  2, respectively 
(see Ref. \cite{PTP.119.59} for their explicit expressions). 
We define the quadrupole deformation variables ($\beta, \gamma$) 
in terms of the expectation values of the quadrupole operators: 
\begin{align}
 \beta\cos\gamma &=  \eta D^{(+)}_{20} 
= \eta \bra{\phi(\bg)} \Dhatp_{20} \ket{\phi(\bg)},  \label{eq:definition1}  \\
 \frac{1}{\sqrt{2}} \beta\sin\gamma &= \eta D^{(+)}_{22} 
= \eta \bra{\phi(\bg)} \Dhatp_{22} \ket{\phi(\bg)}, \label{eq:definition2} 
\end{align}
where $\eta$ is a scaling factor (to be discussed in subsection \ref{sec:collHresult:model}). 

The moving frame QRPA equations, (\ref{eq:ASCC1}) and (\ref{eq:ASCC2}), then 
reduce to  
\begin{align}
 \delta & \bra{\phi(\bg)} [ \Hhat_{\rm CHFB}(\bg), \Qhat^i(\bg) ] \nonumber \\
 &- \frac{1}{i} \Phat_i(\bg) \ket{\phi(\bg)}  = 0, \quad\quad (i=1, 2) 
\label{eq:LQRPA1}
\end{align}
and
\begin{align} \delta & \bra{\phi(\bg)} [ \Hhat_{\rm CHFB}(\bg), \frac{1}{i} \Phat_i(\bg)]
 \nonumber \\
 &- C_i(\bg) \Qhat^i(\bg) \ket{\phi(\bg)} = 0.  \quad\quad (i=1, 2)  
\label{eq:LQRPA2}
\end{align}
Here the infinitesimal generators, $\Qhat^i(\bg)$ and $\Phat_i(\bg)$, 
are local operators defined at $(\bg)$ with respect to 
the CHFB state $\ket{\phi(\bg)}$. These equations are solved at
each point of $(\bg)$ to determine $\Qhat^i(\bg)$, $\Phat_i(\bg)$,
and $C_i(\bg)=\omega_i^2(\bg)$. 
Note that  these equations
are valid also for regions with negative curvature 
($C_i(\bg)<0$) where the QRPA frequency $\omega_i(\bg)$ takes an imaginary value. 
We call the above equations `local QRPA (LQRPA) equations'.
There exist more than two solutions of LQRPA equations 
(\ref{eq:LQRPA1}) and (\ref{eq:LQRPA2}), and we need to select relevant solutions.
A useful criterion for selecting two collective modes among many LQRPA modes will be 
given in subsection \ref{sec:collHresult:LQRPAmodes} with numerical examples. 
Concerning the accuracy of the CHFB+LQRPA approximation, 
some arguments will be given in subsection \ref{sec:collHresult:comparison}.  

\subsection{Derivation of the vibrational masses}\label{sec:theory:5dcollD:mass}

Once the infinitesimal generators $\Qhat^i(\bg)$ and $\Phat_i(\bg)$ are
obtained,
we can derive the vibrational masses appearing in the 5D quadrupole 
collective Hamiltonian (\ref{eq:collH_BM1}).
We rewrite the vibrational kinetic energy $T_{\rm vib}$ given by Eq.~(\ref{eq:Tvib})    
in terms of the time-derivatives, $\dot{\beta}$ and $\dot{\gamma}$, 
of the quadrupole deformation variables in the following way.  
We first note that an infinitesimal displacement of 
the collective coordinates $(q^1, q^2)$ brings about a corresponding change, 
\begin{align}
d\Dp_{2m}   =& \sum_{i=1,2} \frac{\del \Dp_{2m}}{\del q^i} dq^i, \quad\quad (m=0,2), 
\end{align}
in the expectation values of the quadrupole operators. 
The partial derivatives can be easily evaluated as 
\begin{align}
 \frac{\del \Dp_{20}}{\del q^i} =& \frac{\del}{\del q^i}\bra{\phi(\bg)}
 \Dhatp_{20} \ket{\phi(\bg)} \nonumber \\
  =& \bra{\phi(\bg)} [\Dhatp_{20}, \frac{1}{i}\Phat_i 
 (\bg)] \ket{\phi(\bg)}, \\
 \frac{\del \Dp_{22}}{\del q^i} =& \frac{\del}{\del q^i}\bra{\phi(\bg)}
 \Dhatp_{22} \ket{\phi(\bg)} \nonumber \\
  =& \bra{\phi(\bg)} [\Dhatp_{22}, \frac{1}{i}\Phat_i
 (\bg)] \ket{\phi(\bg)}, 
\end{align}
without need of numerical derivatives. 
Accordingly, the vibrational kinetic energy can be written 
\begin{align}
 T_{\rm vib} = \frac{1}{2} M_{00} [\Ddotp_{20}]^2
  + M_{02} \Ddotp_{20} \Ddotp_{22}+ \frac{1}{2} M_{22} [\Ddotp_{22}]^2,
\label{eq:TvibD}
\end{align}
with 
\begin{align}
 M_{mm'}(\bg) = \sum_{i=1,2} \frac{\del q^i}{\del
 \Dp_{2m}} \frac{\del q^i}{\del \Dp_{2m'}}. \label{eq:M_mm}
\end{align} 
Taking time-derivative of the definitional equations of ($\bg$),  
Eqs.~(\ref{eq:definition1}) and (\ref{eq:definition2}),   
we can straightforwardly transform the above expression (\ref{eq:TvibD}) to 
the form in terms of  ($\dot{\beta}$, $\dot{\gamma}$).  
The vibrational masses ($D_{\beta\beta}$, $D_{\beta\gamma}$,
$D_{\gamma\gamma}$) are then obtained from ($M_{00}$, $M_{02}$, $M_{22}$)  
through the following relations:  
\begin{align}
 D_{\beta\beta}=& \eta^{-2}
\left(
M_{00} \cos^2\gamma + \sqrt{2}M_{02}\sin\gamma\cos\gamma
 \right. \nonumber \\
&+ \left. \frac{1}{2}M_{22}\sin^2\gamma \right), \\ 
D_{\beta\gamma}=& 
 \beta \eta^{-2}
\left[ -M_{00}\sin\gamma\cos\gamma \right. \nonumber \\ 
&+ \left. \frac{1}{\sqrt{2}}M_{02}(\cos^2\gamma- \sin^2\gamma) + \frac{1}{2} M_{22} \sin\gamma\cos\gamma \right], \\
D_{\gamma\gamma} =& \beta^2 \eta^{-2} 
\left( M_{00} \sin^2\gamma
- \sqrt{2} M_{02} \sin\gamma\cos\gamma \right. \nonumber \\
&+  \left. \frac{1}{2} M_{22} \cos^2\gamma \right).
\end{align}

\subsection{Calculation of the rotational moments of inertia} \label{sec:theory:5dcollH:MOI}

We calculate the rotational moments of inertia $\Jc_k(\bg)$   
using the LQRPA equation for the collective rotation \cite{PTP.119.59} 
at each CHFB state, 
\begin{align}
 \delta \bra{\phi(\bg)} [\Hhat_{\rm CHFB}, \hat{\Psi}_k]
- \frac{1}{i} (\Jc_k)^{-1} \Ihat_k \ket{\phi(\bg)} = 0, \\ 
\bra{\phi(\bg)} [\hat{\Psi}_k(\bg), \Ihat_{k'}]\ket{\phi(\bg)} = i \delta_{kk'},
\end{align}
where $\hat{\Psi}_k(\bg)$ and $\Ihat_k$ represent the rotational angle
and the angular momentum operators with respect to the principal axes
associated with the CHFB state $\ket{\phi(\bg)}$.
This is an extension of the Thouless-Valatin equation \cite{Thouless1962211} 
for the HFB equilibrium state to non-equilibrium CHFB states.  
The three moments of inertia can be written as 
\begin{align}
 \Jc_k(\bg)  = 4 \beta^2 D_k(\bg) \sin^2 \gamma_k  \quad (k = 1, 2, 3) \label{eq:MOI}
\end{align}
with $\gamma_k = \gamma - (2\pi k/3)$.
If the inertial functions $D_k(\bg)$ above are replaced with a constant, 
then $\Jc_k(\bg)$ reduce to the well-known irrotational moments of inertia. 
In fact, however, we shall see that their  ($\bg$) dependence is very important. 
We call  $\Jc_k(\bg)$ and $D_k(\bg)$ determined by the above equation  
`LQRPA moments of inertia' and `LQRPA rotational masses', respectively.

\subsection{Collective Schr\"odinger equation}

Quantizing the collective Hamiltonian (\ref{eq:collH_BM1}) with the Pauli prescription,  
we obtain the collective Schr\"odinger equation \cite{Belyaev196517} 
\begin{align}
 \{\hat{T}_{\rm vib} + \hat{T}_{\rm rot} + V \}
 \Psi_{\alpha IM}(\bg,\Omega) = E_{\alpha I} \Psi_{\alpha IM}(\bg,\Omega), 
\label{eq:Schroedinger}
\end{align}
where  
\begin{align}
 \hat{T}_{\rm vib} =& -\frac{1}{2\sqrt{WR}} 
 \left\{ 
   \frac{1}{\beta^4} 
 \left[
 \left(
   \del_\beta 
   \beta^2 \sqrt{\frac{R}{W}} D_{\gamma\gamma} \del_\beta 
 \right)  
 \right. 
 \right.
 \nonumber \\
  & \quad \quad \quad \quad 
 \left.
 \left.
  - 
   \del_\beta 
  \left(
   \beta^2 \sqrt{\frac{R}{W}} D_{\beta\gamma} \del_\gamma
 \right)
 \right] 
 \right. 
 \nonumber \\
 & 
 \left.
 + \frac{1}{\beta^2 \sin 3\gamma} 
\left[
-\del_\gamma 
\left( 
\sqrt{\frac{R}{W}} \sin 3\gamma D_{\beta\gamma} \del_\beta 
\right) 
\right.
\right.
\nonumber \\
& \quad \quad \quad \quad 
\left.
\left.
+ \del_\gamma 
\left(
\sqrt{\frac{R}{W}} \sin 3\gamma
 D_{\beta\beta} \del_\gamma 
\right)
\right] 
\right\},
\\
\hat{T}_{\rm rot} =& \sum_{k=1}^3 \frac{ \Ihat_k^2}{2\Jc_k} 
\end{align}
with 
\begin{align}
 R(\bg) =& D_1(\bg) D_2(\bg) D_3(\bg), \\
 W(\bg) =& \left\{
 D_{\beta\beta}(\bg) D_{\gamma\gamma}(\bg) - [ D_{\beta\gamma}(\bg)]^2
\right\} \beta^{-2}.  \label{eq:Wmetric}
\end{align}

The collective wave function in the laboratory frame, 
$\Psi_{\alpha IM}(\bg,\Omega)$, 
is a function of $\beta$, $\gamma$, and a set of three Euler angles $\Omega$. 
It is specified by the total angular momentum $I$, its projection onto the
$z$-axis in the laboratory frame $M$, and $\alpha$ that distinguishes the eigenstates
possessing the same values of $I$ and $M$. 
With the rotational wave function $\Dc^I_{MK}(\Omega)$, it is written as
\begin{align}
 \Psi_{\alpha IM}(\bg,\Omega) = 
 \sum_{K={\rm even}}\Phi_{\alpha IK}(\bg)\langle\Omega|IMK\rangle,
\end{align}
where
\begin{align}
\langle \Omega | IMK \rangle = 
\sqrt{\frac{2I+1}{16\pi^2 (1+\delta_{k0})}} [\Dc^{I}_{MK}(\Omega) + (-)^I \Dc^I_{M-K}(\Omega)].
\end{align}
The vibrational wave functions in the body-fixed frame, $\Phi_{\alpha IK}(\bg)$,  
are normalized as
\begin{align}
 \int d\beta d\gamma  |\Phi_{\alpha I}(\bg)|^2 |G(\bg)|^{\frac{1}{2}} = 1, 
\end{align}
where  
\begin{align}
 |\Phi_{\alpha I}(\bg)|^2 \equiv \sum_{K={\rm even}} |\Phi_{\alpha IK}(\bg)|^2,
\end{align}
and the volume element $|G(\bg)|^{\frac{1}{2}}d\beta d\gamma$ is given by
\begin{align} \label{eq:metric}
 |G(\bg)|^{\frac{1}{2}}d\beta d\gamma = 2\beta^4 \sqrt{W(\bg)R(\bg)}
 \sin 3\gamma d\beta d\gamma.
\end{align}
Thorough discussions of their symmetries and the boundary conditions 
for solving the collective Schr\"odinger equation are given in 
Refs.~\cite{BMvol2,Belyaev196517,Kumar1967608}.

\section{Calculation of the collective potential and the collective masses
\label{sec:collHresult}}

\subsection{Details of numerical calculation} \label{sec:collHresult:model}

The CHFB+LQRPA method outlined in the preceding section may be used 
in conjunction with any effective interaction, e.g., 
density-dependent effective interactions like Skyrme forces, 
or modern nuclear density functionals.  
In this paper, as a first step toward such calculations,  we use  
a version of the P+Q force model \cite{Bes-Sorensen,Baranger1968490}   
that includes the quadrupole-pairing in addition to the monopole-pairing interaction.  
Inclusion of the quadrupole-pairing is essential, because 
neither the monopole-pairing nor the quadrupole particle-hole interaction 
contributes to the time-odd mean-field effects on the collective masses 
\cite{Dobaczewski1981123}; that is, only the quadrupole-pairing induces 
the time-odd contribution in the present model.  
Note that the quadrupole-pairing effects were not considered 
in Ref.~\cite{Dobaczewski1981123}. 
In the numerical calculation for $^{68,70,72}$Se presented below,  
we use the same notations and parameters 
as in our previous work \cite{hinohara:014305}.
The shell model space consists of two major shells ($N_{\rm sh}=3,4$) 
for neutrons and protons and the spherical single-particle energies are 
calculated using the modified oscillator potential
\cite{Bengtsson198514,Nilsson-Ragnarsson}.
The monopole-pairing interaction strengths (for neutrons and protons), $G_0^{(\tau)}$, 
and the quadrupole-particle-hole interaction strength,  $\chi$,  are determined 
such that the magnitudes of the quadrupole deformation
$\beta$ and the monopole-pairing gaps (for neutrons and protons) 
at the oblate and prolate local minima in $^{68}$Se approximately reproduce those 
obtained in the Skyrme-HFB calculations \cite{Yamagami2001579}.
The interaction strengths for $^{70}$Se and $^{72}$Se are then determined 
assuming simple mass-number dependence \cite{Baranger1968490};  
$G_0^{(\tau)}\sim A^{-1}$ and $\chi'\equiv\chi b^4 \sim A^{-\frac{5}{3}}$ 
($b$ denotes the oscillator-length parameter). 
For the quadrupole-pairing interaction strengths (for neutrons and protons),  
we use the Sakamoto-Kishimoto prescription \cite{Sakamoto1990321} 
to derive the self-consistent values.  
Following the conventional treatment of the P+Q model \cite{Baranger1965113}, 
we ignore the Fock term, so that
we use the abbreviation HB (Hartree-Bogoliubov) in place of HFB in the following. In the case of the conventional P+Q model,
the HB equation reduces to 
a simple Nilsson + BCS equation (see, e.g., Ref.~\cite{Ring-Schuck}).
The presence of the quadrupole-pairing interaction in our case 
does not allow such a reduction, however, 
and we directly solve the HB equation. 
In the P+Q model, the scaling factor $\eta$ in
Eqs. (\ref{eq:definition1}) and (\ref{eq:definition2}) 
is given by $\eta=\chi'/\hbar\omega_0 b^2$, where 
$\omega_0$ denotes the frequency of the harmonic-oscillator potential.
Effective charges, $(e_n,e_p)=(0.4,1.4)$, are used   
in the calculation of quadrupole transitions and moments.

To solve the CHB + LQRPA equations on the $(\bg)$ plane,  
we employ a two-dimensional mesh consisting of 
3600 points in the region $0<\beta<0.6$ and $0^\circ <\gamma < 60^\circ$.
Each mesh point $(\beta_i,\gamma_j)$ is represented as
\begin{align}
 \beta_i  = (i - 0.5) \times 0.01, \quad (i = 1, \cdots 60), \\
 \gamma_j = (j - 0.5) \times 1^\circ, \quad (j = 1, \cdots 60). 
\end{align}
One of the advantages of the present approach is 
that we can solve the CHB + LQRPA equations independently at 
each mesh point on the $(\bg)$ plane, so that it is suited to  
parallel computation. 

Finally, we summarize the most important differences 
between the present approach and the Baranger-Kumar approach
\cite{Baranger1968241}. 
First, as repeatedly emphasized,  
we introduce the LQRPA collective massess in place of the cranking masses. 
Second, we take into account the quadrupole-pairing force 
(in addition to the monopole-pairing force),  
which brings about the time-odd effects on the collective masses.       
Third, we exactly solve the CHB self-consistent problem, Eq.~(\ref{eq:CHFB}),  
at every point on the $(\beta, \gamma)$ plane using the gradient method, 
while in the Baranger-Kumar works the CHB Hamiltonian is replaced 
with a Nilsson-like single-particle model Hamiltonian. 
Fourth, we do not introduce the so-called core contributions to the collective masses,    
although we use the effective charges to renormalize 
the core polarization effects (outside of the model space 
consisting of two major shells) into the quadrupole operators, 
We shall see that we can well reproduce the major characteristics 
of the experimental data without introducing 
such core contributions to the collective masses.  
Fifth, most importantly, the theoretical framework developed in this paper 
is quite general, that is, it can be used in conjunction with modern density 
functionals going far beyond the P+Q force model.

\subsection{Collective potentials and pairing gaps}

\begin{figure}
\includegraphics[width=80mm]{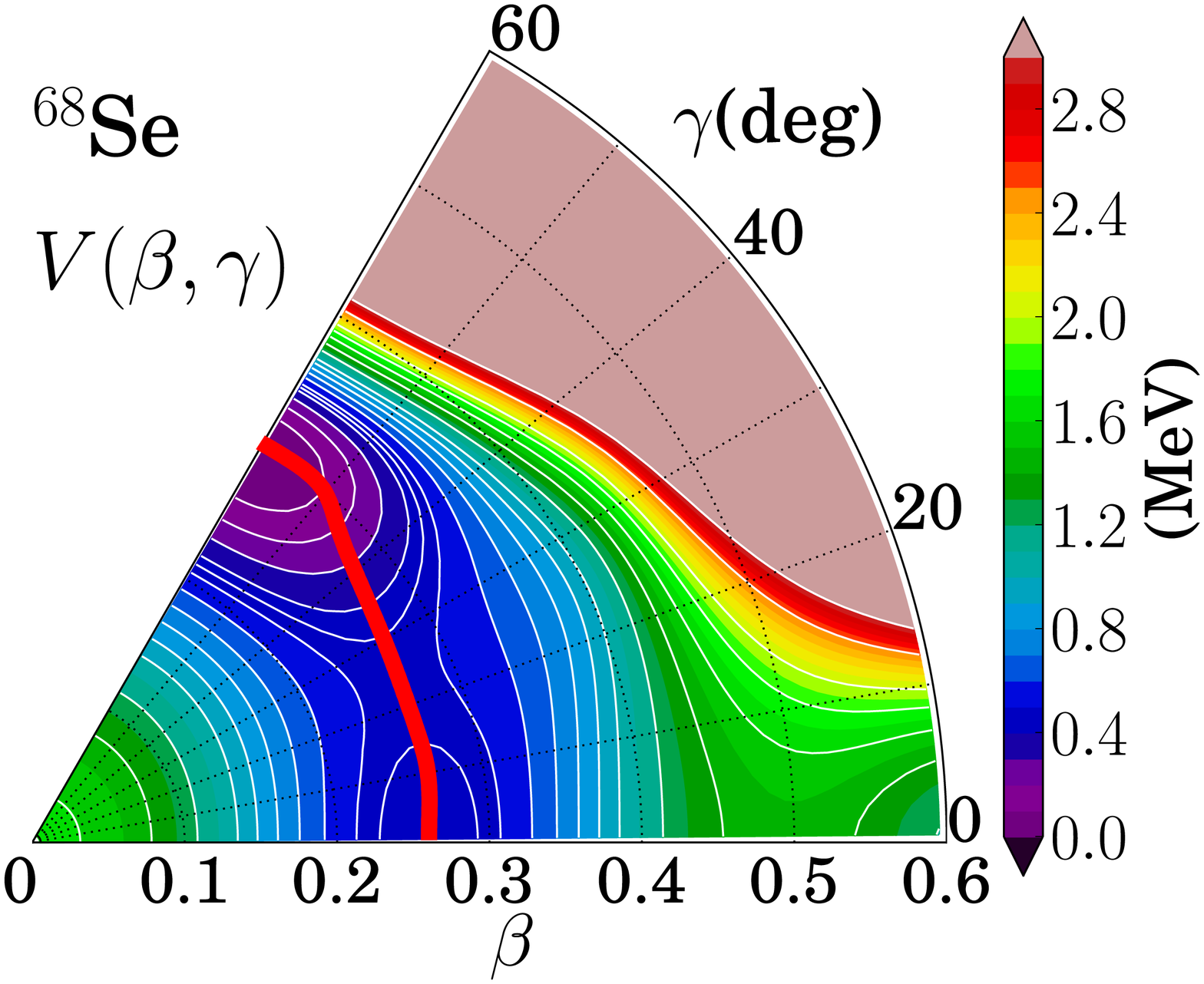} \\
\includegraphics[width=80mm]{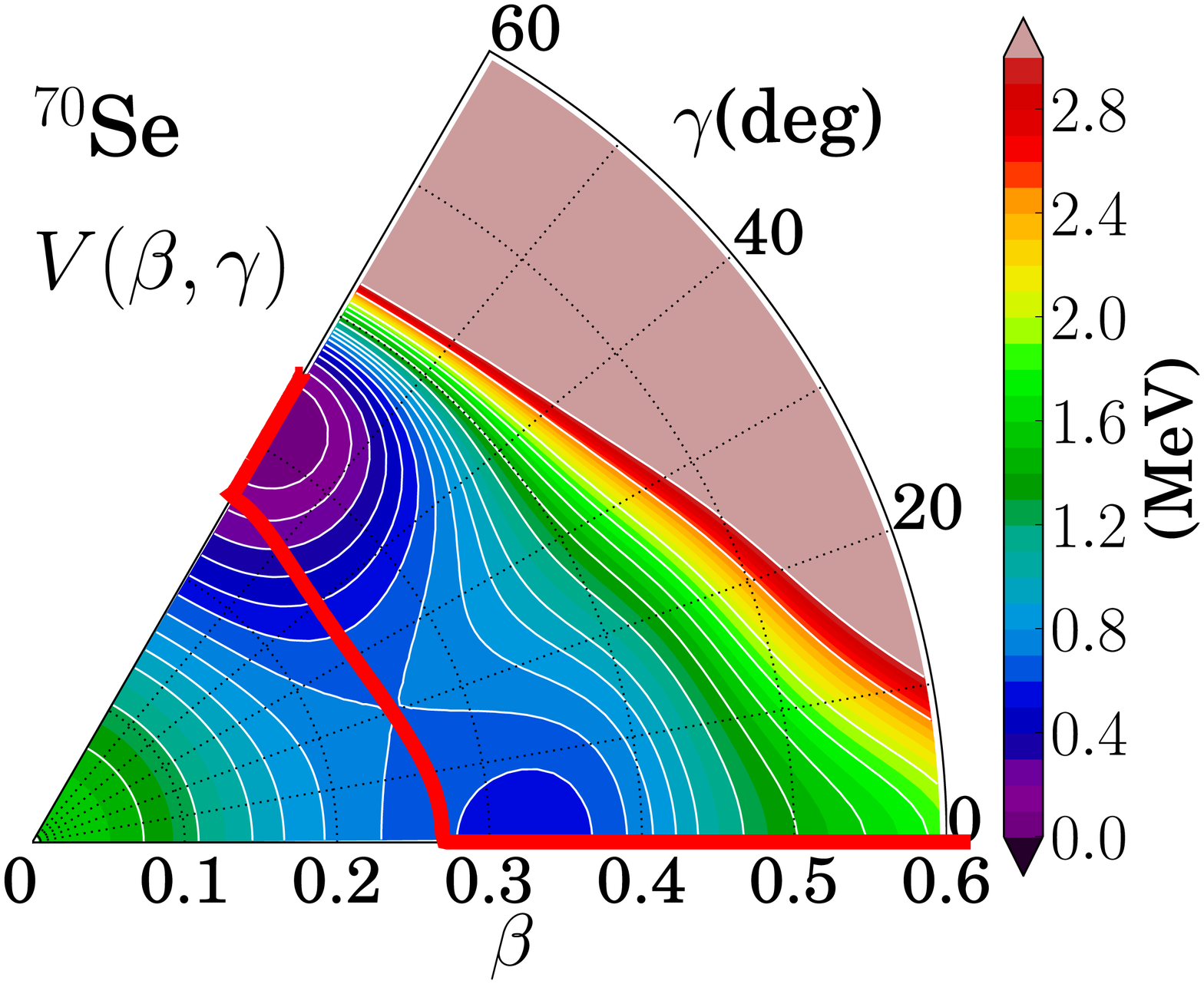} \\
\includegraphics[width=80mm]{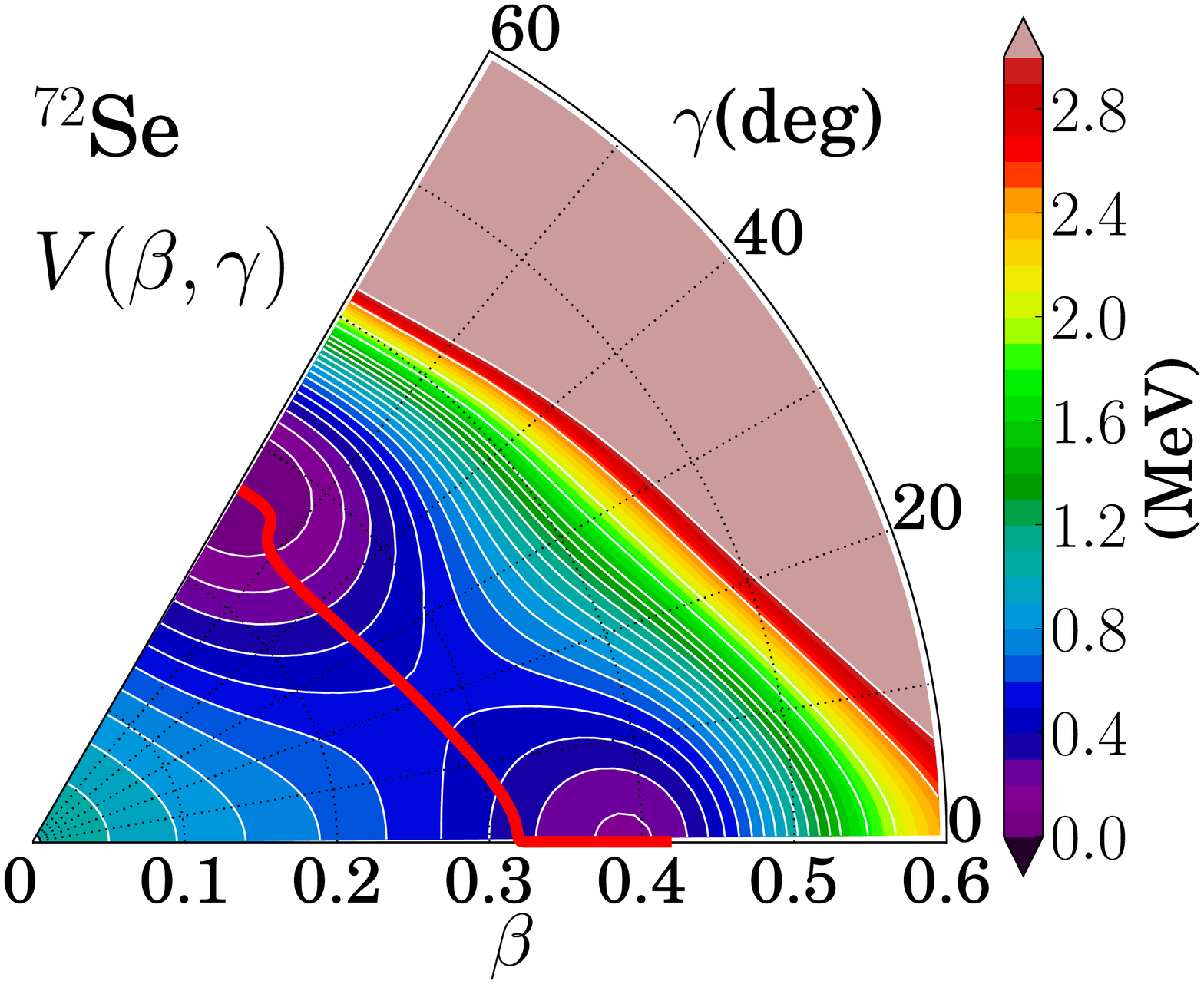}
\caption{\label{fig:V}(Color online) Collective potential $V(\bg)$ for $^{68,70,72}$Se. 
The regions higher than 3 MeV (measured from the oblate HB minima) 
are drawn by rosybrown color.  
One-dimensional collective paths connecting the oblate and prolate 
local minima are determined by using the ASCC method and depicted
with bold red lines.}
\end{figure}

We show in Fig.~\ref{fig:V} 
the collective potentials $V(\bg)$ calculated for $^{68,70,72}$Se.
It is seen that two local minima always appear both  
at the oblate ($\gamma=60^\circ$) and prolate ($\gamma=0^\circ$) shapes, 
and, in all these nuclei, the oblate minimum is lower than the prolate minimum.
The energy difference between them is, however, only several hundred keV 
and the potential barrier is low in the direction of triaxial shape
(with respect to $\gamma$) indicating $\gamma$-soft character of 
these nuclei. 
In Fig.~\ref{fig:V} 
we also show the collective paths (connecting the oblate and prolate minima) 
determined by using the 1D version of the ASCC method \cite{hinohara:014305}. 
It is seen that they always run through the triaxial valley and never 
go through the spherical shape.  

\begin{figure}
\begin{tabular}{cc}
\includegraphics[width=45mm]{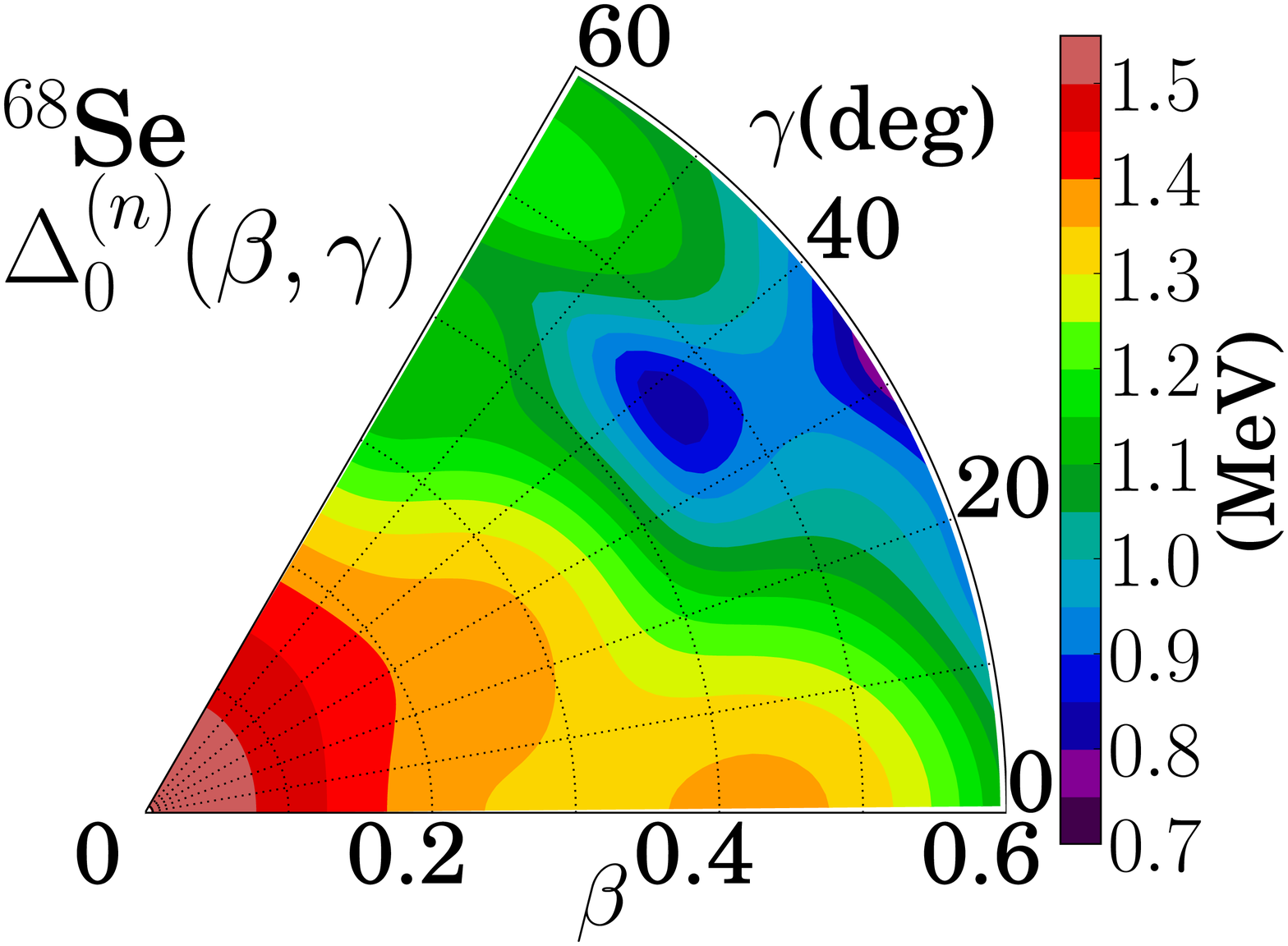} \\
\includegraphics[width=45mm]{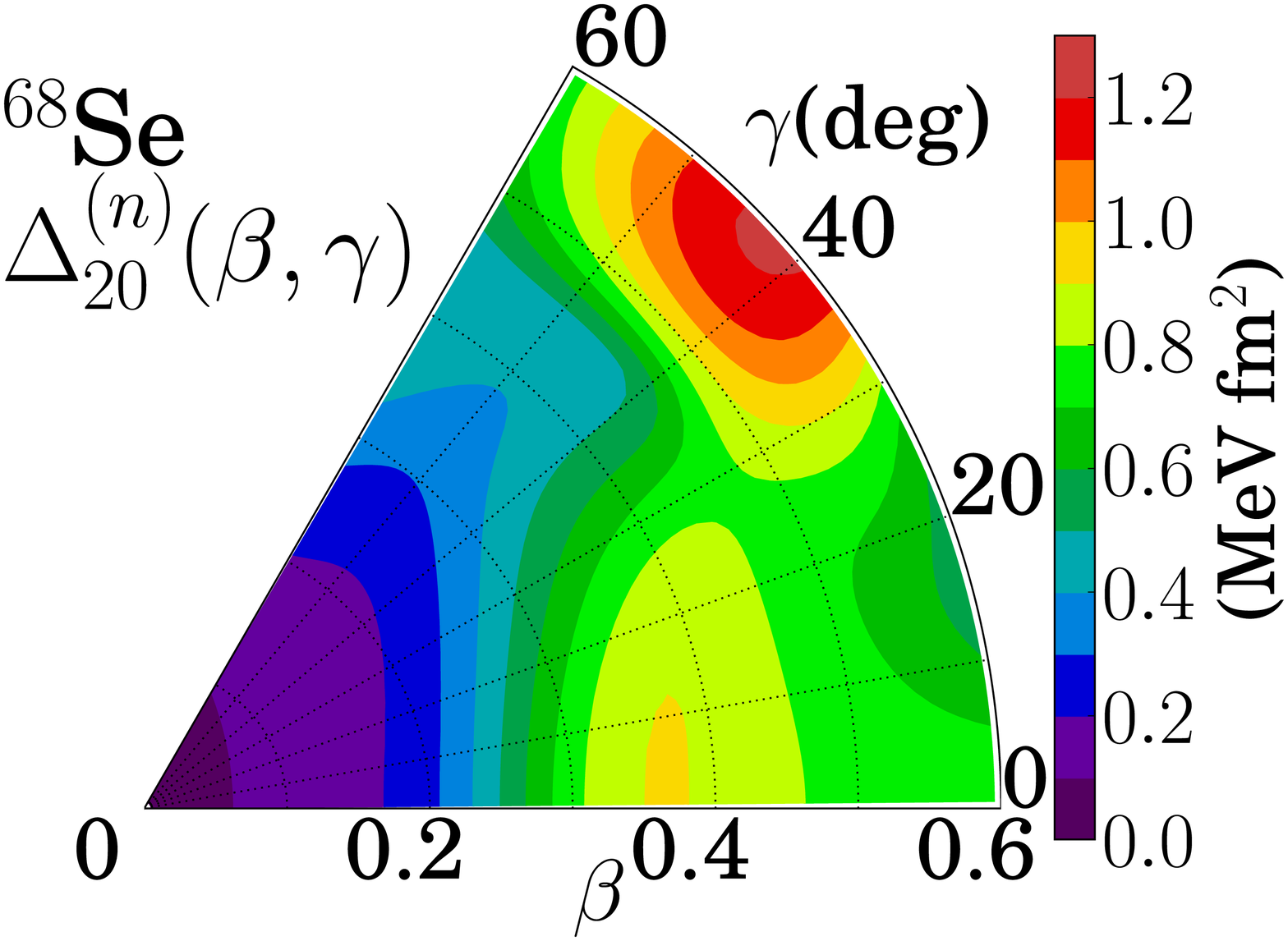} &
\includegraphics[width=45mm]{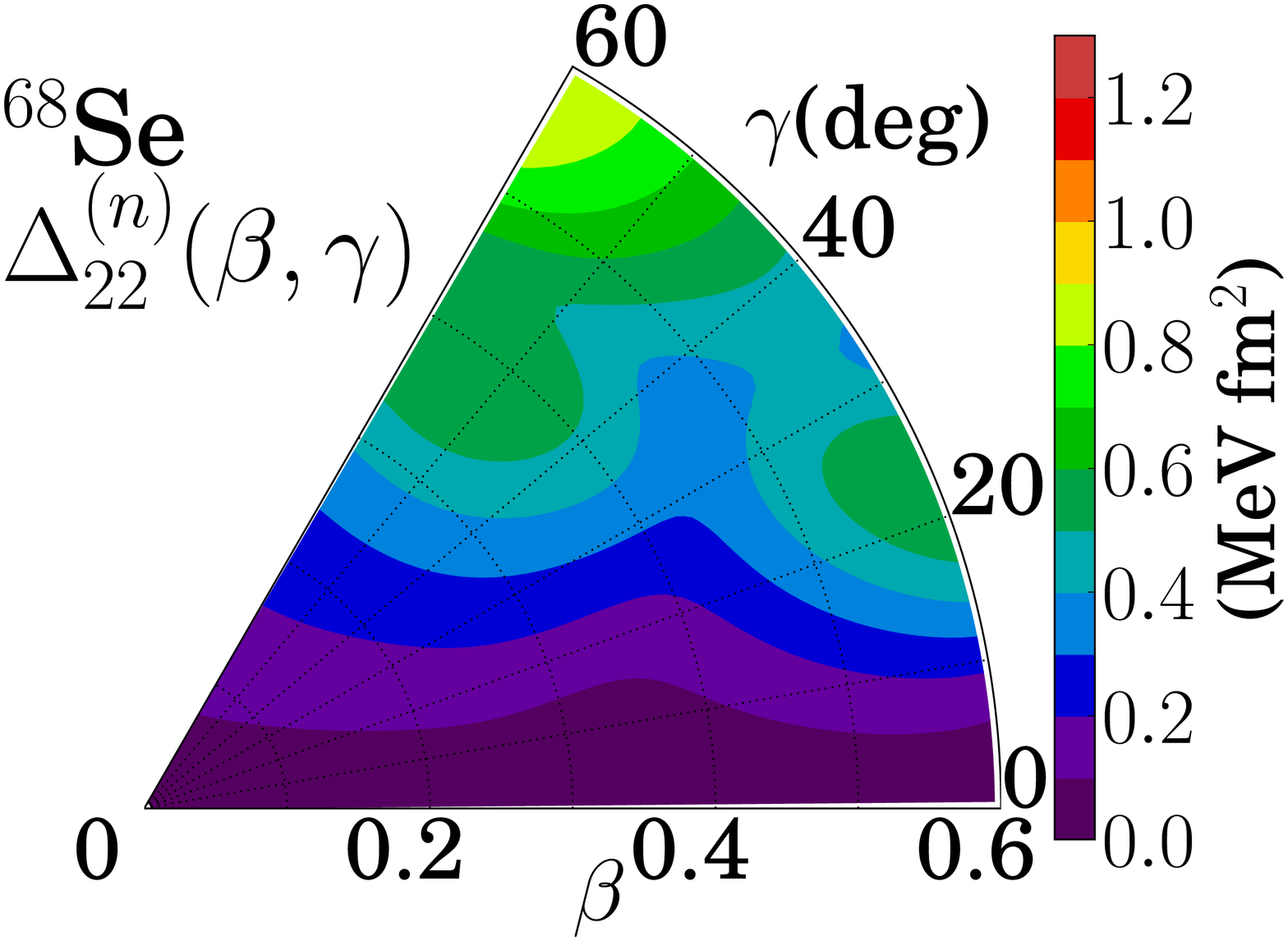}
\end{tabular}
\caption{\label{fig:gap} (Color online)
Monopole- and quadrupole-pairing gaps for neutrons of $^{68}$Se 
are plotted in the $(\bg)$ deformation plane. 
({\it upper left})~Monopole pairing gap $\Delta_{0}^{(n)}$. 
({\it lower left})~Quadrupole pairing gap $\Delta_{20}^{(n)}$.
({\it lower right})~Quadrupole pairing gap $\Delta_{22}^{(n)}$.
See Ref.~\cite{PTP.119.59} for definitions of 
$\Delta_0^{(n)}, \Delta_{20}^{(n)}$, and $\Delta_{22}^{(n)}$.}
\end{figure}

In Fig.~\ref{fig:gap}, 
the monopole- and quadrupole-pairing gaps calculated for $^{68}$Se   
are displayed. They show a significant $(\bg)$ dependence. 
Broadly speaking, the monopole pairing decreases while the quadrupole pairing 
increases as $\beta$ increases. 

\subsection{Properties of the LQRPA modes} \label{sec:collHresult:LQRPAmodes}

\begin{figure}
\includegraphics[width=70mm]{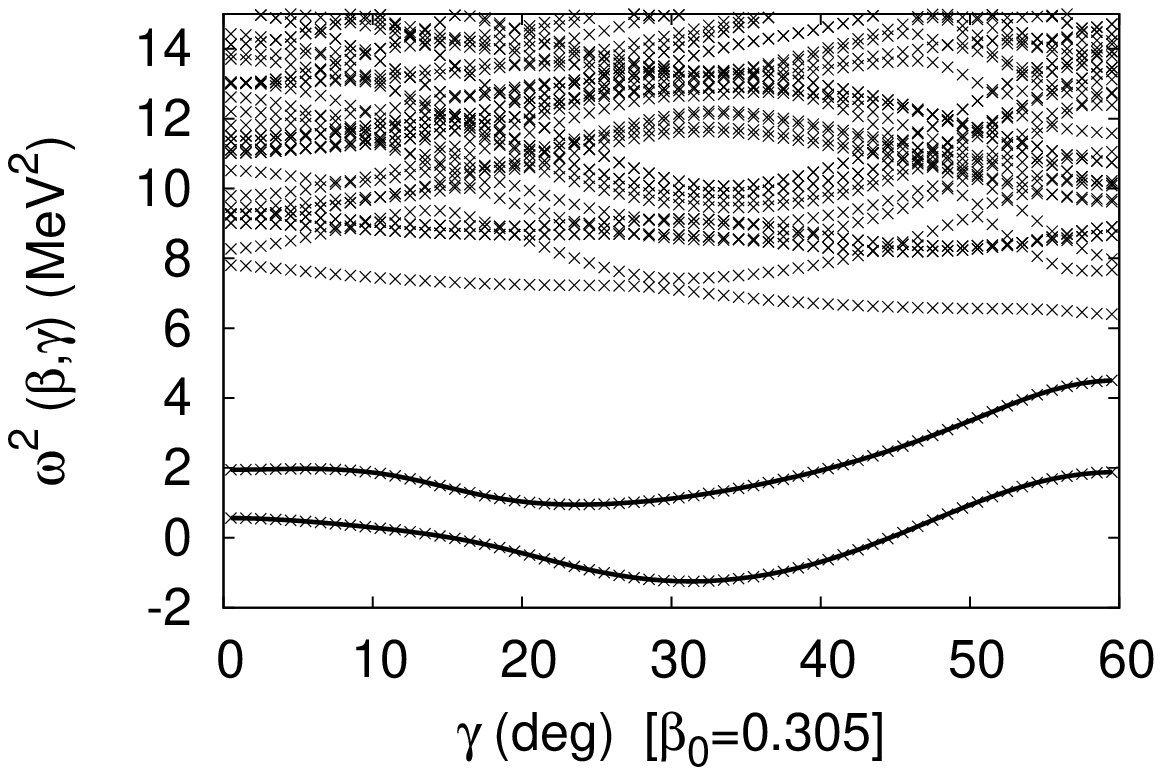} \\
\includegraphics[width=70mm]{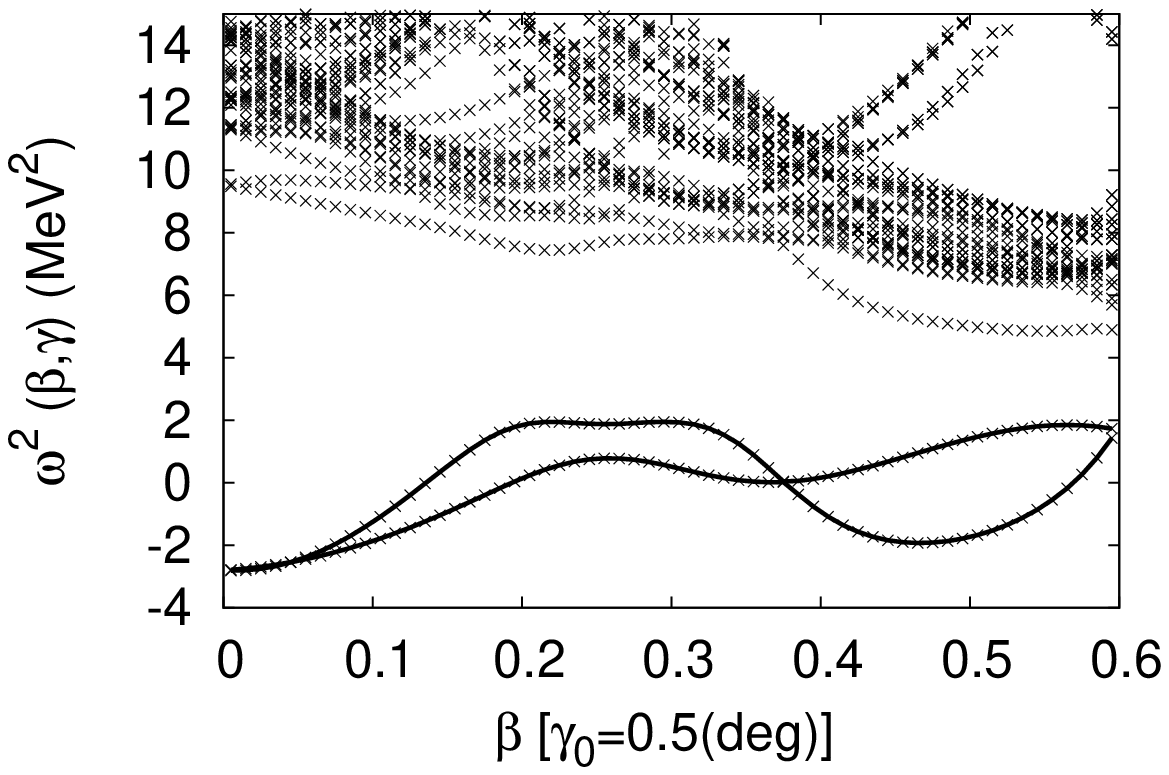} \\
\includegraphics[width=70mm]{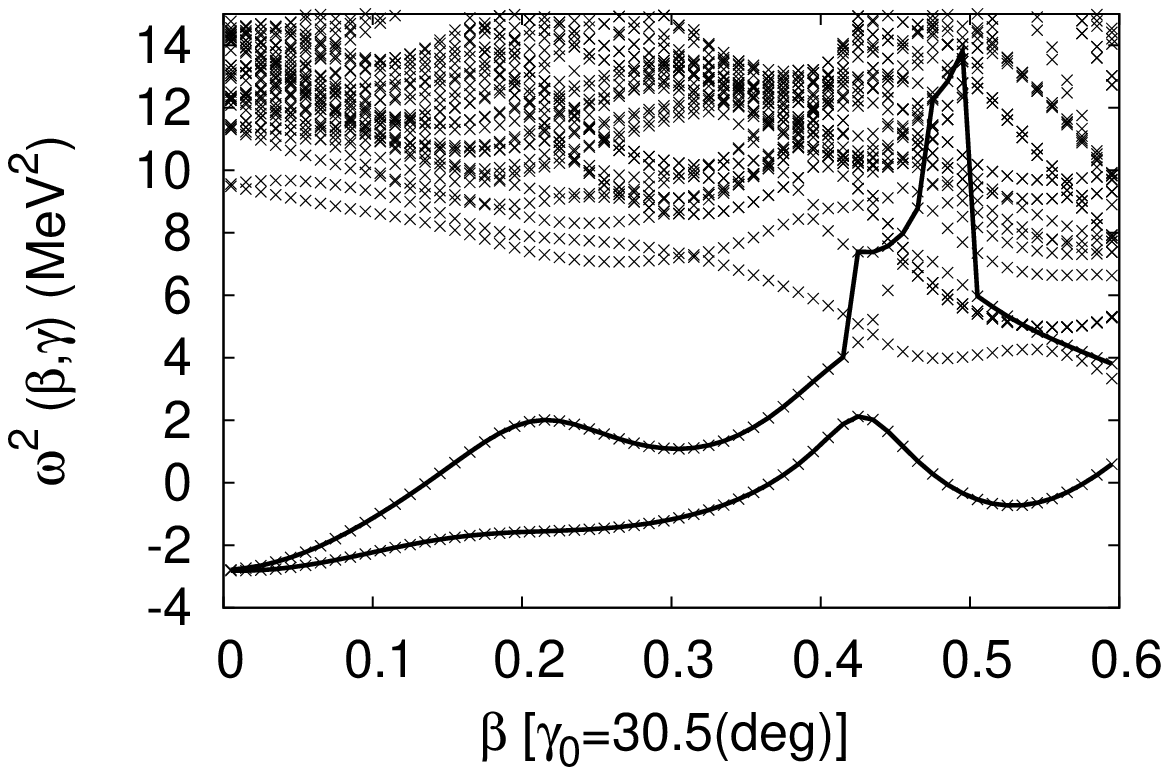}
\caption{\label{fig:omega2}
Frequencies squared $\omega^2$ of the LQRPA modes 
calculated for $^{68}$Se are plotted as functions of $\beta$ or $\gamma$.  
The LQRPA modes adopted for calculation of the vibrational masses are 
connected with solid lines. 
({\it top}) Dependence on $\gamma$ at $\beta=0.3$.  
({\it middle}) Dependence on $\beta$ along the $\gamma=0.5^\circ$ line. 
({\it bottom}) Dependence on $\beta$ along the $\gamma=30.5^\circ$ line.}
\end{figure}

In Fig.~\ref{fig:omega2} 
the frequencies squared, $\omega_i^2(\bg)$,
of various LQRPA modes calculated for $^{68}$Se are plotted 
as functions of $\beta$ and $\gamma$. 
In the region of the $(\bg)$ plane 
where the collective potential energy is less than about 5 MeV, 
we can easily identify two collective modes among many LQRPA modes,  
whose  $\omega_i^2(\bg)$ are much lower than those of other modes. 
Therefore we adopt the two lowest frequency modes to derive the collective 
Hamiltonian. This result of numerical calculation supports our assumption that 
there exists a 2D hypersurface associated with large-amplitude quadrupole shape vibrations,  
which is approximately decoupled from other degrees of freedom. 
The situation changes when the collective potential energy 
exceeds about 5 MeV and/or the monopole-pairing gap becomes small. 
A typical example is presented in the bottom panel of Fig.~\ref{fig:omega2}. 
It becomes hard to identify two collective modes well-separated
from other modes when $\beta>0.4$,  
where the collective potential energy is high (see Fig.~\ref{fig:V})
and the monopole-pairing gap becomes small (see Fig.~\ref{fig:gap}).
In this example, the second-lowest LQRPA mode in the $0.4<\beta<0.5$ 
region has pairing-vibrational character but becomes non-collective 
for $\beta>0.5$. In fact, many non-collective two-quasiparticle modes 
appear in its neighborhood. 
This region in the $(\bg)$ plane is not important, however,  because 
only tails of the collective wave function enter into this region.

\begin{figure}
 \includegraphics[width=70mm]{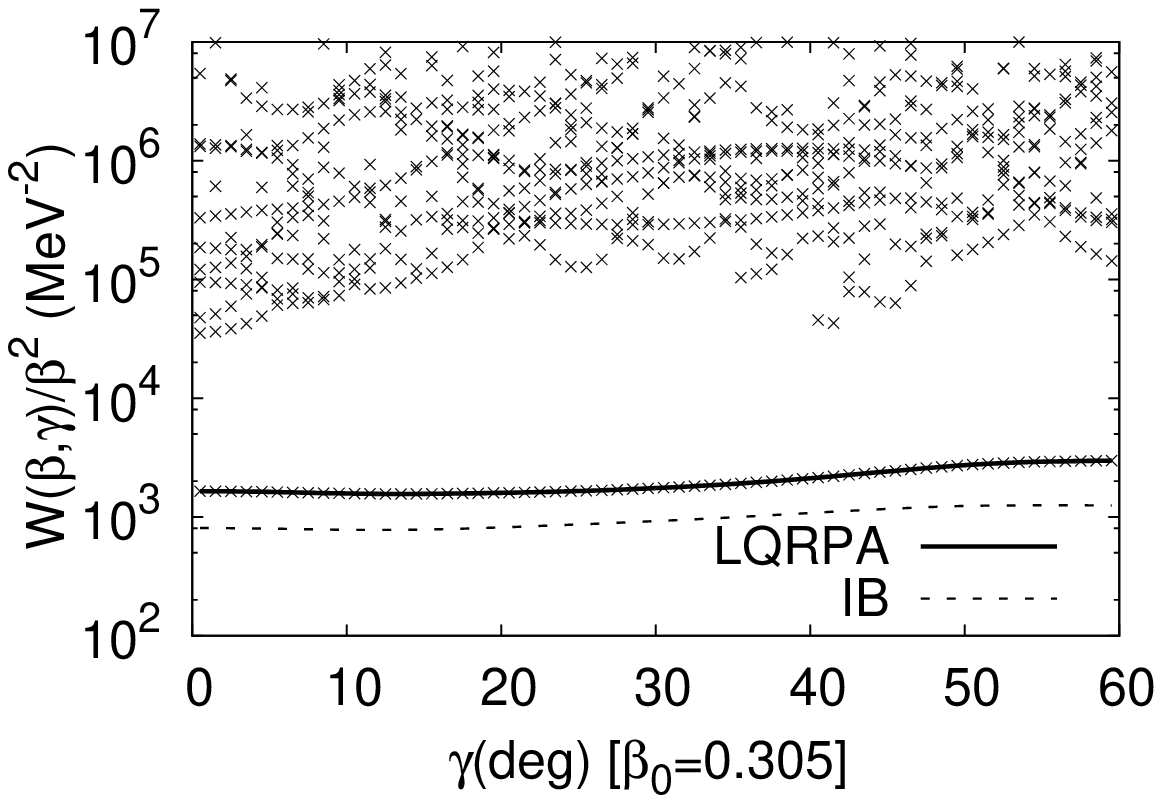} \\
 \includegraphics[width=70mm]{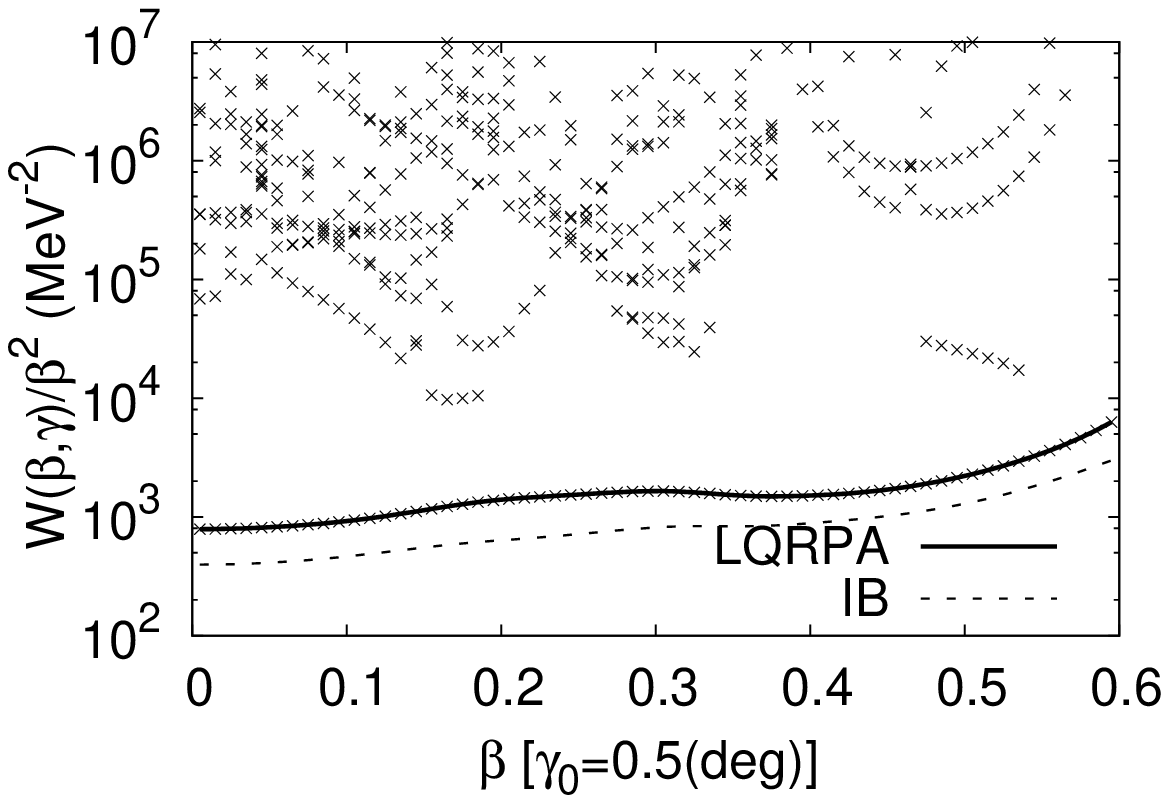} \\
 \includegraphics[width=70mm]{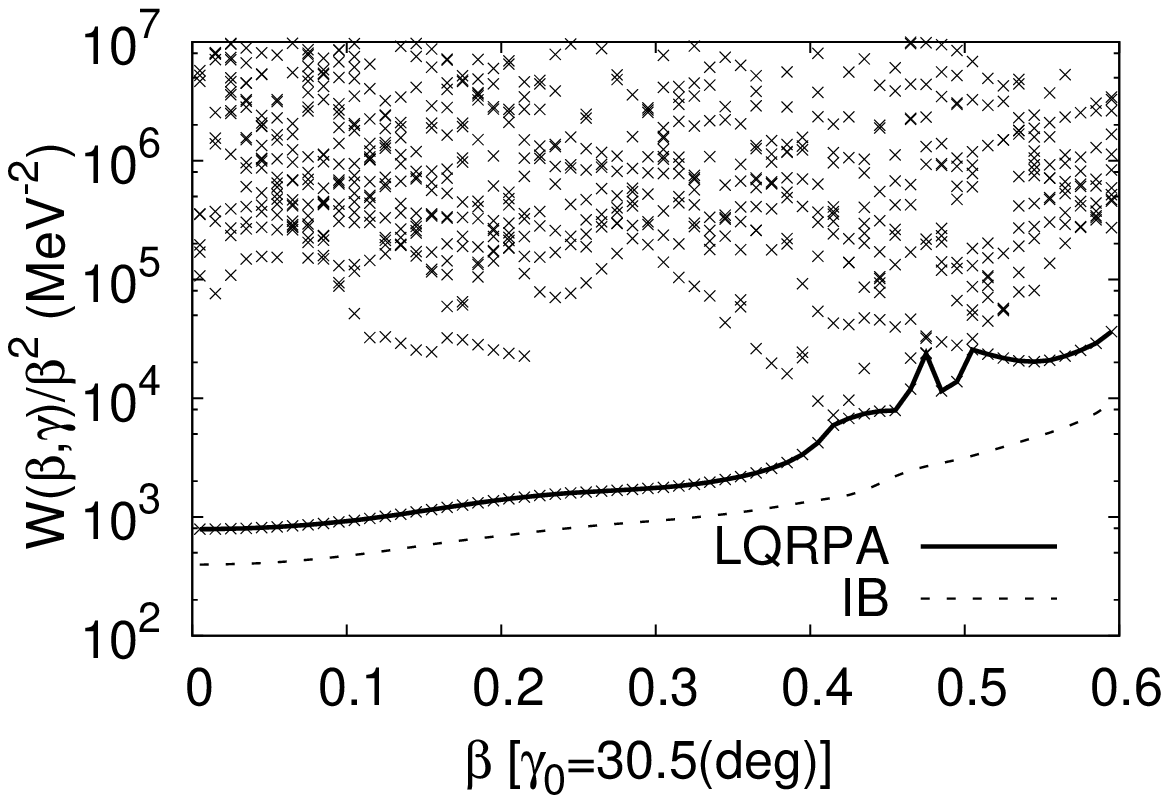}
\caption{\label{fig:metric}
Dependence on $\beta$ and $\gamma$ 
of the vibrational part of the metric $W(\bg)$ calculated for $^{68}$Se.
({\it top}) Dependence on $\gamma$ at $\beta=0.3$.  
({\it middle}) Dependence on $\beta$ along the $\gamma=0.5^\circ$ line.
({\it bottom}) Dependence on $\beta$ along the $\gamma=30.5^\circ$ line.
The cross symbols indicate values of the vibrational metric calculated 
for various choices of two LQRPA modes from among the lowest 40 LQRPA modes; 
the lowest mode is always chosen and the other is from the remaining 39 modes.
The smallest vibrational metric is shown by solid line. 
For reference, 
the vibrational metric calculated using the IB vibrational mass is 
indicated by broken lines.}
\end{figure}

It may be useful to set up a prescription that works even in a difficult situation 
where it is not apparent how to choose two collective LQRPA modes.   
We find that the following prescription always works well for selecting 
two collective modes among many LQRPA modes.   
This may be called a minimal metric criterion. 
At each point on the $(\bg)$ plane,  
we evaluate the vibrational part of the metric $W(\bg)$
given by Eq.~(\ref{eq:Wmetric})  
for all combinations of two LQRPA modes, 
and find the pair that gives the minimum value.  
We show in Fig.~\ref{fig:metric} 
how this prescription actually works. 
In this figure, the $W(\bg)$ values are plotted as functions of $\beta$ and $\gamma$ 
for many pairs of the LQRPA modes. 
In the situations where the two lowest-frequency LQRPA modes are well separated 
from other modes, this prescription gives the same results as  
choosing the two lowest-frequency modes (see the top and middle panels). 
On the other hand, a pair of the LQRPA modes different from the lowest two 
modes is chosen by this prescription in the region mentioned above (the bottom panel).
This choice may be better than that using the lowest-frequency criterion,
because we often find that a normal mode of pairing vibrational character  
becomes the second lowest LQRPA mode when the monopole-pairing gap 
significantly decreases in the region of large $\beta$. 
The small values of the vibrational metric implies that 
the direction of the infinitesimal displacement associated with 
the pair of the LQRPA modes has a large projection onto the  ($\bg$) plane.  
Therefore, this prescription may be well suited to our purpose of deriving 
the collective Hamiltonian for the $(\bg)$ variables.
It remains as an interesting open question for future to examine 
whether or not the explicit inclusion of the pairing vibrational degree 
of freedom as another collective variable will give us a better description 
in such situations.

\subsection{Vibrational masses}

\begin{figure}
 \includegraphics[width=70mm]{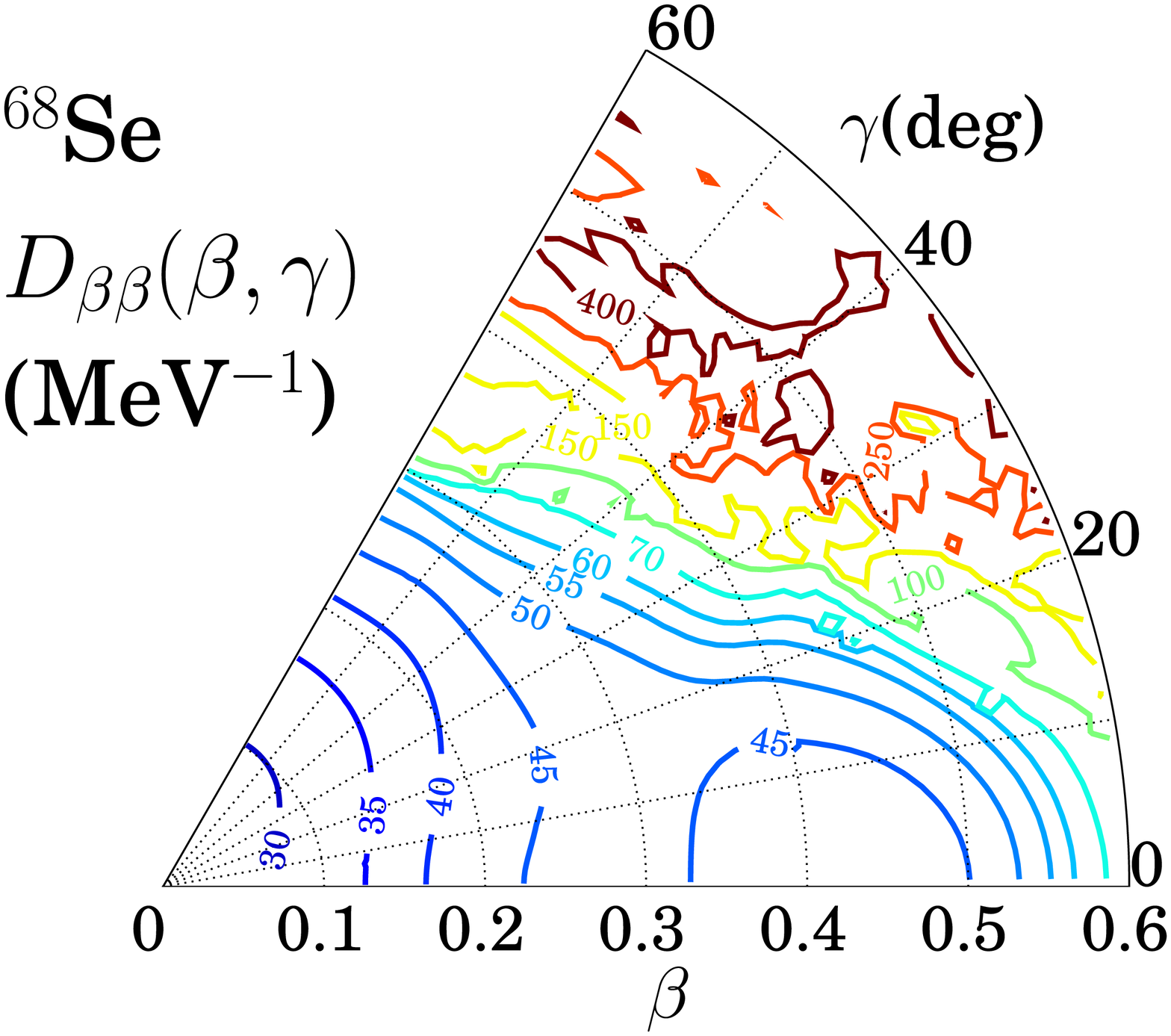} \\ \mbox{} \\
 \includegraphics[width=70mm]{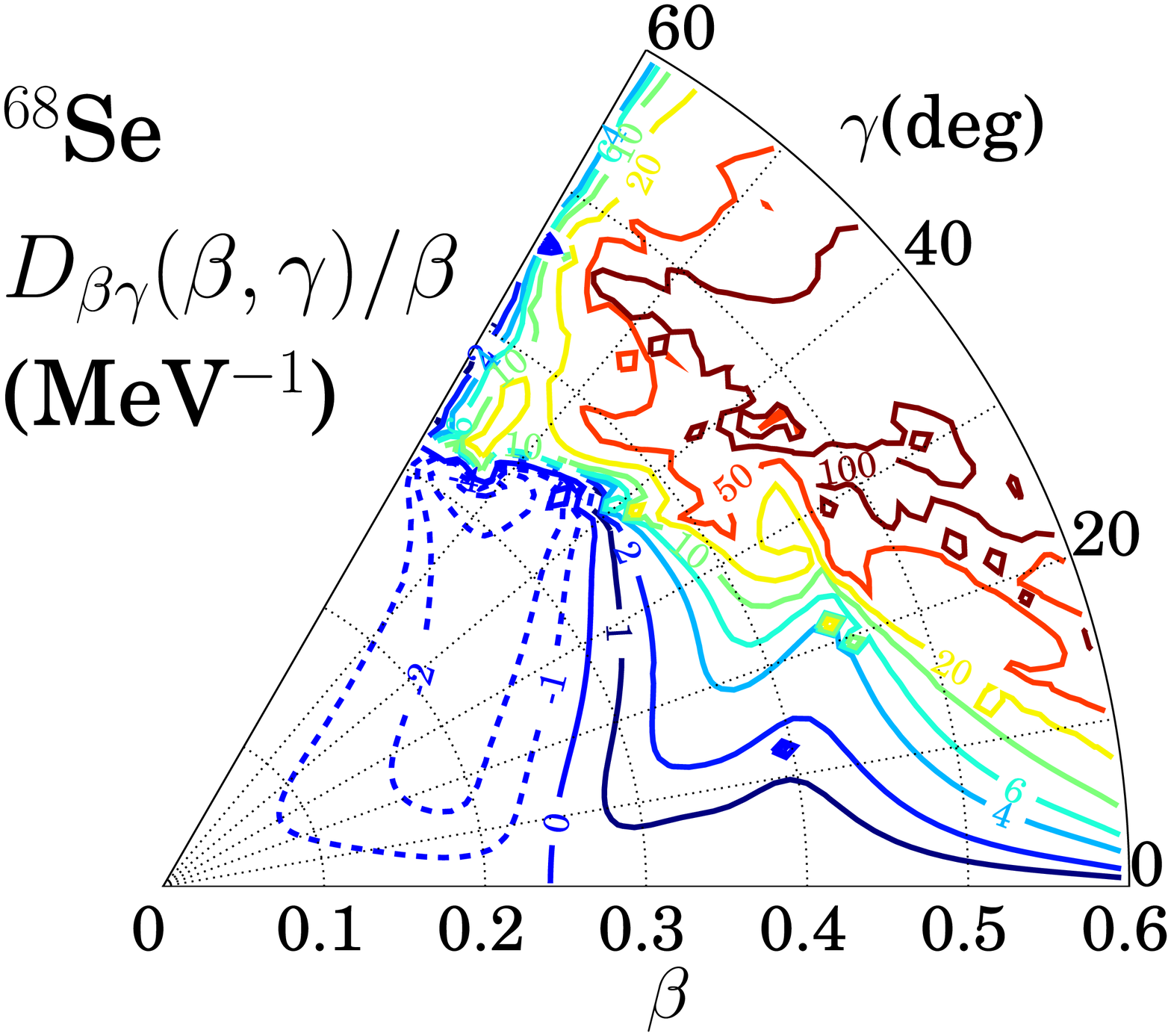} \\ \mbox{} \\
 \includegraphics[width=70mm]{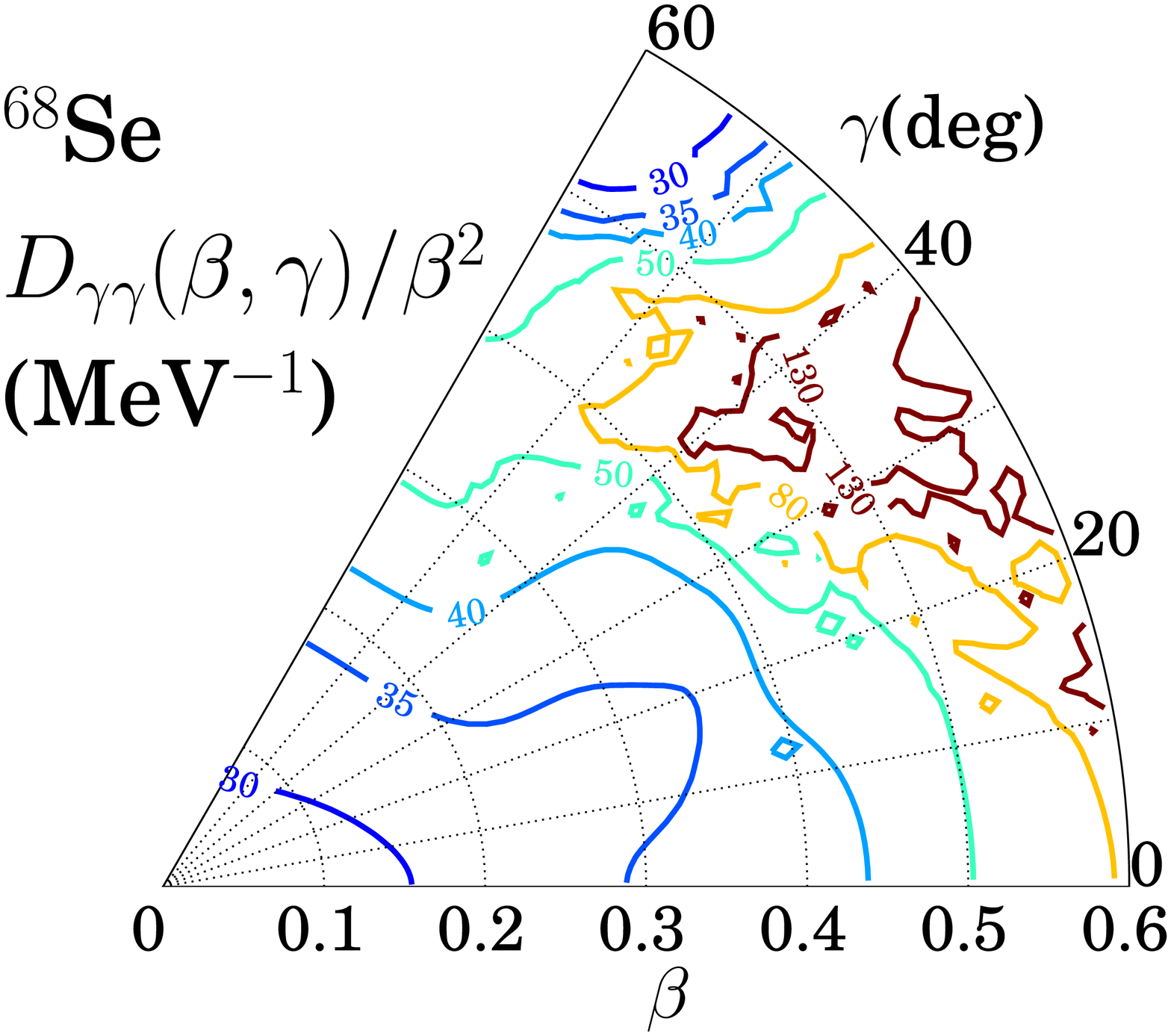} 
\caption{\label{fig:mass}(Color online)
Vibrational masses, $D_{\beta\beta}(\bg)$, $D_{\beta\gamma}(\bg)/\beta$,
and $D_{\gamma\gamma}(\bg)/\beta^2$, in unit of MeV$^{-1}$ 
calculated for $^{68}$Se.}
\end{figure}

In Fig.~\ref{fig:mass} 
the vibrational masses calculated for $^{68}$Se are displayed.
We see that their values exhibit a significant variation in the $(\bg)$ plane. 
In particular, the increase in the large $\beta$ region is remarkable.  

\begin{figure}
 \includegraphics[width=70mm]{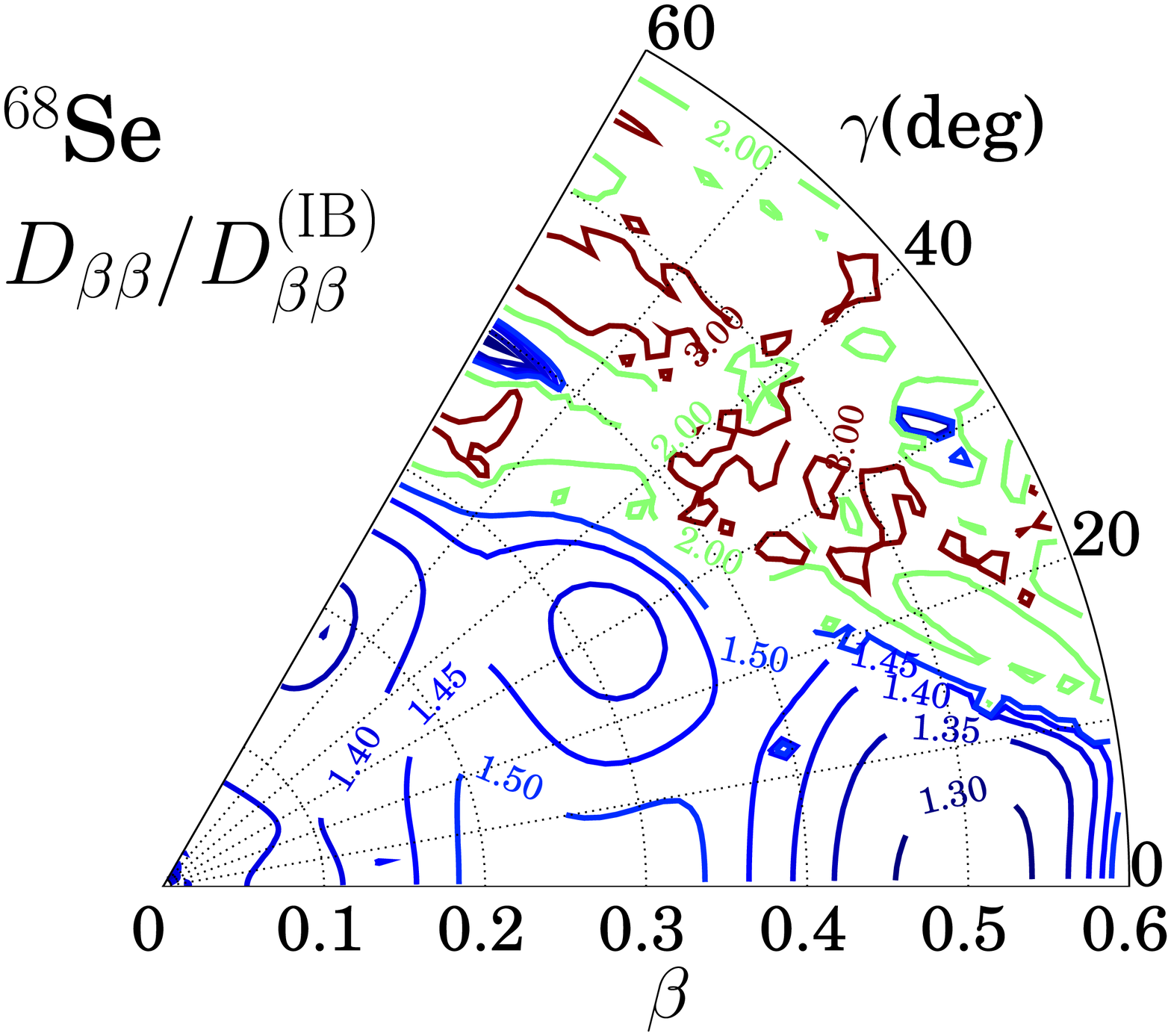} \\ \mbox{} \\
 \includegraphics[width=70mm]{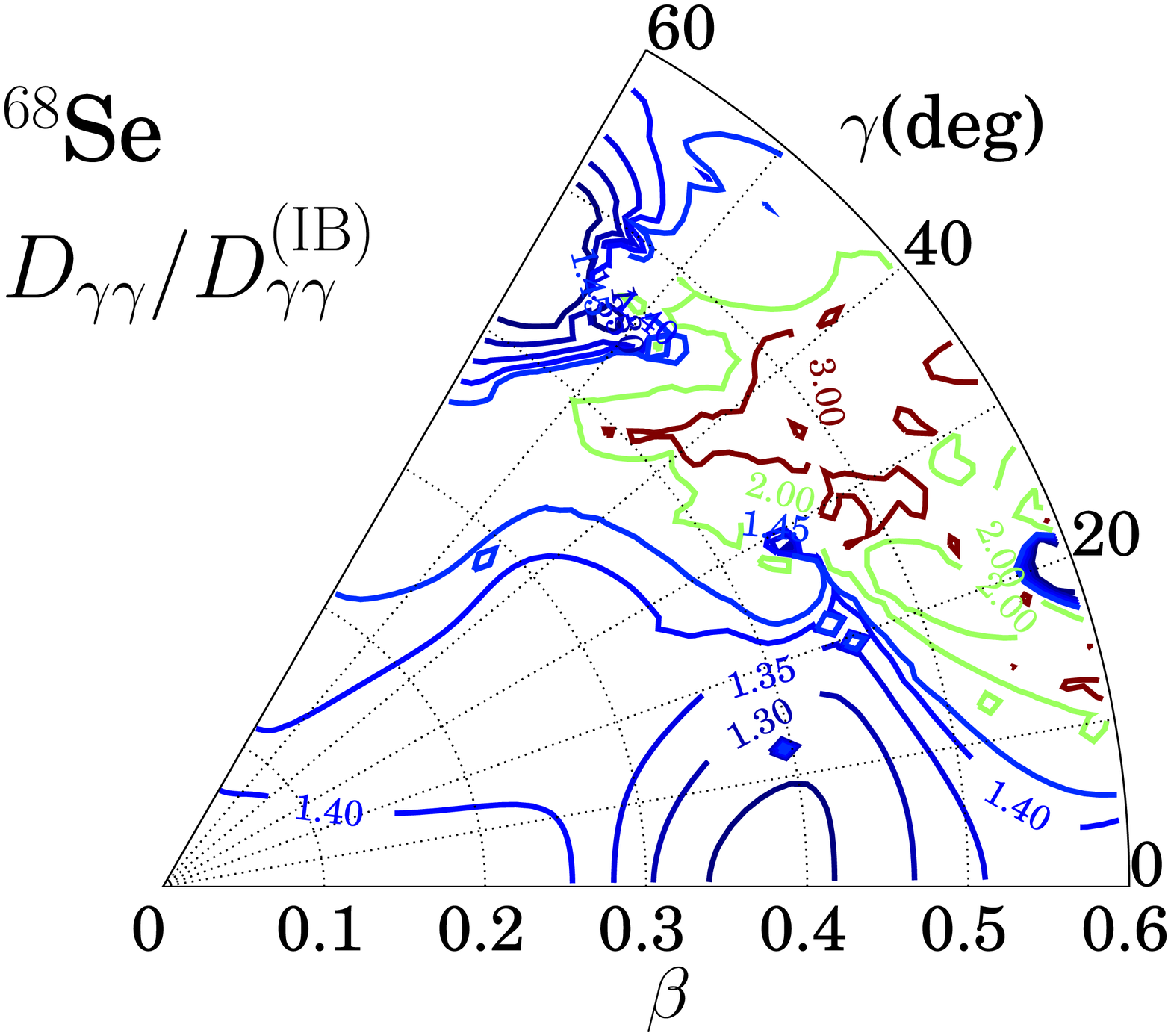}
\caption{\label{fig:massratio}(Color online) 
Ratios of the LQRPA vibrational masses to the IB vibrational masses, 
$D_{\beta\beta}   / D^{({\rm IB})}_{\beta\beta}$ and 
$D_{\gamma\gamma} / D^{({\rm IB})}_{\gamma\gamma}$, calculated for $^{68}$Se.  }
\end{figure}

Figure~\ref{fig:massratio} shows how the ratios of the LQRPA vibrational masses 
to the IB vibrational masses vary on the $(\bg)$ plane.
It is clearly seen that the LQRPA vibrational masses are considerably  
larger than the IB vibrational masses and   
their ratios change depending on $\beta$ and $\gamma$.
In this calculation, the IB vibrational masses are evaluated 
using the well-known formula: 
\begin{align}
  D^{(\rm IB)}_{\xi_i \xi_j}(\bg) = 2\sum_{\mu\bar{\nu}} \frac
{\bra{\mu\bar{\nu}} \displaystyle\frac{\del\Hhat_{\rm CHB}}{\del \xi_i}\ket{0 }
 \bra{0           } \displaystyle\frac{\del\Hhat_{\rm CHB}}{\del \xi_j}\ket{\mu\bar{\nu}} }
{[ E_\mu(\bg) + E_{\bar{\nu}}(\bg)]^3 }, \\ 
(\xi_i=\beta~{\rm or}~\gamma) \nonumber
\end{align}
where 
$E_\mu(\bg)$, $\ket{0}$, and $\ket{\mu\bar{\nu}}$ denote 
the quasiparticle energy,  the CHB state $\ket{\phi(\bg)}$ and 
the two-quasiparticle state~~$\adag_{\mu}\adag_{\bar{\nu}} \ket{\phi(\bg)}$, respectively 
(see Ref.~\cite{PTP.119.59} for the meaning of the indices $\mu$ and $\bar{\nu}$). 

The vibrational masses calculated for $^{70,72}$Se exhibit behaviors similar  
to those for $^{68}$Se.

\subsection{Rotational masses}

\begin{figure}
\begin{tabular}{ccc}
 \includegraphics[width=70mm]{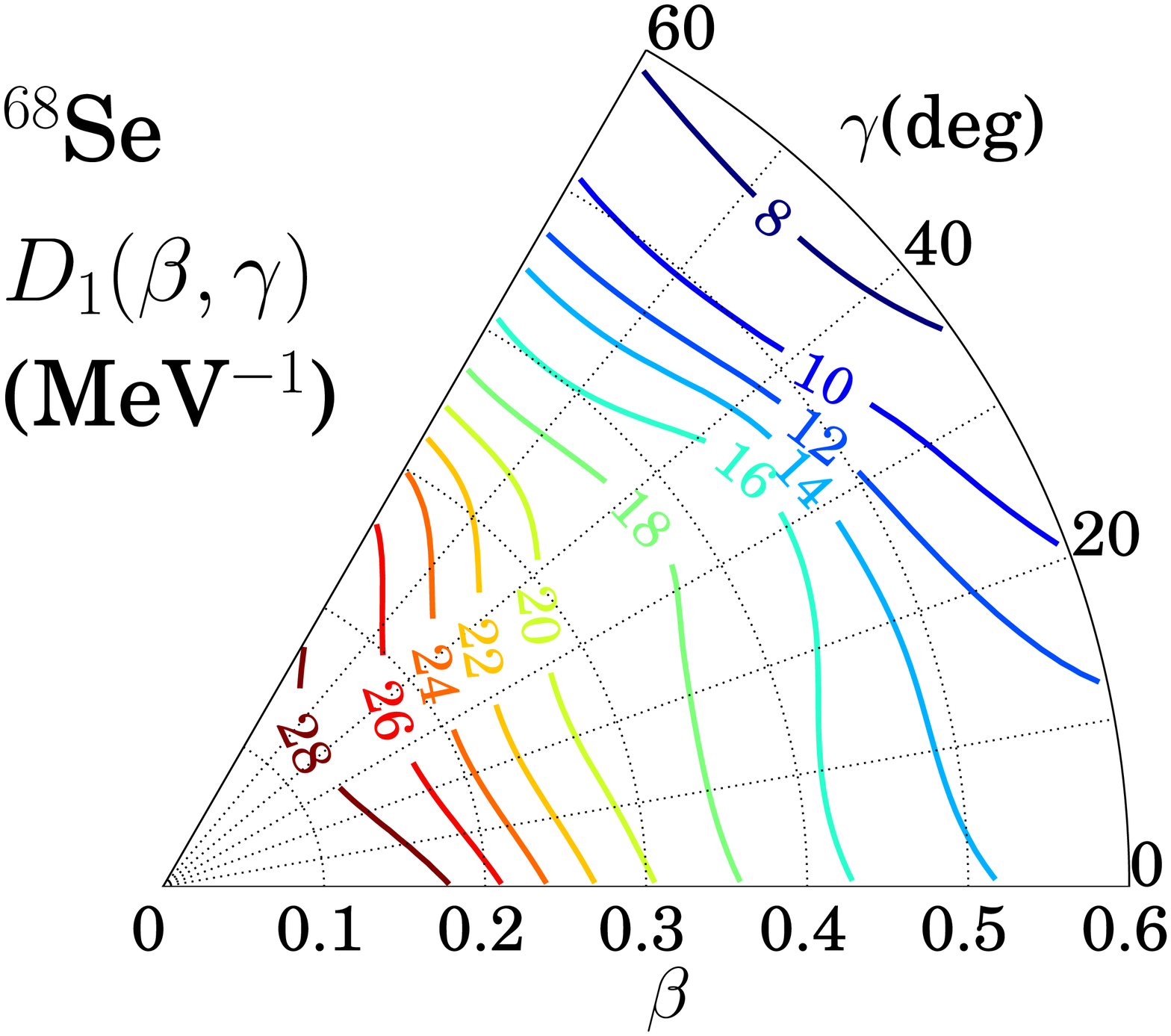} \\ \mbox{} \\
 \includegraphics[width=70mm]{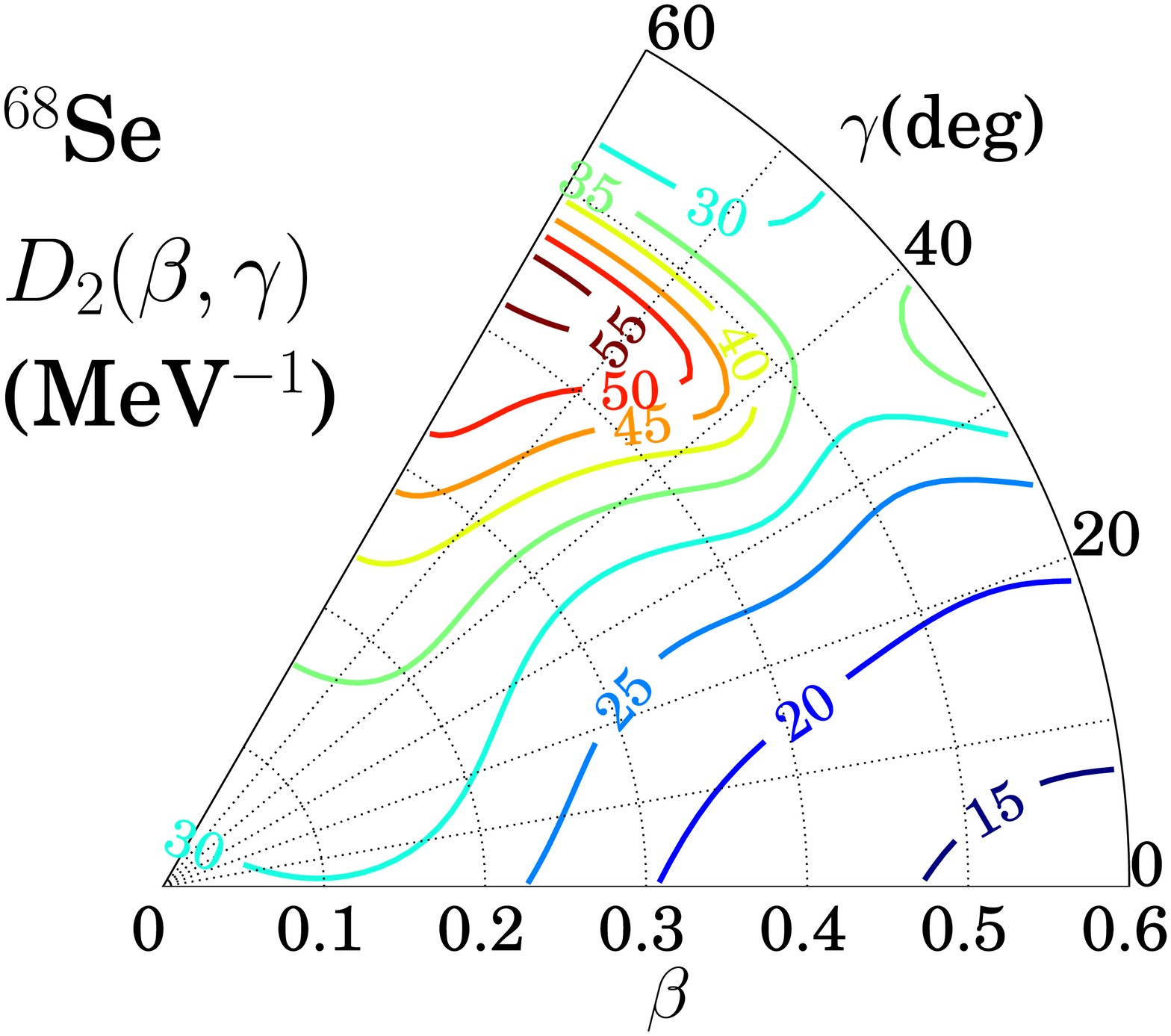} \\ \mbox{} \\
 \includegraphics[width=70mm]{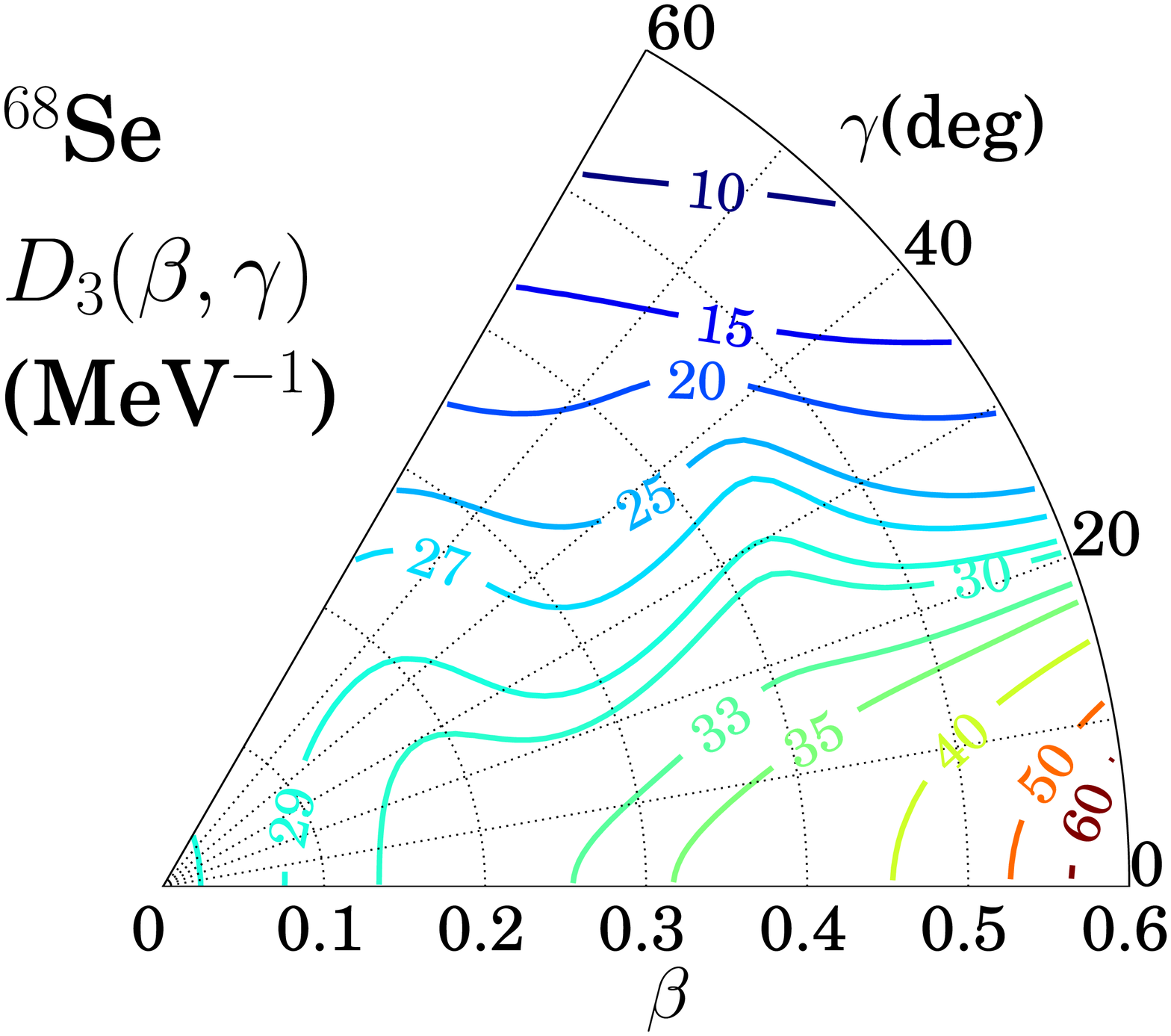}
\end{tabular}
\caption{\label{fig:MOI}(Color online) 
Rotational masses $D_k(\bg)$ in unit of MeV$^{-1}$, calculated for
 $^{68}$Se.
See Eq.~(\ref{eq:MOI}) for the relation with the rotational moments of inertia $\Jc_k(\bg)$.}
\end{figure}

In Fig.~\ref{fig:MOI}, 
the LQRPA rotational masses $D_k(\bg)$  calculated for $^{68}$Se
are displayed. 
Similarly to the vibrational masses discussed above, 
the LQRPA rotational masses also exhibit a remarkable variation 
over the $(\bg)$ plane, 
indicating a significant deviation from the irrotational property.  

\begin{figure}
 \includegraphics[width=70mm]{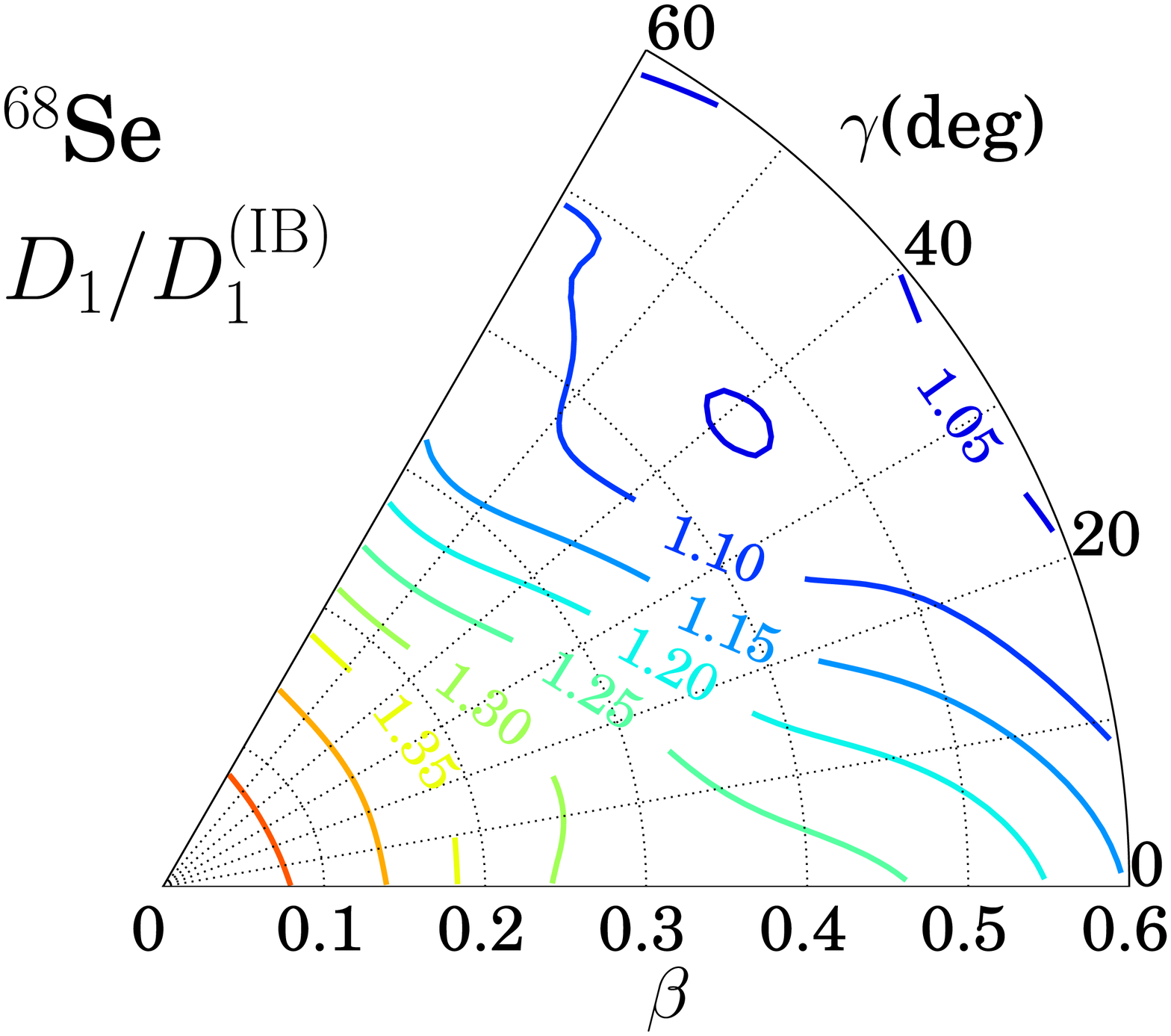} \\ \mbox{} \\
 \includegraphics[width=70mm]{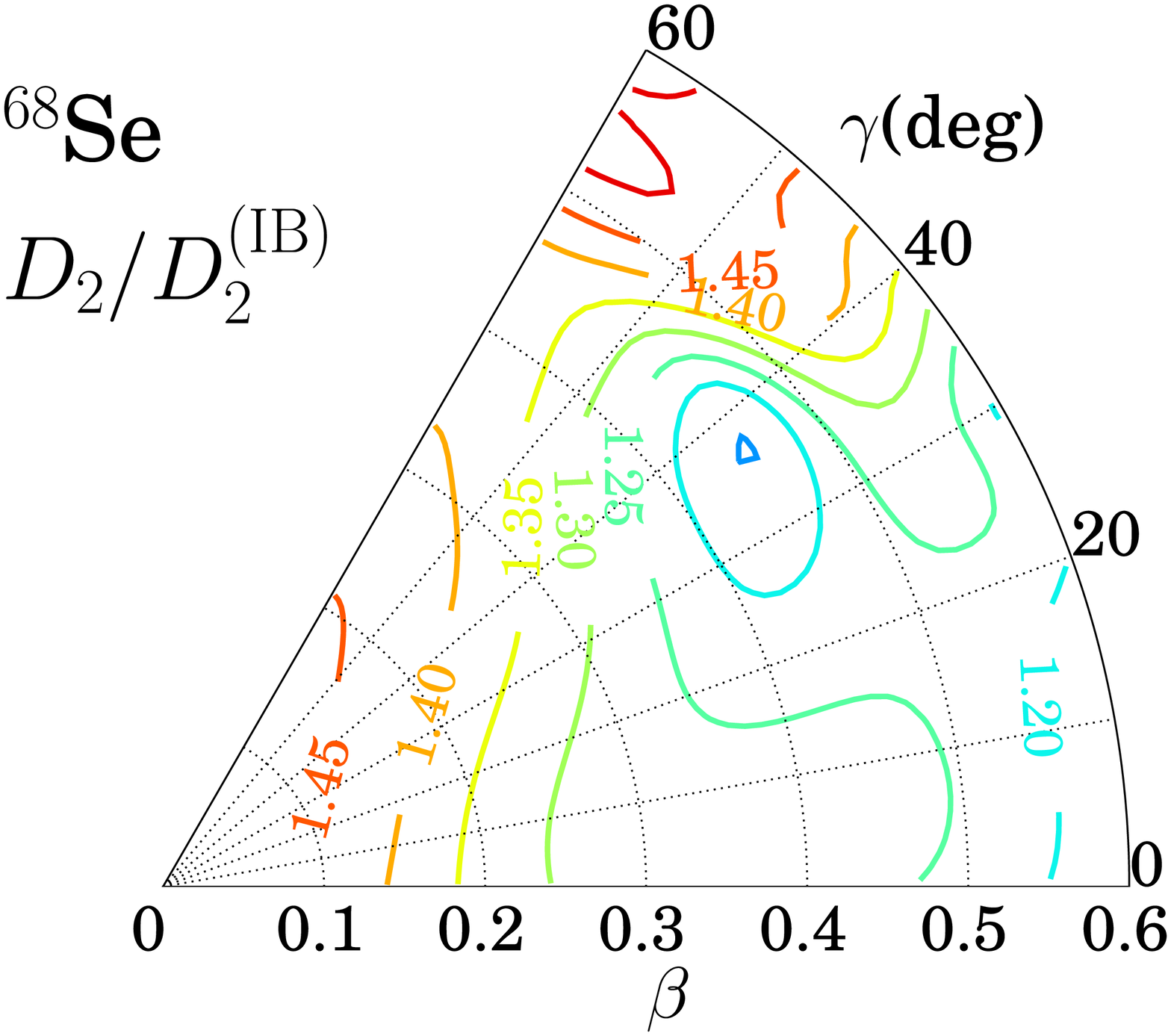} \\ \mbox{} \\
 \includegraphics[width=70mm]{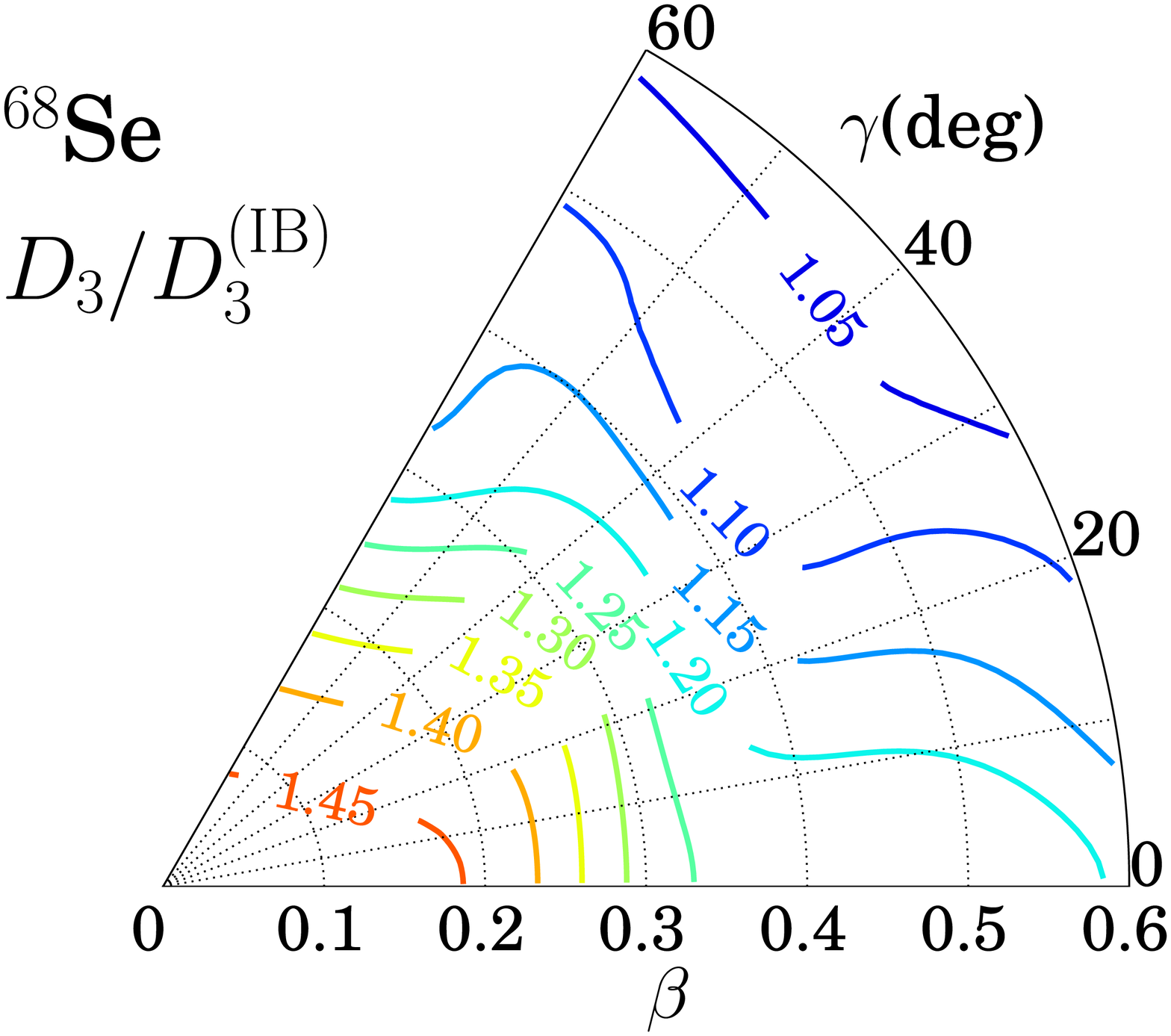}
\caption{\label{fig:MOIratio}(Color online) 
Ratios of the LQRPA rotational masses to the IB rotational masses,  
$D_k(\bg)/D^{(\rm IB)}_k(\bg)$, calculated for $^{68}$Se.}
\end{figure}

Figure~\ref{fig:MOIratio}  
shows how the ratios of the LQRPA rotational masses $D_k(\bg)$ to 
the IB cranking masses $D^{(\rm IB)}_k(\bg)$ vary on the $(\bg)$ plane.
The rotational masses calculated for $^{70,72}$Se exhibit behaviors similar  
to those for $^{68}$Se. 

As we have seen in Figs.~\ref{fig:mass}--\ref{fig:MOIratio}, 
not only the vibrational and rotational masses 
but also their ratios to the IB cranking masses  
exhibit an intricate dependence on $\beta$ and $\gamma$. 
For instance, it is clearly seen that the ratios, 
$D_k(\bg)/D^{(\rm IB)}_k(\bg)$, gradually increase as $\beta$ decreases. 
This result is consistent with the calculation by Hamamoto and Nazarewicz   
\cite{PhysRevC.49.2489}, 
where it is shown that the ratio of the Migdal term to the cranking term 
in the rotational moment of inertia (about the 1st axis) 
increases as $\beta$ decreases.  
Needless to say, the Migdal term 
(also called the Thouless-Valation correction) 
corresponds to the time-odd mean-field contribution 
taken into account in the LQRPA rotational masses, 
so that the result of Ref.~\cite{PhysRevC.49.2489} implies that the ratio 
$D_1(\bg)/D^{(\rm IB)}_1(\bg)$, increases as $\beta$ decreases, 
in agreement with our result. 
To understand this behavior, it is important to note that,  
in the present calculation, 
the dynamical effect of the time-odd mean-field on $D_1(\bg)$
is associated with the $K=1$ component of the quadrupole-pairing interaction 
and it always works and increase the rotational masses,  
in contrast to the behavior of the static quantities like  
the magnitude of the quadrupole-pairing gaps, 
$\Delta_{20}$ and $\Delta_{22}$, 
which diminish in the spherical shape limit. 
Obviously, this qualitative feature holds true 
irrespective of details of our choice of 
the monopole and quadrupole pairing interaction strengths. 

The above results of calculation obviously indicate 
the need to take into account the time-odd contributions to 
the vibrational and rotational masses 
by going beyond the IB cranking approximation. 
In Refs.~\cite{li:054301,niksic:092502,niksic:034303,PhysRevC.81.034316}, 
a phenomenological prescription is adopted 
to remedy the shortcoming of the IB cranking masses;    
that is, a constant factor in the range 1.40-1.45 is multiplied to the 
IB rotational masses. 
This prescription is, however, insufficient in the following points.
First, the scaling only of the rotational masses (leaving the vibrational masses aside)   
violates the symmetry requirement for the 5D collective quadrupole Hamiltonian 
\cite{BMvol2,Belyaev196517,Kumar1967608} 
(a similar comment is made in Ref.~\cite{0954-3899-36-12-123101}).  
Second, the ratios take different values for different LQRPA collective masses 
($D_{\beta\beta}, D_{\beta\gamma}, D_{\gamma\gamma}, D_1, D_2$, and $D_3$). 
Third, for every collective mass, the ratio exhibits an intricate dependence 
on $\beta$ and $\gamma$. 
Thus, it may be quite insufficient to simulate the time-odd mean-field 
contributions to the collective masses by scaling the IB cranking masses 
with a common multiplicative factor. 

\subsection{Check of self-consistency along the collective path} 
\label{sec:collHresult:comparison}

As discussed in Sec.~\ref{sec:theory}, 
the CHB+LQRPA method is a practical approximation to the ASCC method. 
It is certainly desirable to examine the accuracy of this approximation by 
carrying out a fully self-consistent calculation. 
Although, at the present time, such a calculation is too demanding to carry out 
for a whole region of the $(\bg)$ plane,   
we can check the accuracy at least along the 1D collective path. 
This is because the 1D collective path is determined by carrying out 
a fully self-consistent ASCC calculation for a single set of collective 
coordinate and momentum. 
The 1D collective paths projected onto the $(\bg)$ plane are displayed 
in Fig.~\ref{fig:V}. 
Let us use a notation $\ket{\phi(q)}$ for the moving-frame HB state   
obtained by self-consistently solving the ASCC equations 
for a single collective coordinate $q$ \cite{PTP.119.59,hinohara:014305}. 
To distinguish from it, we write the CHB state as 
$\ket{\phi(\beta(q),\gamma(q))}$. 
This notation means that the values of $\beta$ and $\gamma$ 
are specified by the collective coordinate $q$ along the collective path. 
In other words, $\ket{\phi(\beta(q),\gamma(q))}$ has the same   
expectation values of the quadrupole operator as those of $\ket{\phi(q)}$. 
It is important to note, however,  that they are different from each other, 
because $\ket{\phi(\beta(q),\gamma(q))}$ is a solution of the CHB equation 
which is an approximation of the moving-frame HB equation. 
Let us evaluate various physical quantities using the two state vectors 
and compare the results.  

\begin{figure}
\includegraphics[width=80mm]{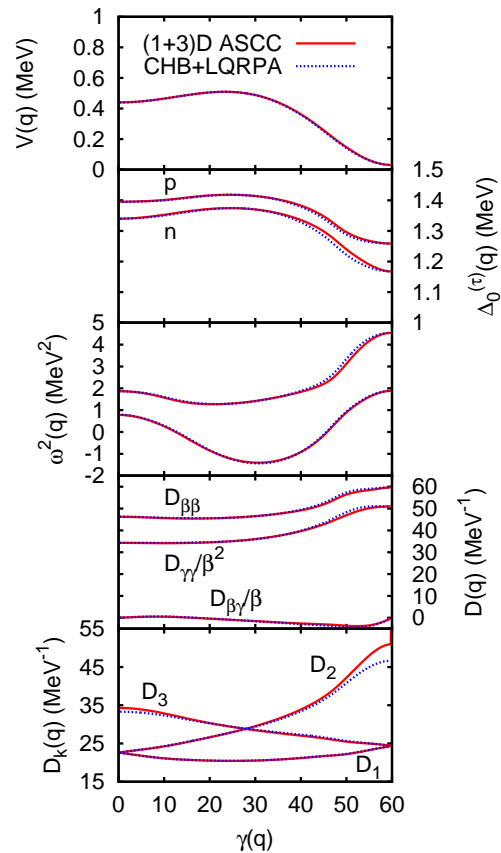}
\caption{\label{fig:comparison}(Color online)
Comparison of physical quantities evaluated with the CHB + LQRPA approximation 
and those with the ASCC method.  
Both calculations are carried out along the 1D collective path for $^{68}$Se 
and the results are plotted as function of $\gamma(q)$. 
From the top to the bottom:  
1) the collective potential,  
2) monopole-pairing gaps, $\Delta_0^{(n))}$ and $\Delta_0^{(p)}$, 
for neutrons and protons, 
3) frequencies squared $\omega^2$ of the lowest and the second-lowest 
modes obtained by solving the moving-frame QRPA and the LQRPA equations,  
4) vibrational masses, $D_{\beta\beta}$,
$D_{\beta\gamma}/\beta$, and $D_{\gamma\gamma}/\beta^2$,  
5) rotational masses $D_k$. 
In almost all cases, results of the two calculations are indistinguishable, 
because they agree within the widths of the line.}
\end{figure}

In Fig.~\ref{fig:comparison}
various physical quantities (the pairing gaps, the collective potential, 
the frequencies of the local normal modes, the rotational masses,
and vibrational masses) calculated 
using the moving-frame HB state $\ket{\phi(q)}$ and 
the CHB state $\ket{\phi(\beta(q),\gamma(q))}$ are presented and compared. 
These calculations are carried out along the 1D collective path for $^{68}$Se.  
Apparently, the results of the two calculations are indistinguishable 
in almost all cases, because they agree within the widths of the line. 
This good agreement implies that the CHB+LQRPA is an excellent 
approximation to the ASCC method along the collective path on the $(\bg)$ plane. 
As we shall see in the next section, collective wave functions distribute  
around the collective path. 
Therefore, it may be reasonable to expect that 
the CHB+LQRPA method is a good approximation to the ASCC method 
and suited, at least, for describing 
the oblate-prolate shape mixing dynamics in $^{68,70,72}$Se.

\section{Large-amplitude shape-mixing properties of $^{68,70,72}$Se
\label{sec:reqresult}}

\begin{figure*}
\includegraphics[width=150mm]{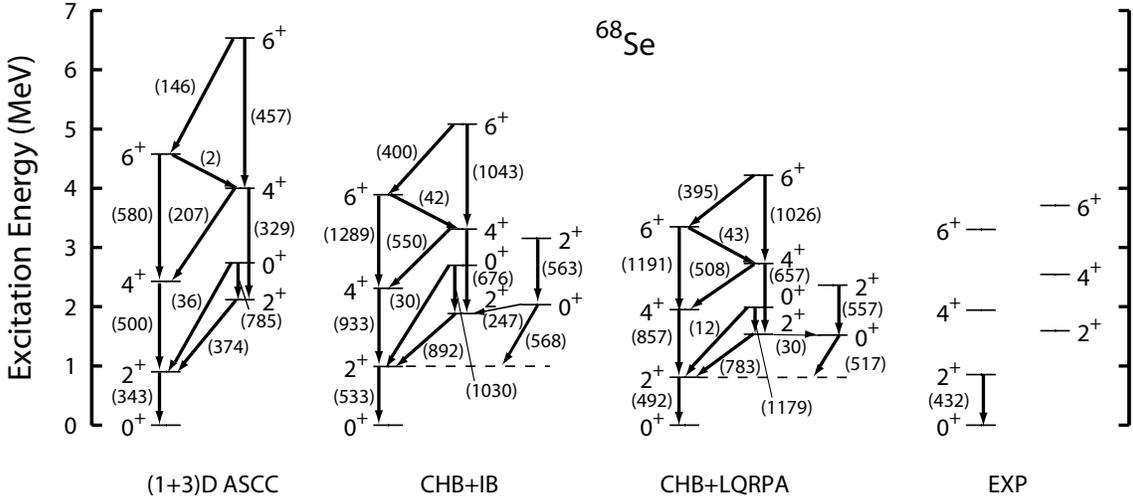}
\caption{\label{fig:68Se_energy}
Excitation spectra and $B(E2)$ values calculated for $^{68}$Se 
by means of the CHB+LQRPA method (denoted CHB+LQRPA) and 
experimental data \cite{PhysRevC.67.064318,PhysRevLett.84.4064,obertelli:031304}.  
For comparison, results calculated using the IB cranking masses (denoted CHB+IB)   
and those obtained using the (1+3)D version of the ASCC method 
(denoted (1+3)D~ASCC) are also shown. 
Only $B(E2)$'s larger than 1 Weisskopf unit 
(in the (1+3)D ASCC and/or the CHB+LQRPA calculations) 
are shown in units of $e^2$fm$^4$. }
\end{figure*}

\begin{figure*}
\begin{tabular}{ccccl}
 \includegraphics[width=30mm]{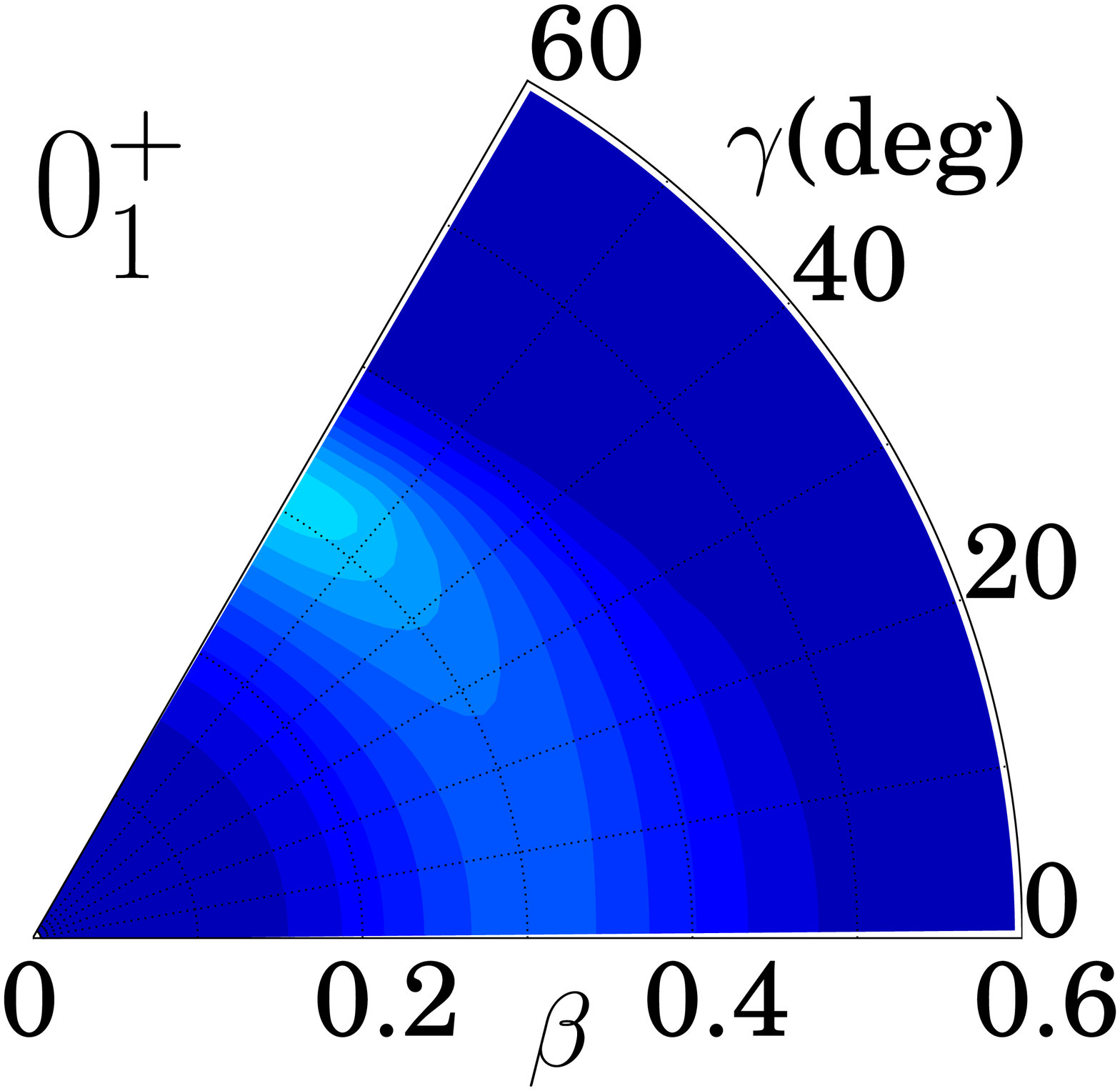} &
 \includegraphics[width=30mm]{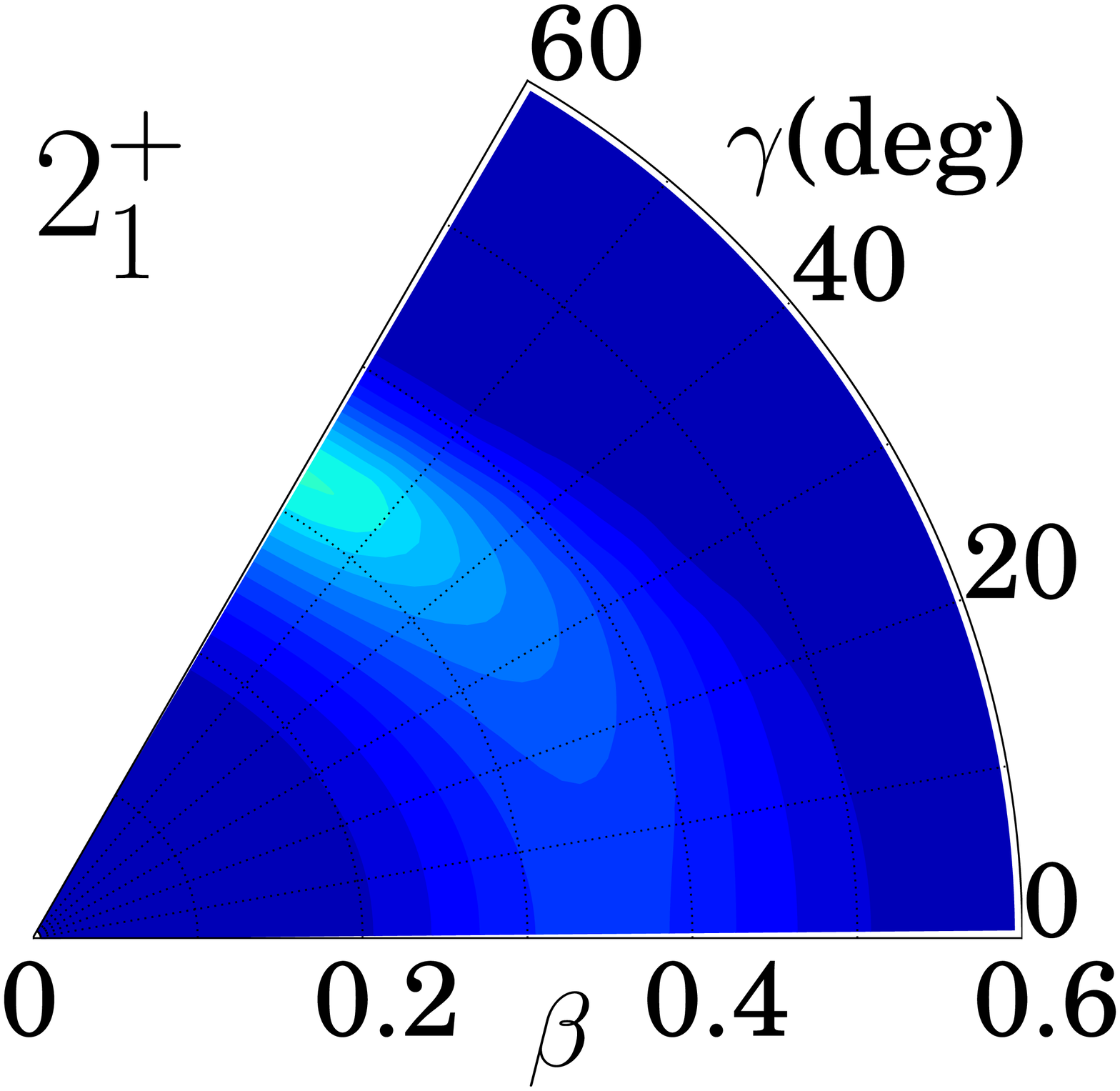} &
 \includegraphics[width=30mm]{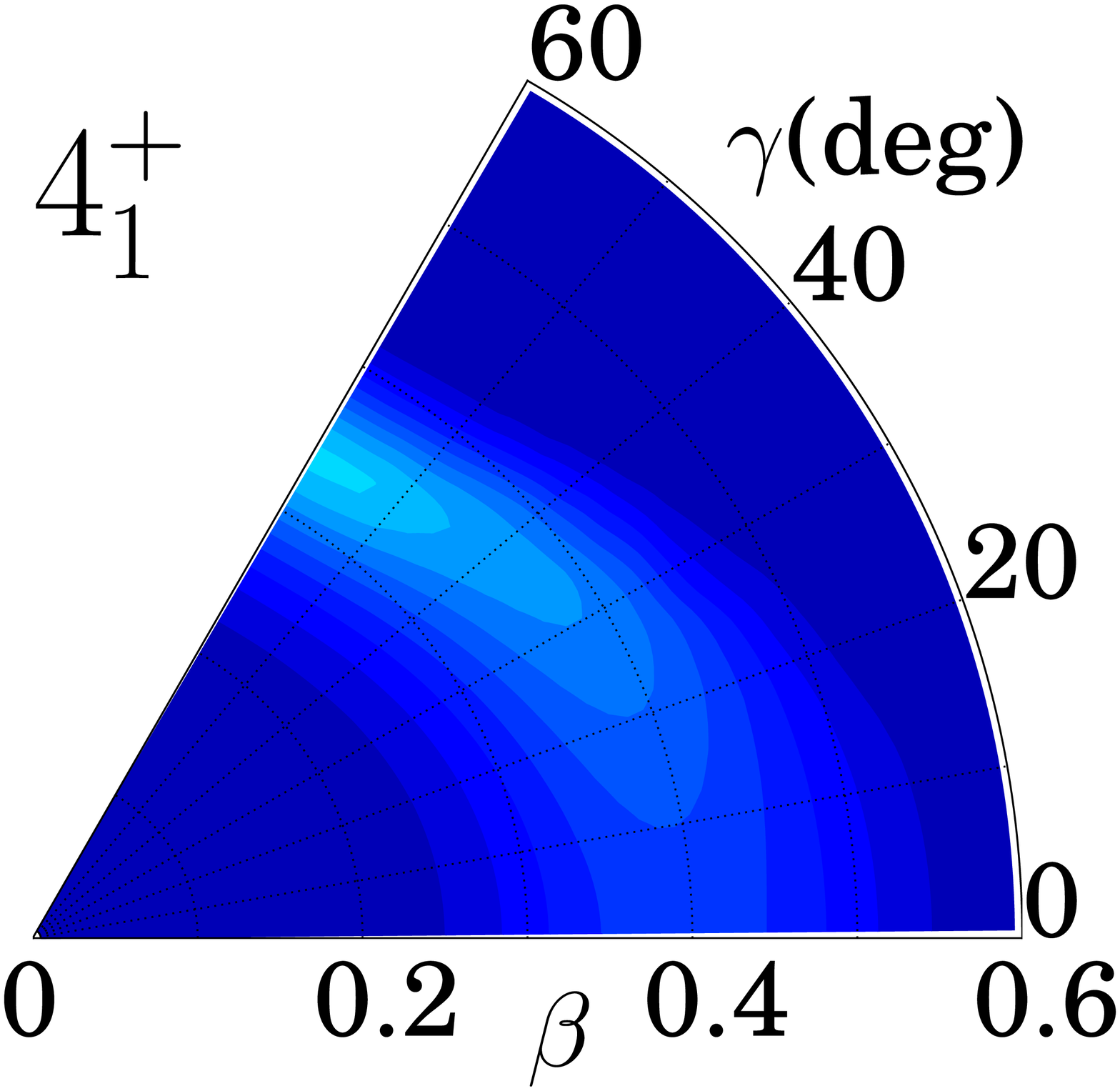} &
 \includegraphics[width=30mm]{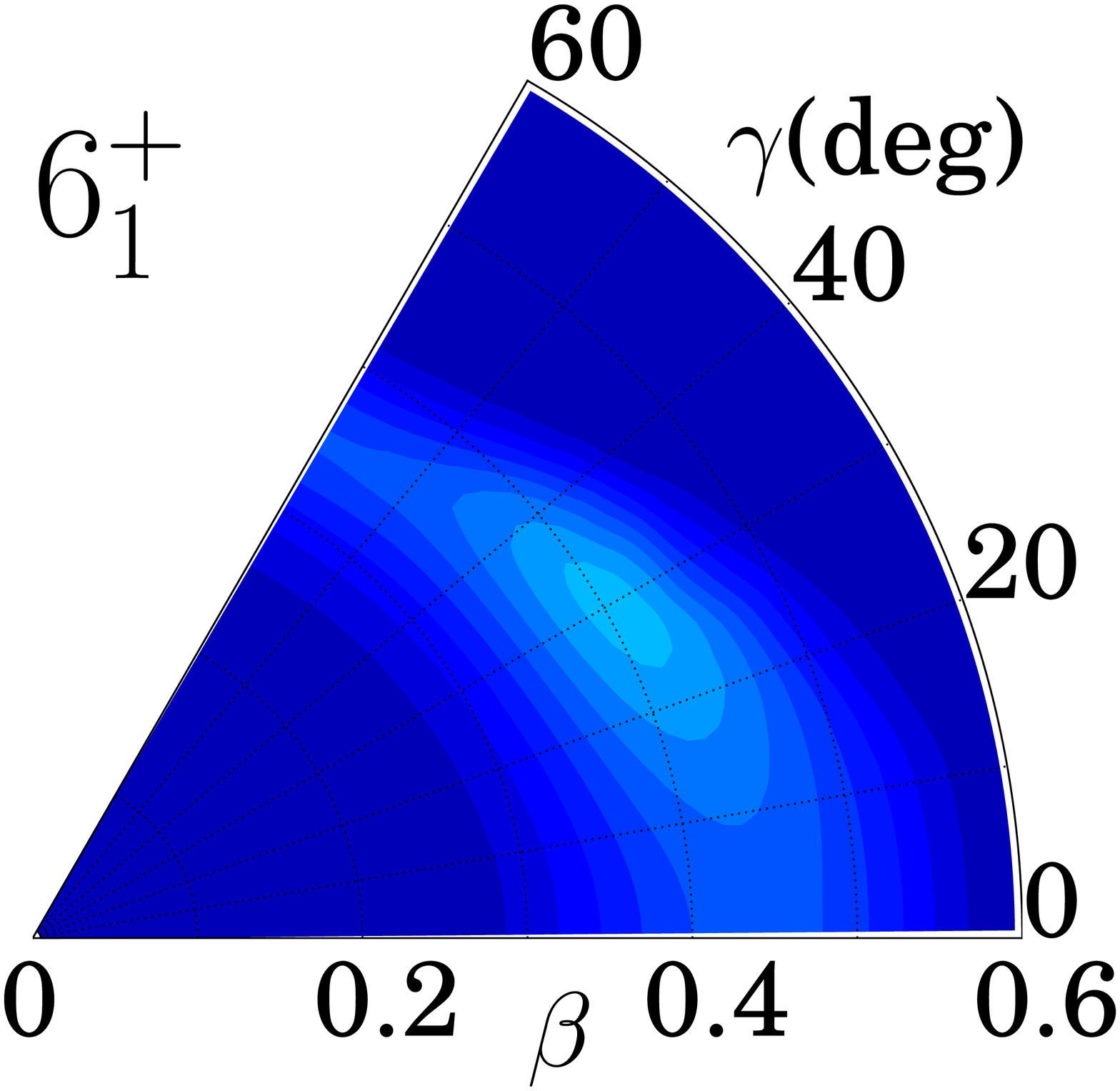} & \\
 \includegraphics[width=30mm]{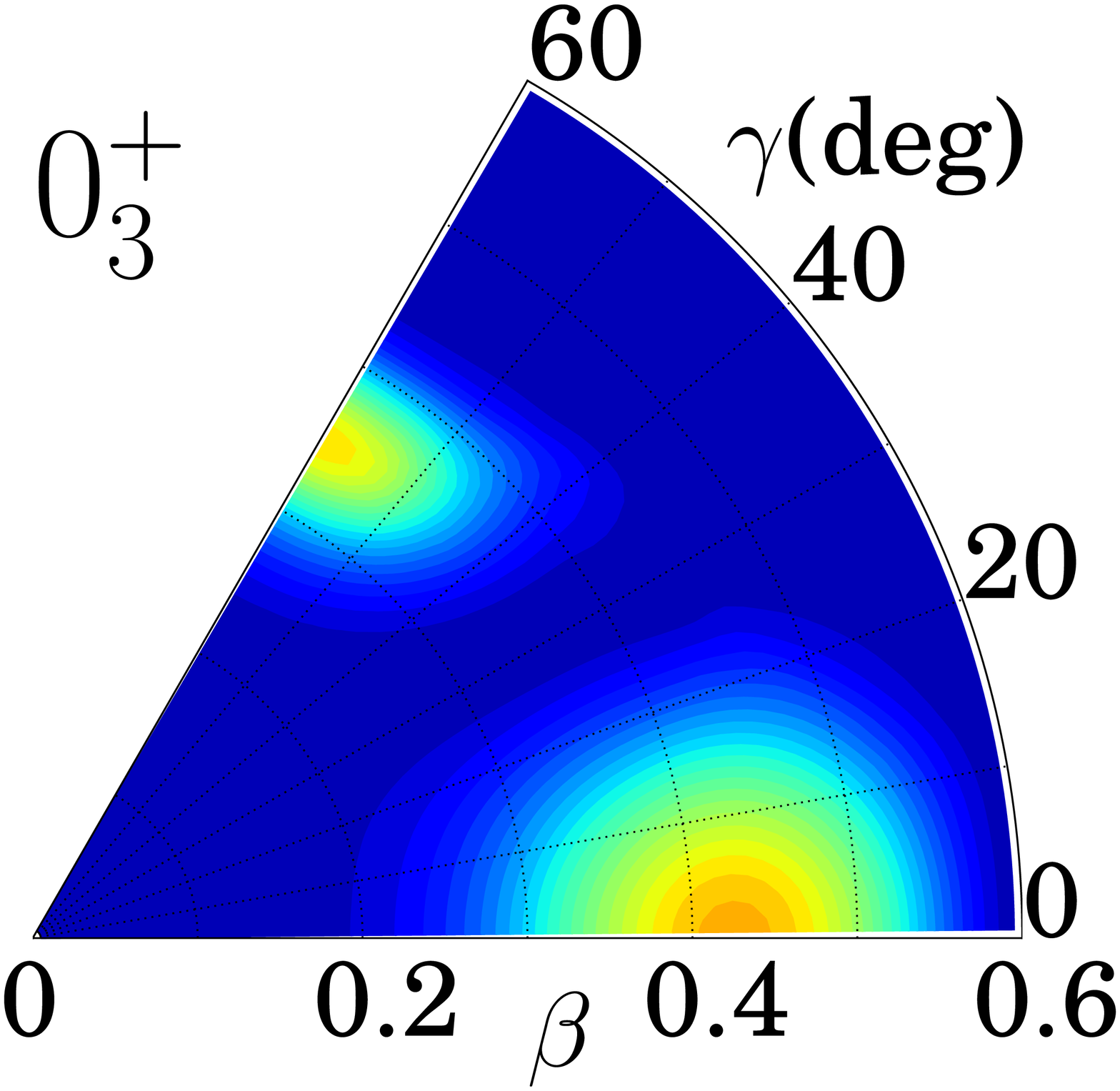} &
 \includegraphics[width=30mm]{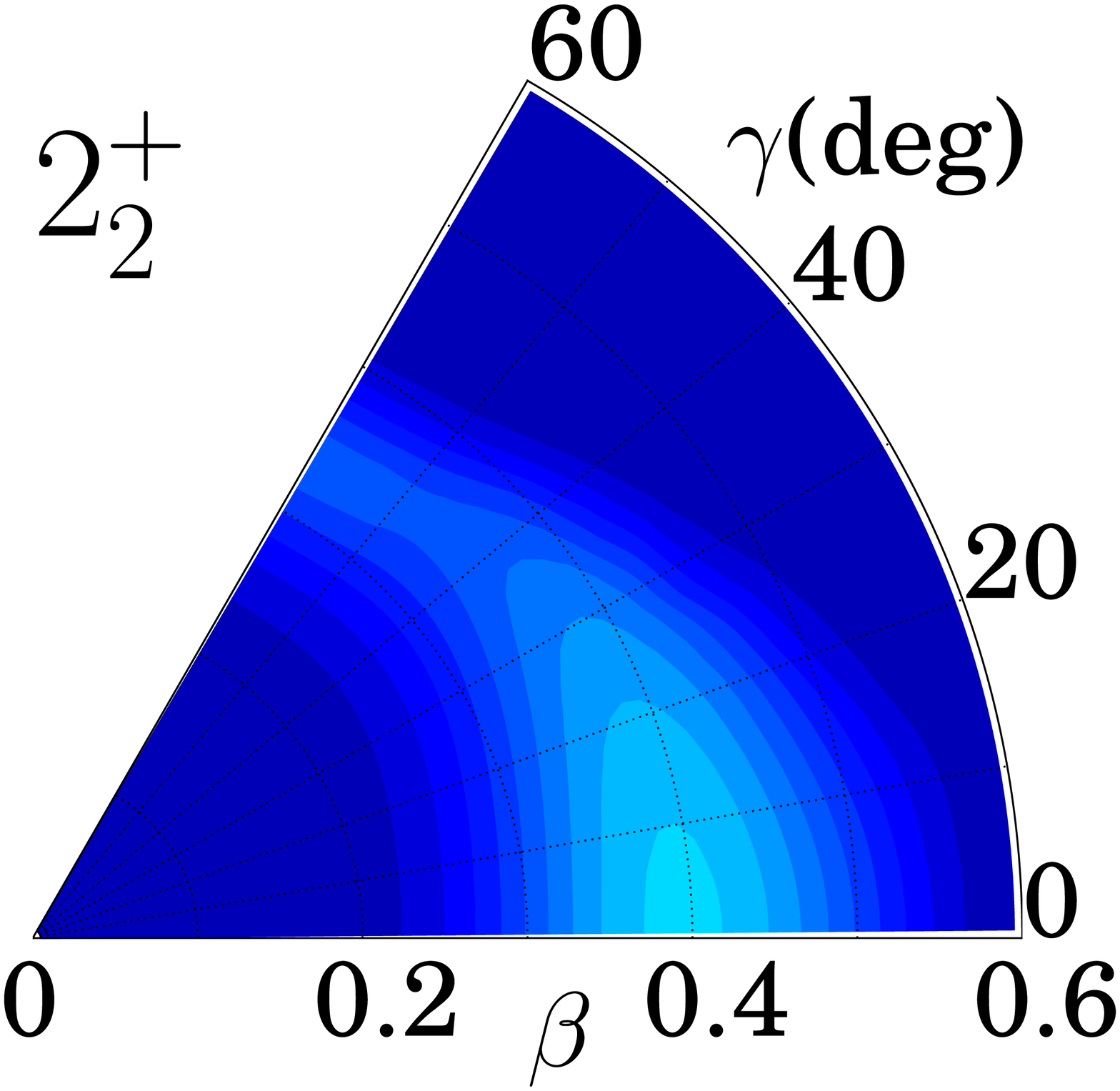} &
 \includegraphics[width=30mm]{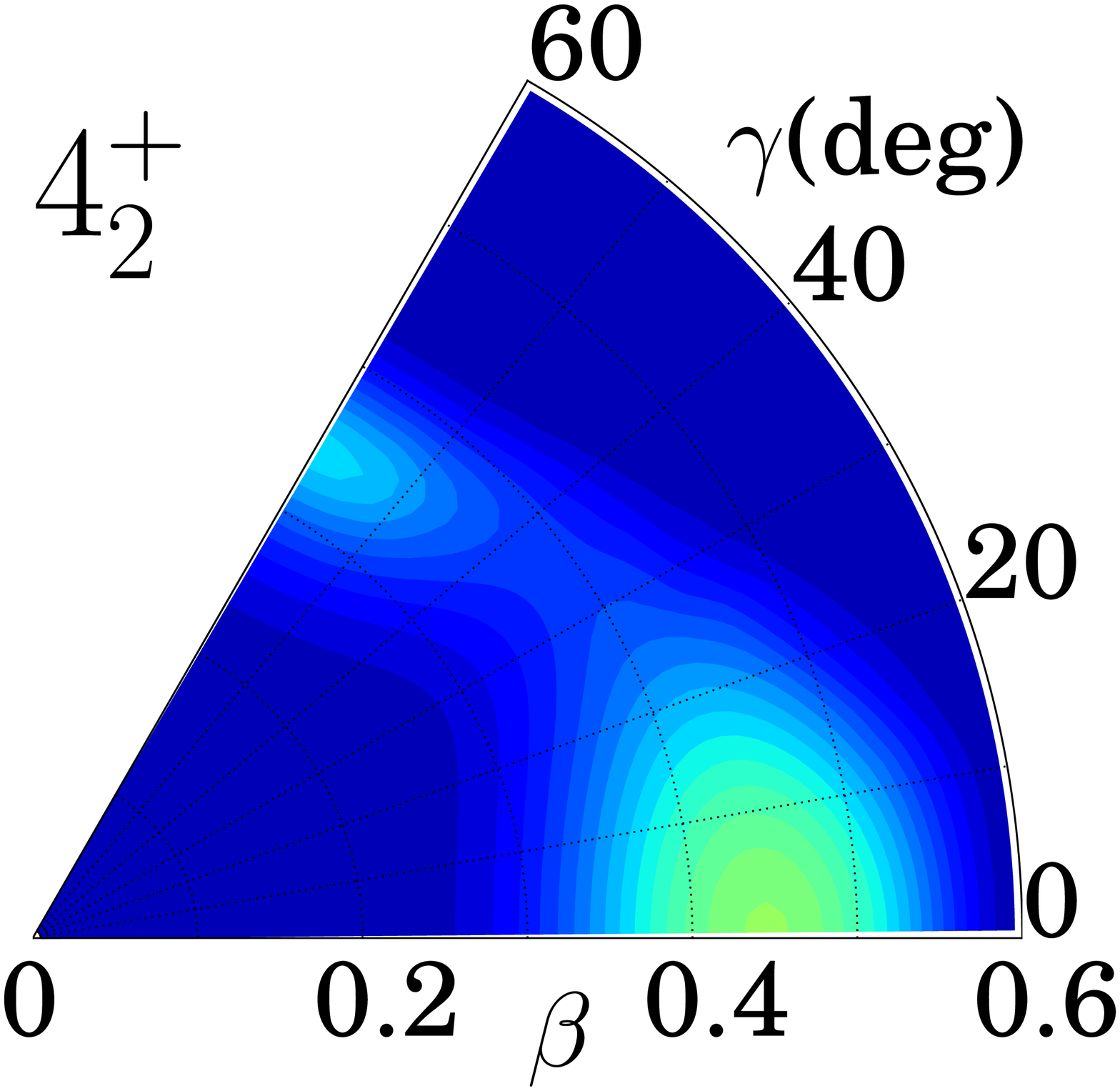} &
 \includegraphics[width=30mm]{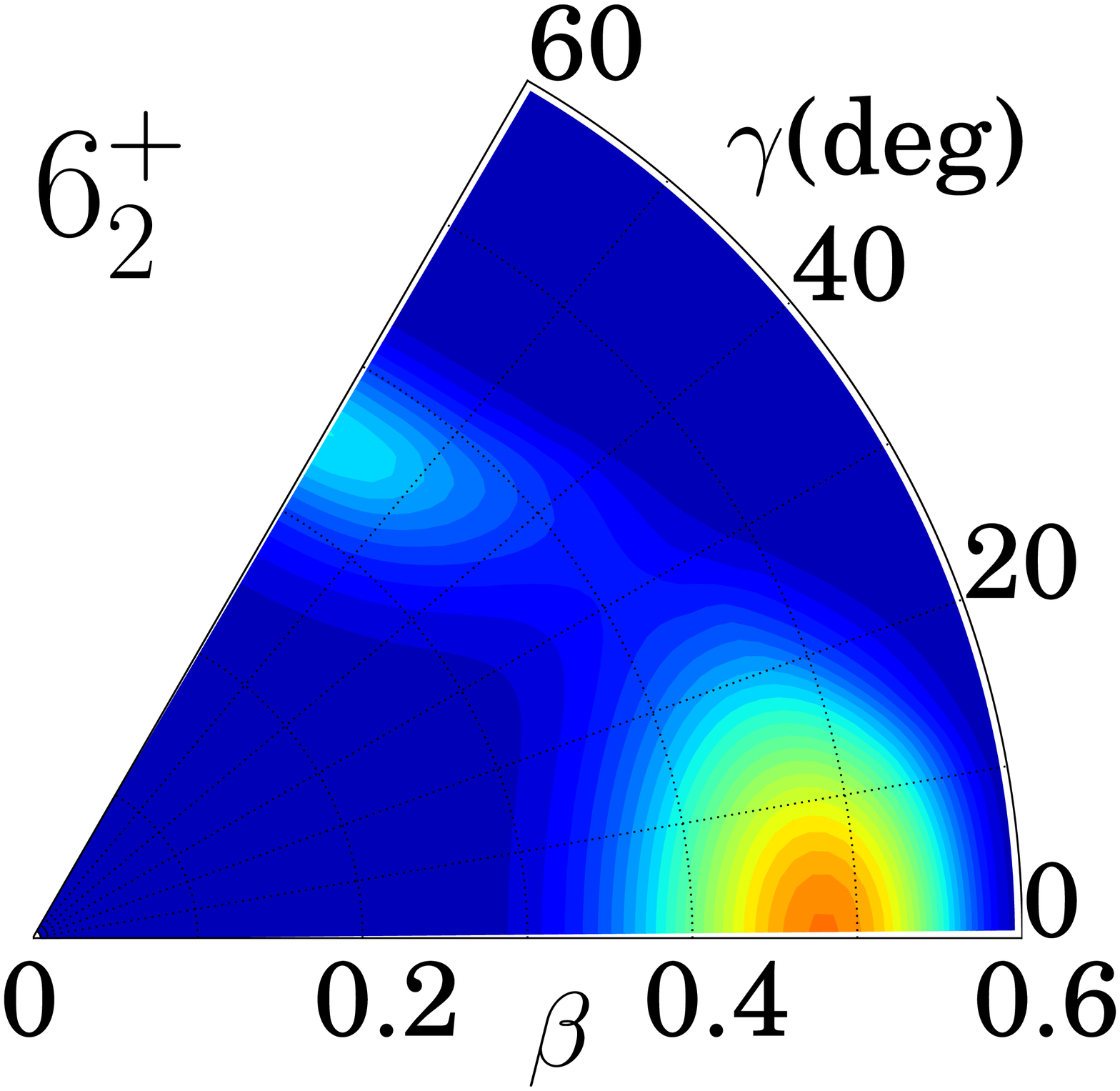} &
 \includegraphics[height=25mm]{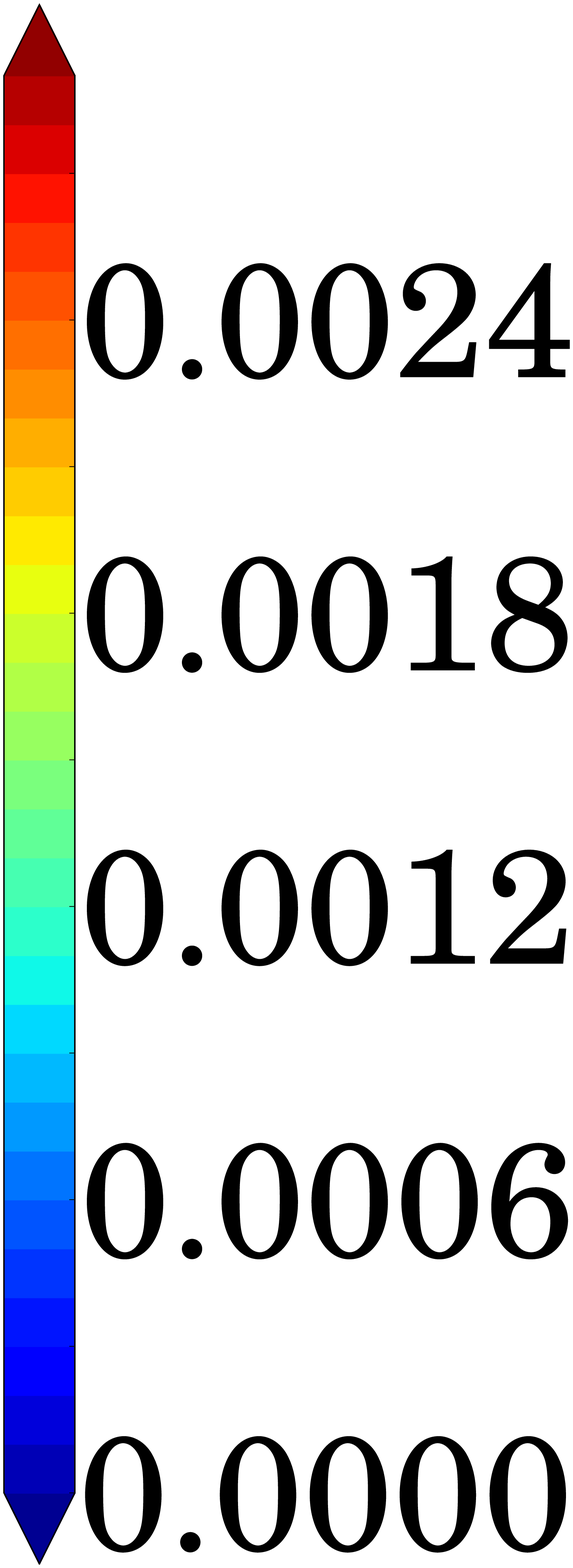} \\
 \includegraphics[width=30mm]{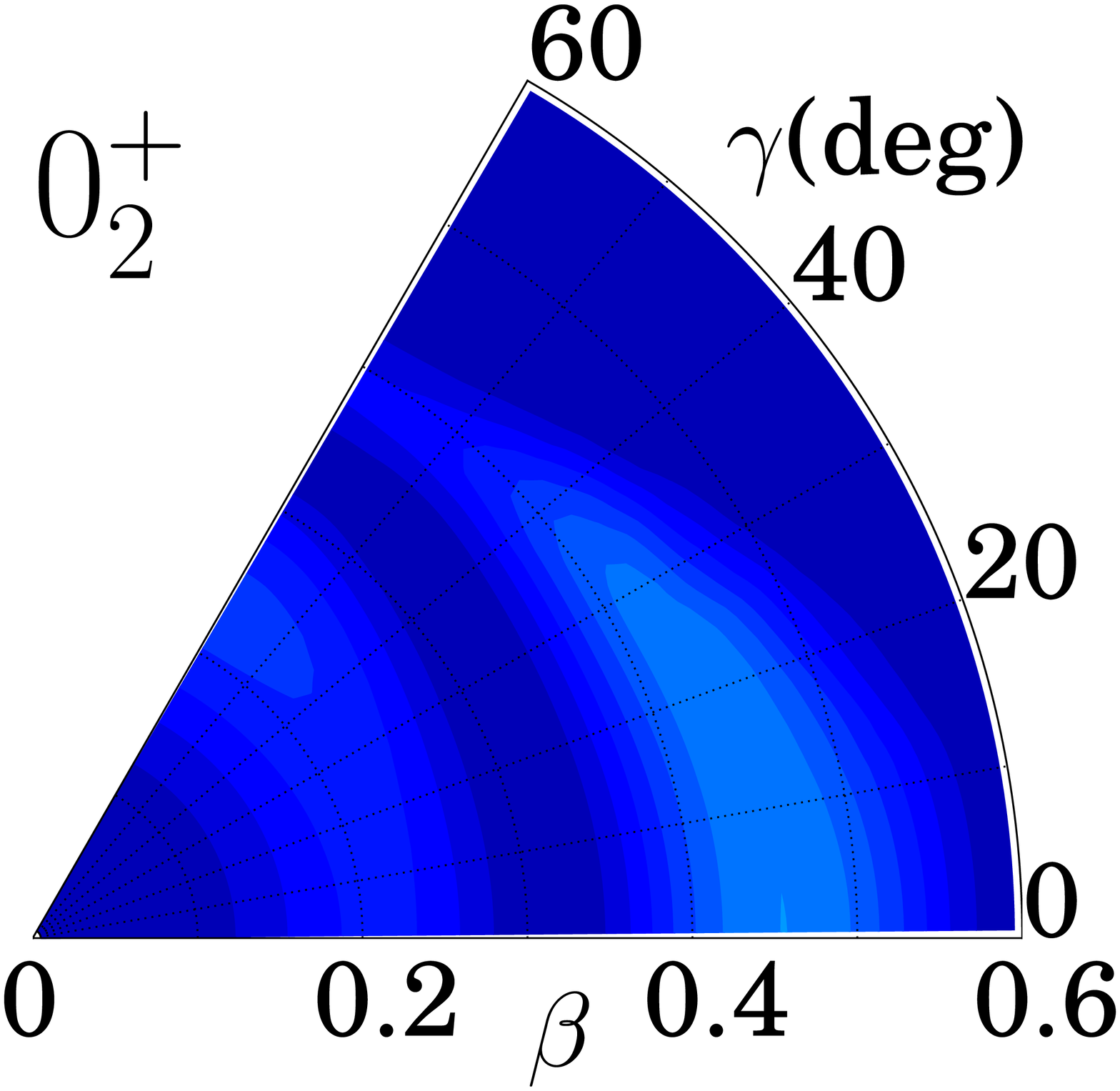} &
 \includegraphics[width=30mm]{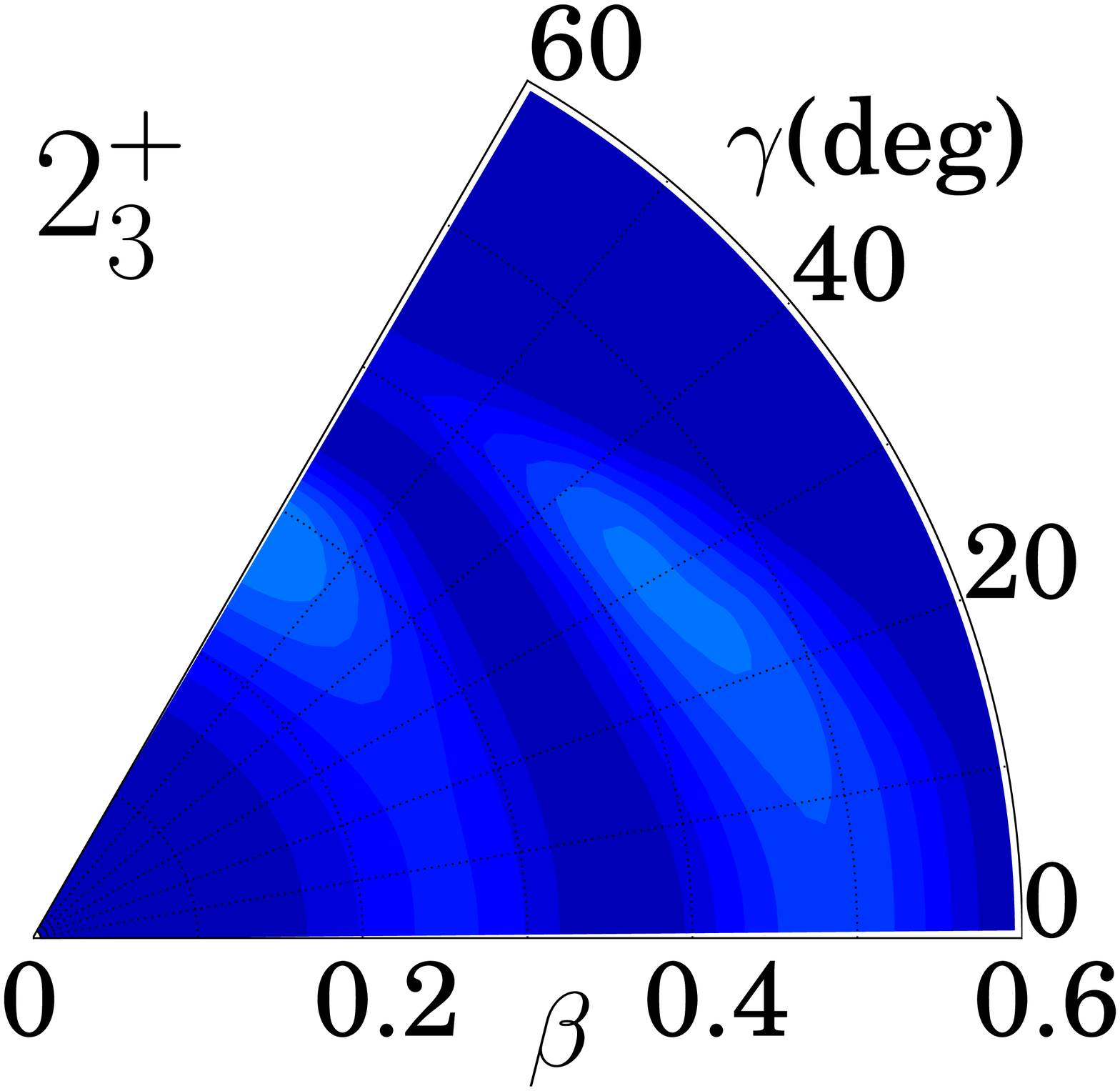} & & &
\end{tabular}
\caption{\label{fig:68Se-wave}
(Color online) Vibrational wave functions squared $\beta^4|\Phi_{Ik}(\bg)|^2$ 
calculated for $^{68}$Se.}
\end{figure*}

\begin{figure*}
\includegraphics[width=150mm]{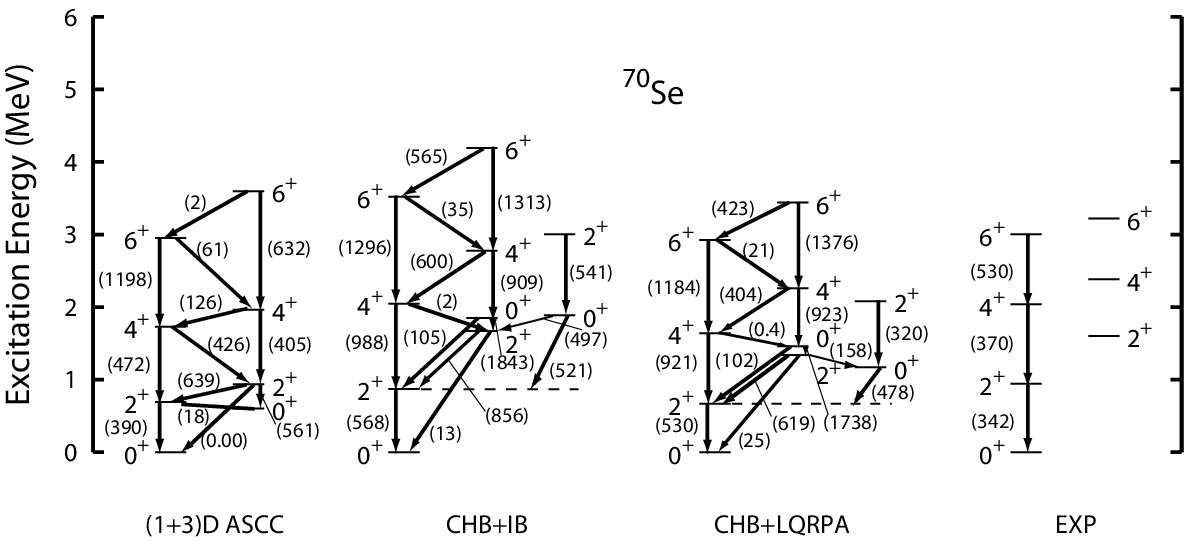}
\caption{\label{fig:70Se_energy}
Same as Fig.~\ref{fig:68Se_energy} but for $^{70}$Se. 
Experimental data is taken from Refs.~\cite{ljungvall:102502,0954-3899-28-10-307}.}
\end{figure*}

\begin{figure*}
\begin{tabular}{ccccl}
 \includegraphics[width=30mm]{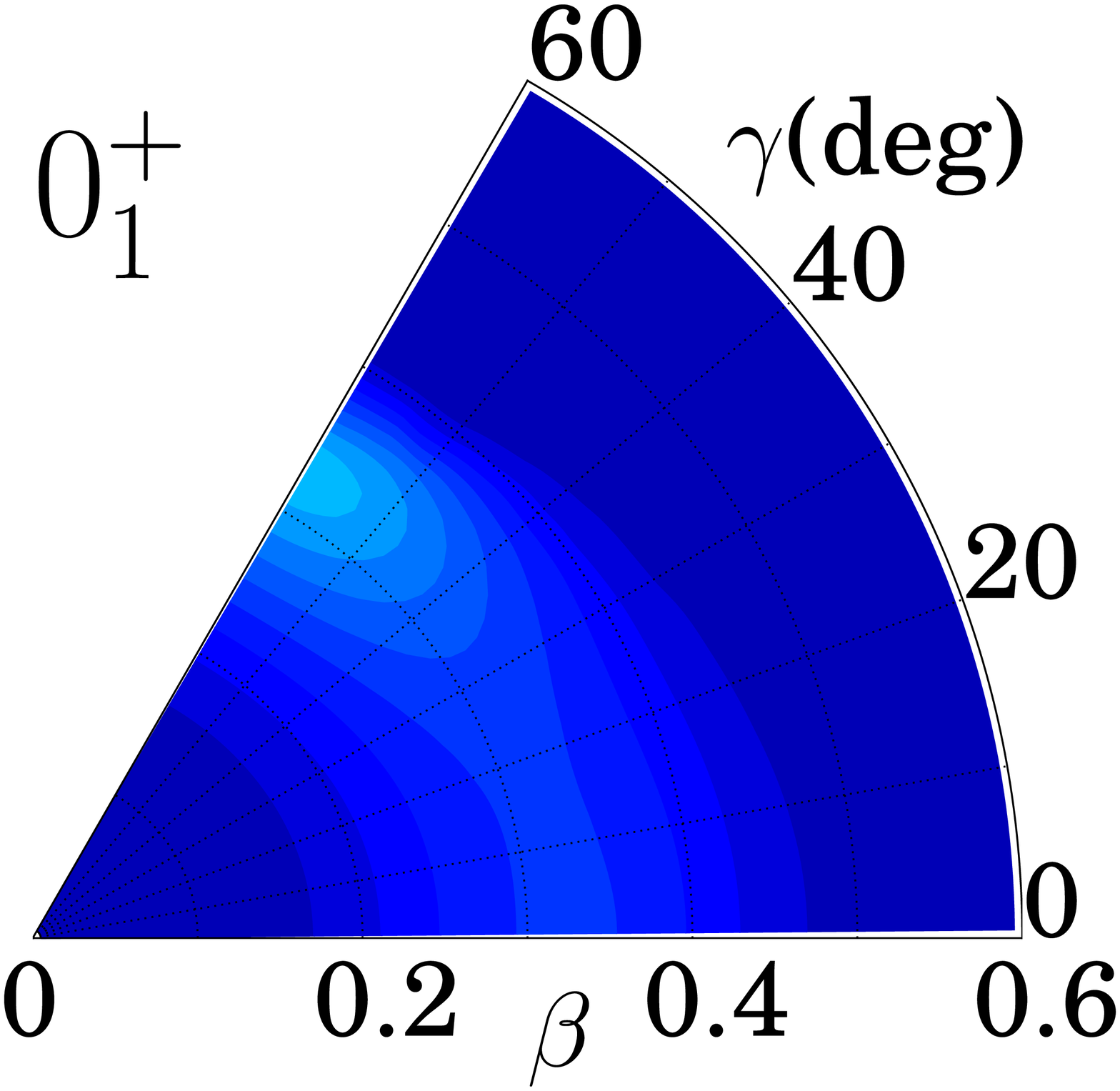} &
 \includegraphics[width=30mm]{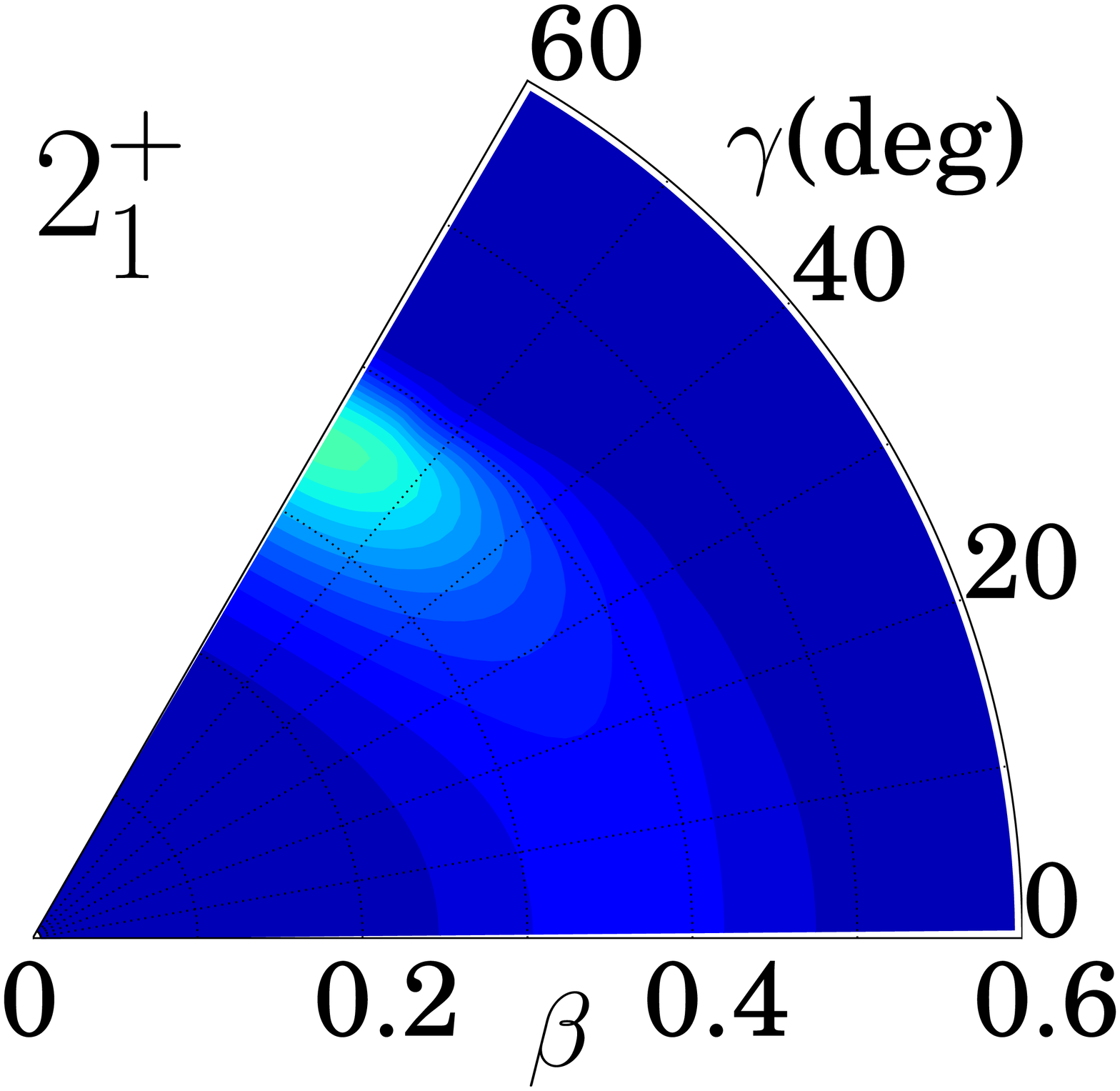} &
 \includegraphics[width=30mm]{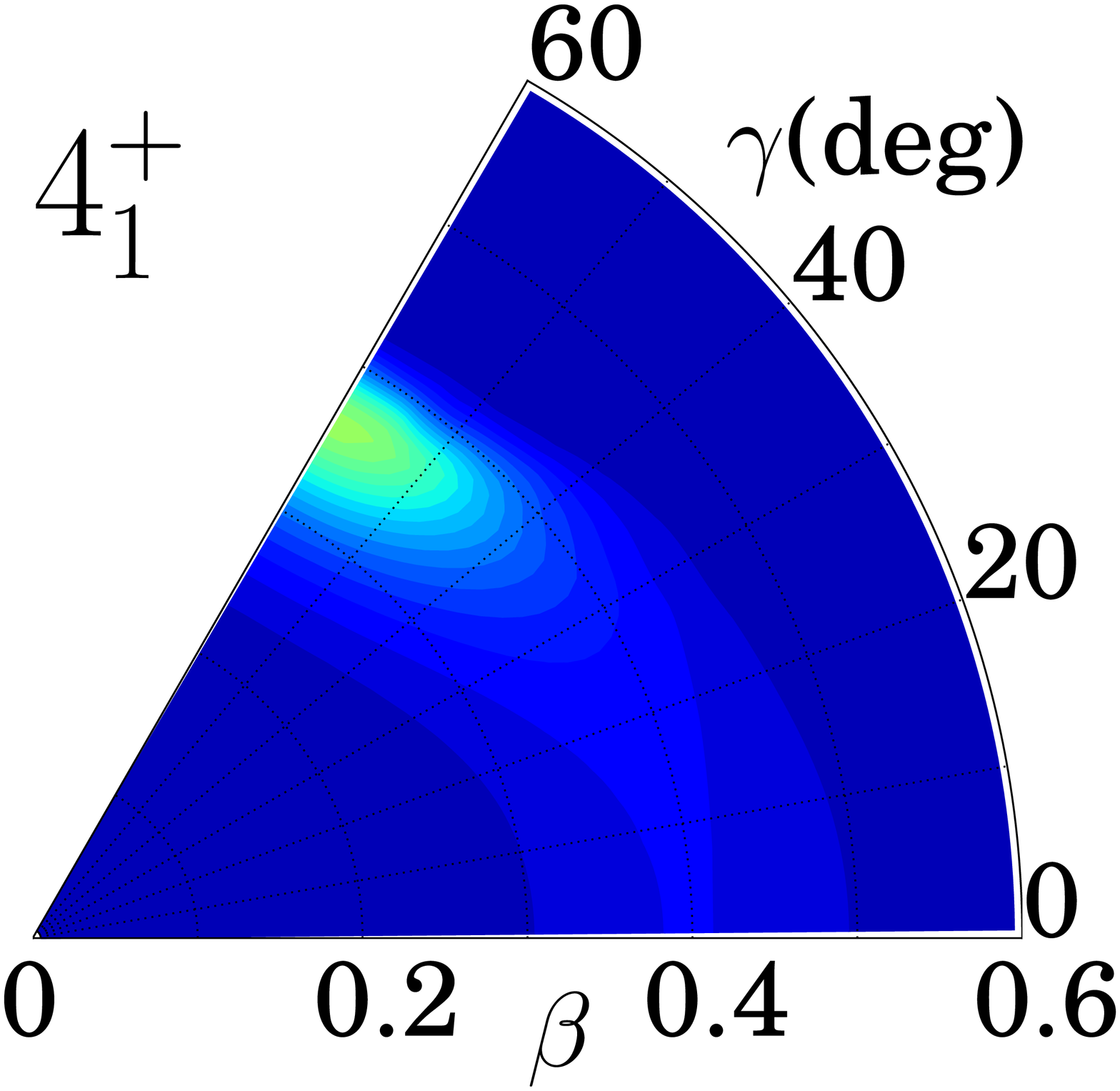} &
 \includegraphics[width=30mm]{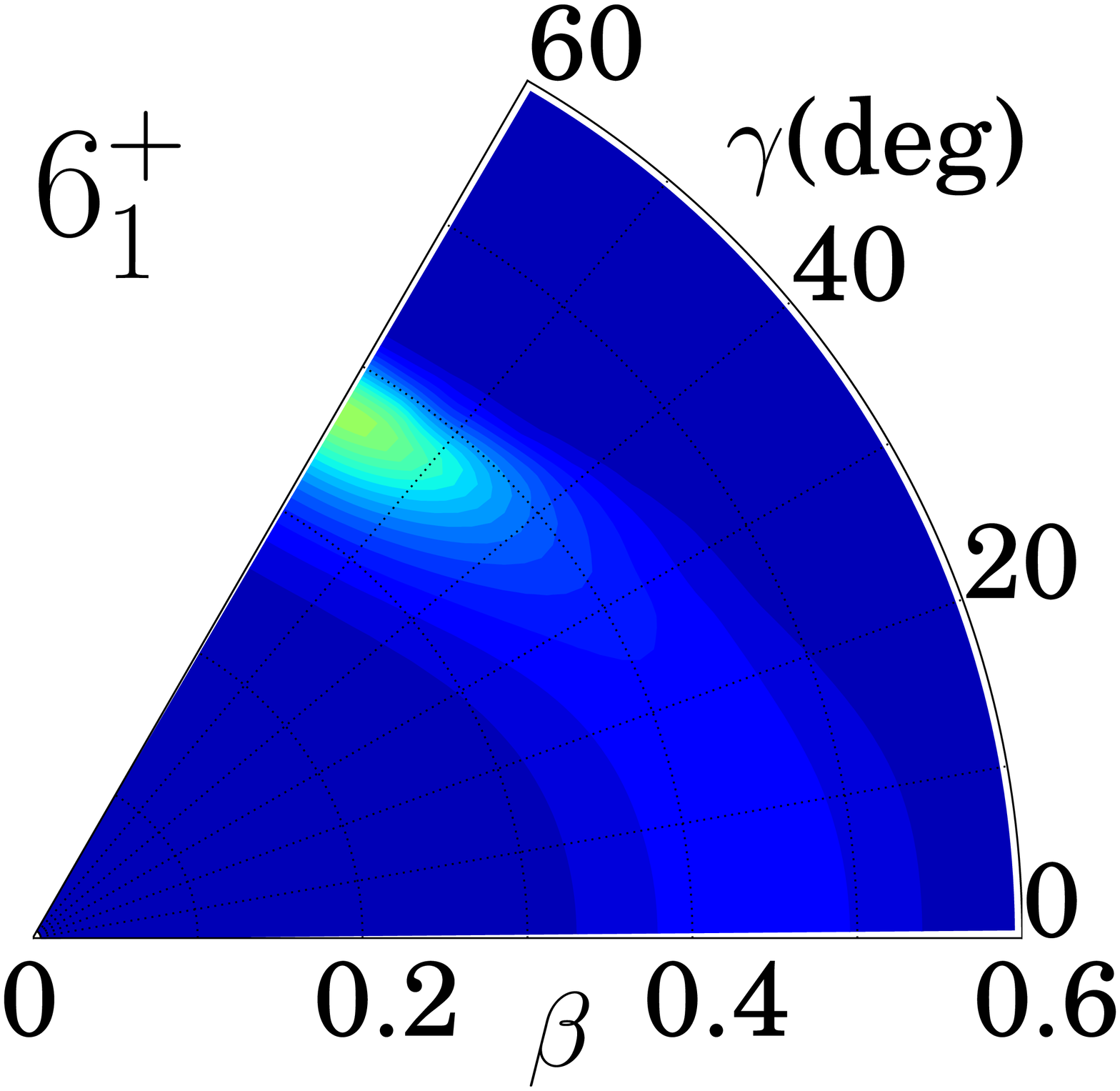} & \\
 \includegraphics[width=30mm]{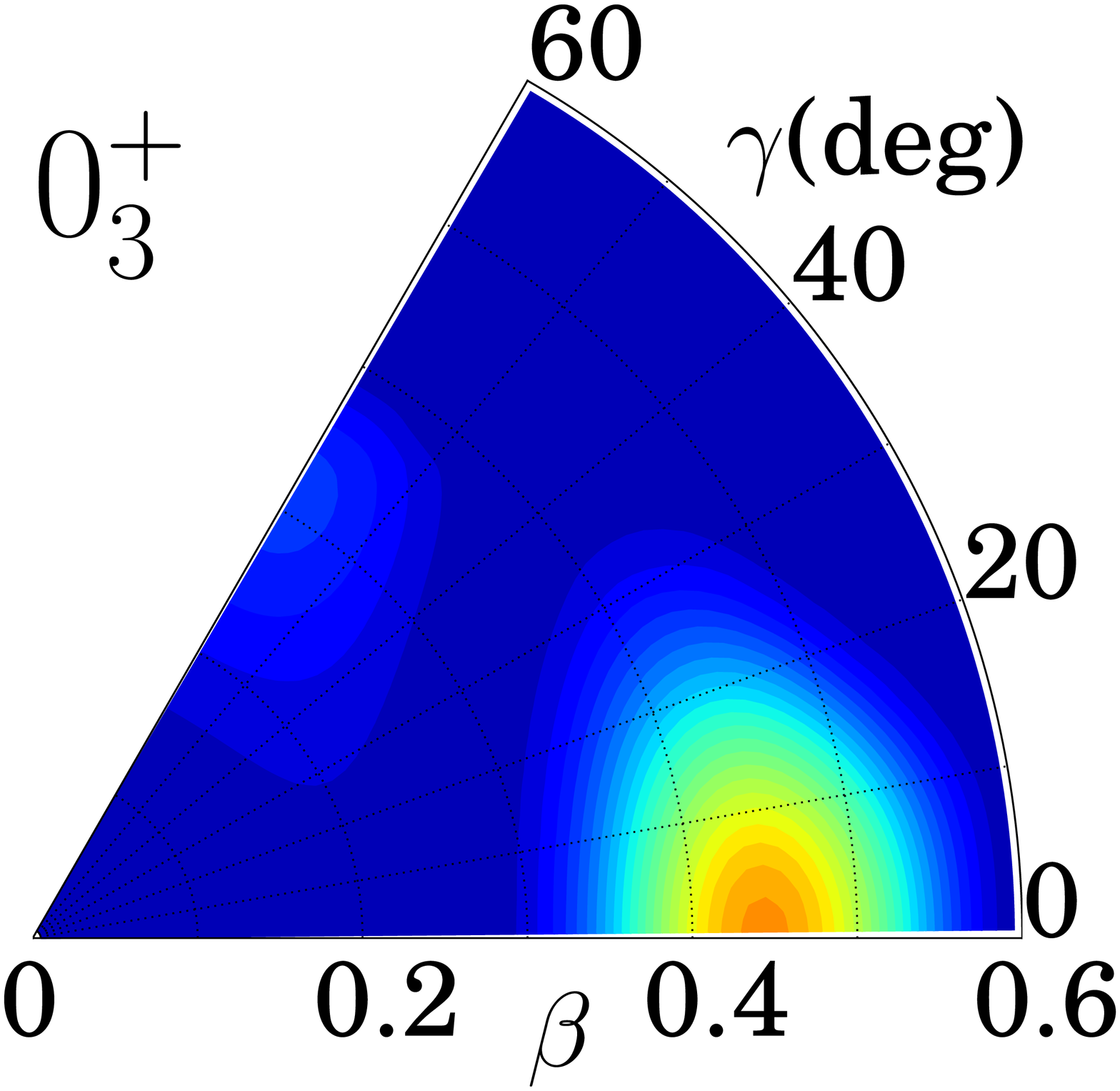} &
 \includegraphics[width=30mm]{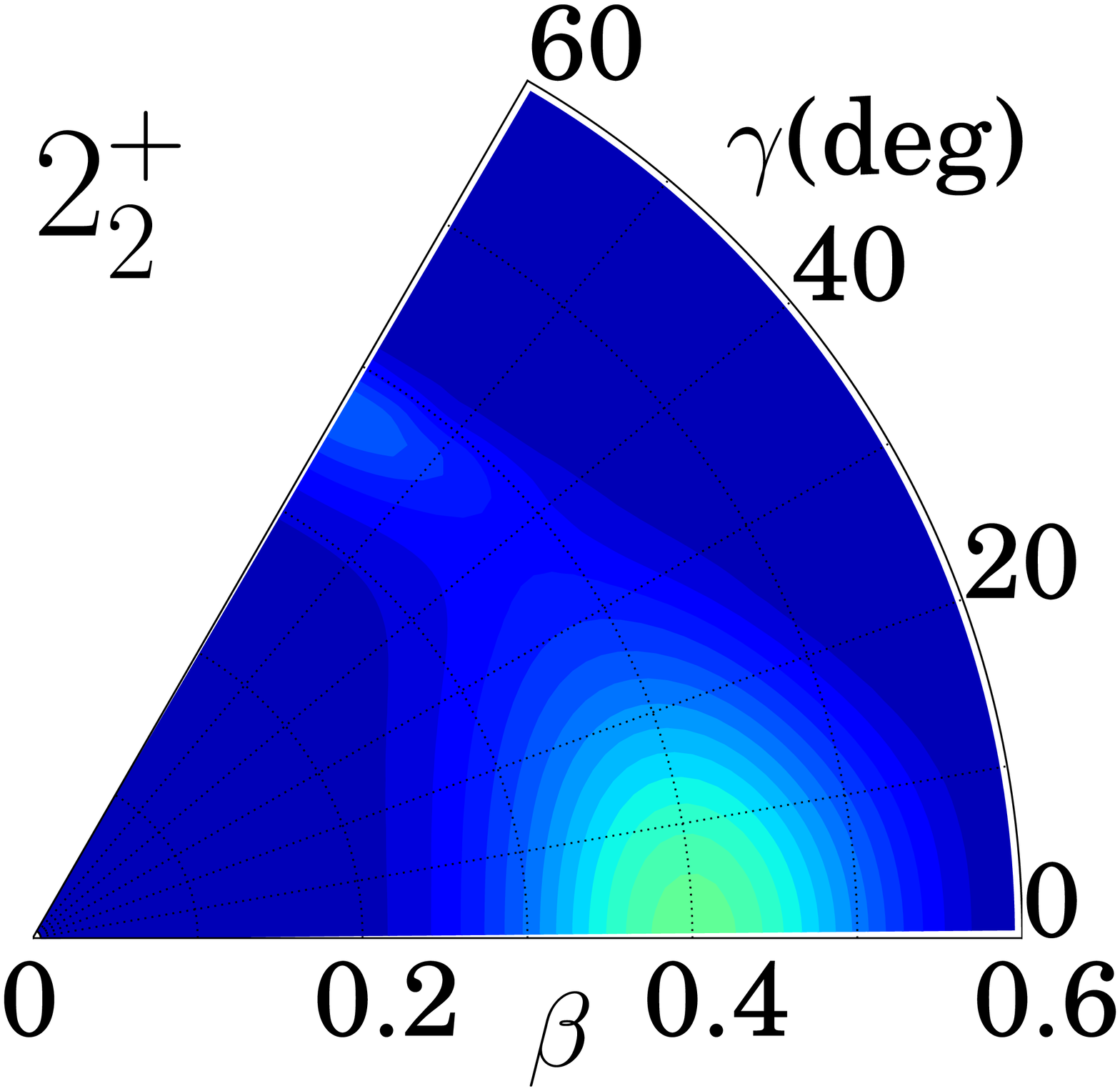} &
 \includegraphics[width=30mm]{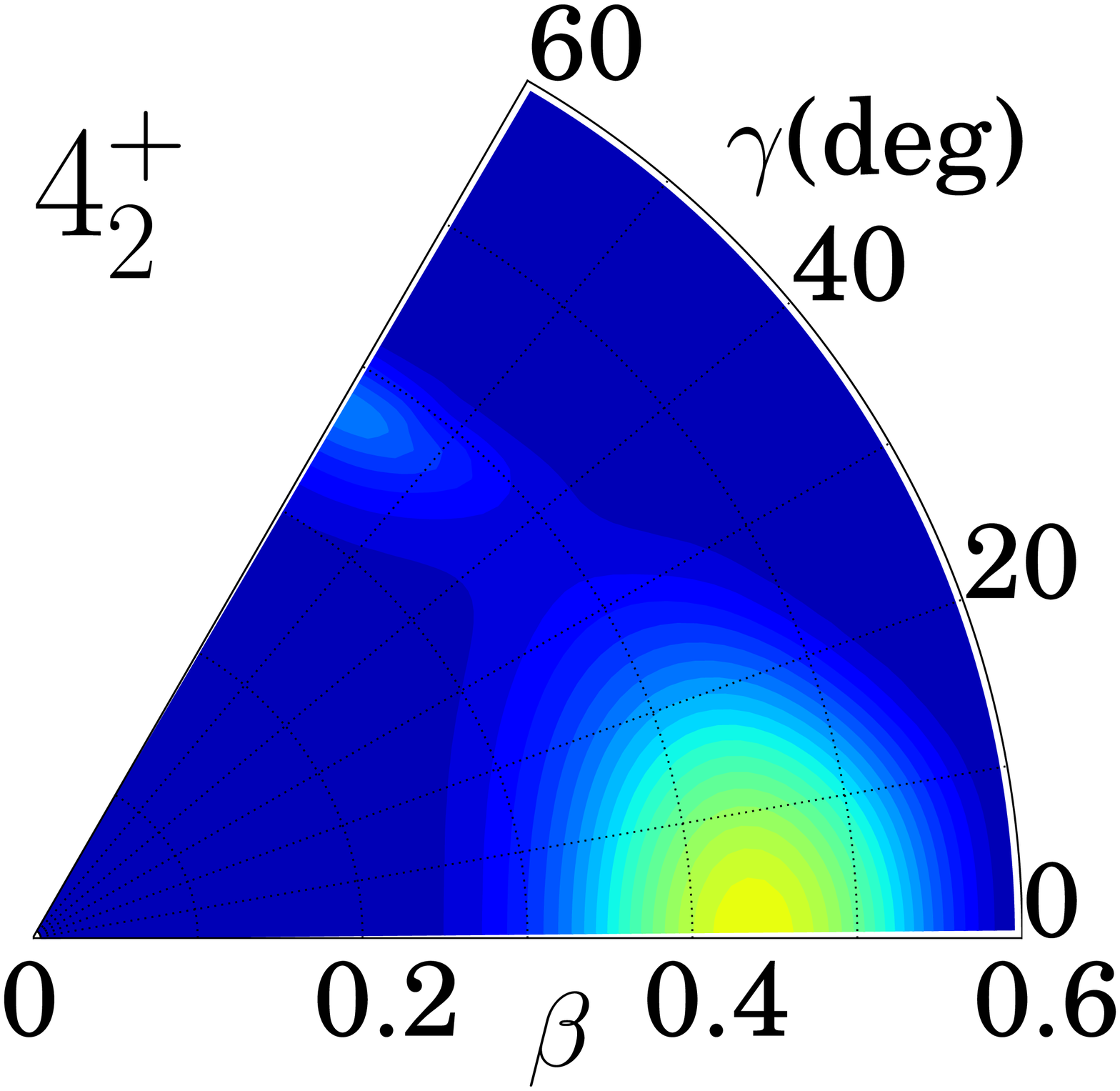} &
 \includegraphics[width=30mm]{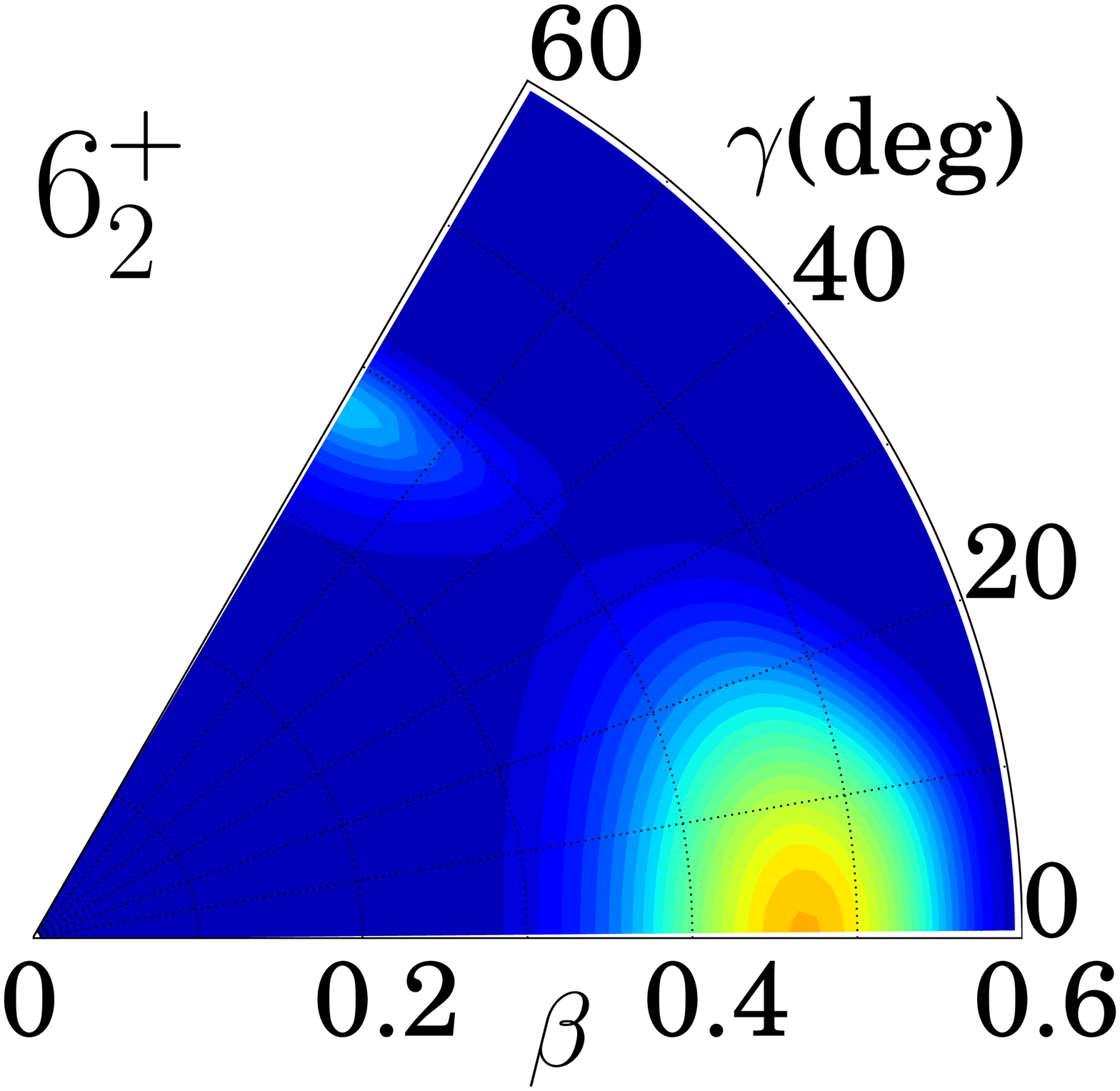} &
 \includegraphics[height=25mm]{colorbar.eps} \\
 \includegraphics[width=30mm]{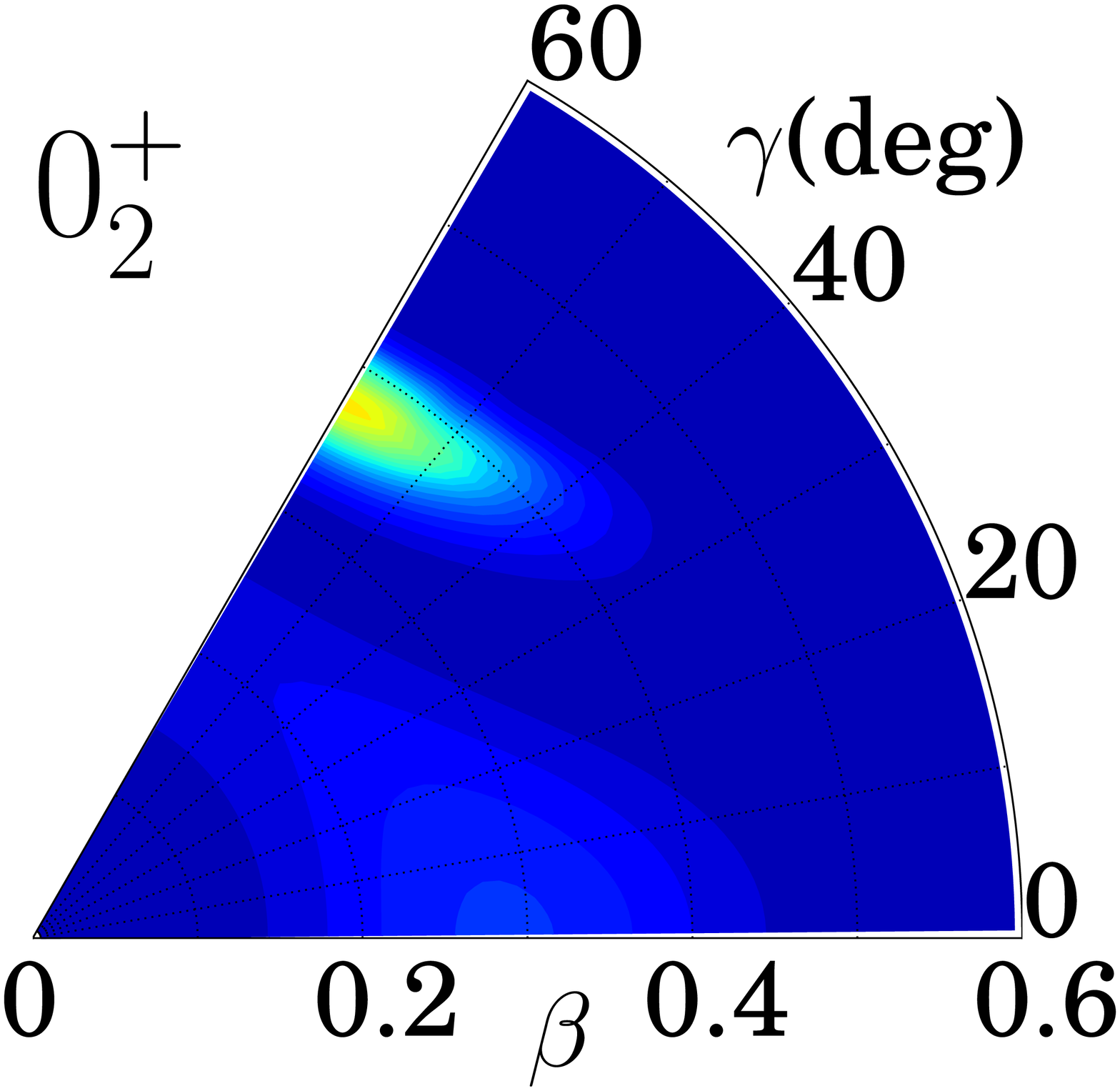} &
 \includegraphics[width=30mm]{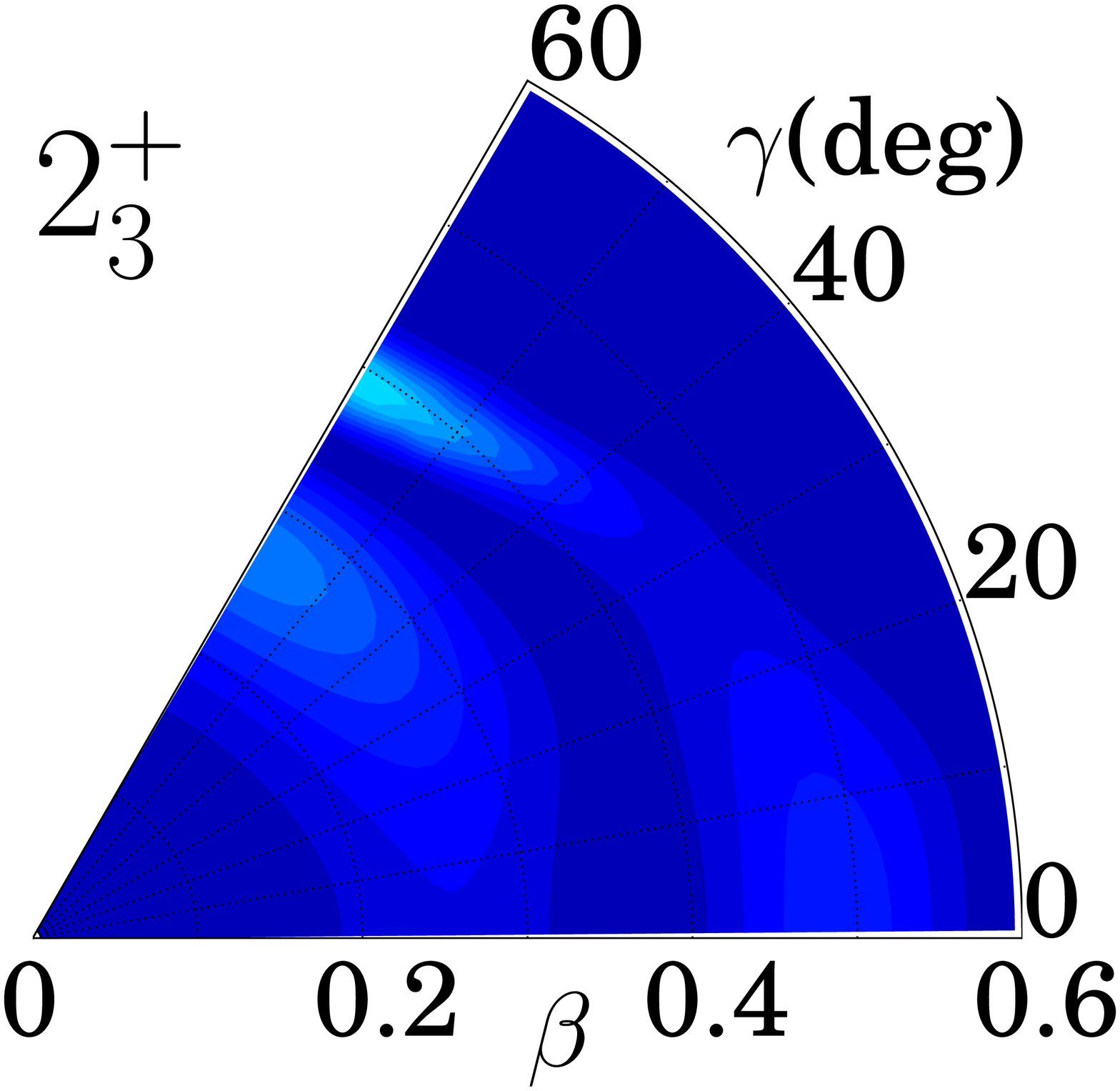} & & &
\end{tabular}
\caption{\label{fig:70Se-wave}(Color online)
Same as Fig.~\ref{fig:68Se-wave} but for $^{70}$Se.}
\end{figure*}

\begin{figure*}
\includegraphics[width=150mm]{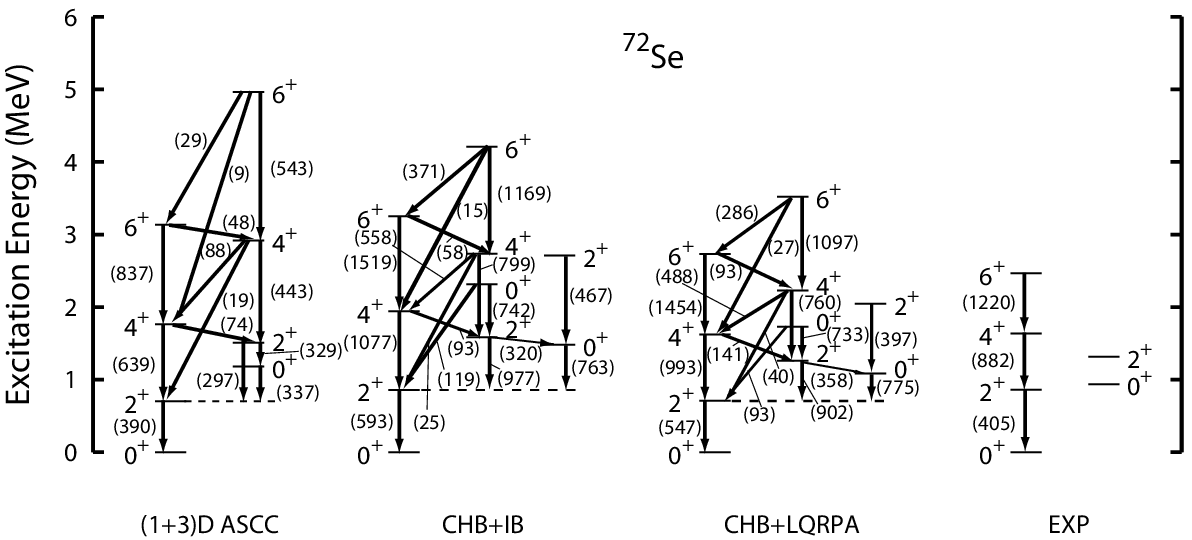}
\caption{\label{fig:72Se_energy}
Same as Fig.~\ref{fig:68Se_energy} but for $^{72}$Se.
Experimental data is taken from Refs.~\cite{ljungvall:102502,PhysRevC.63.024313}.}
\end{figure*}

\begin{figure*}
\begin{tabular}{ccccl}
 \includegraphics[width=30mm]{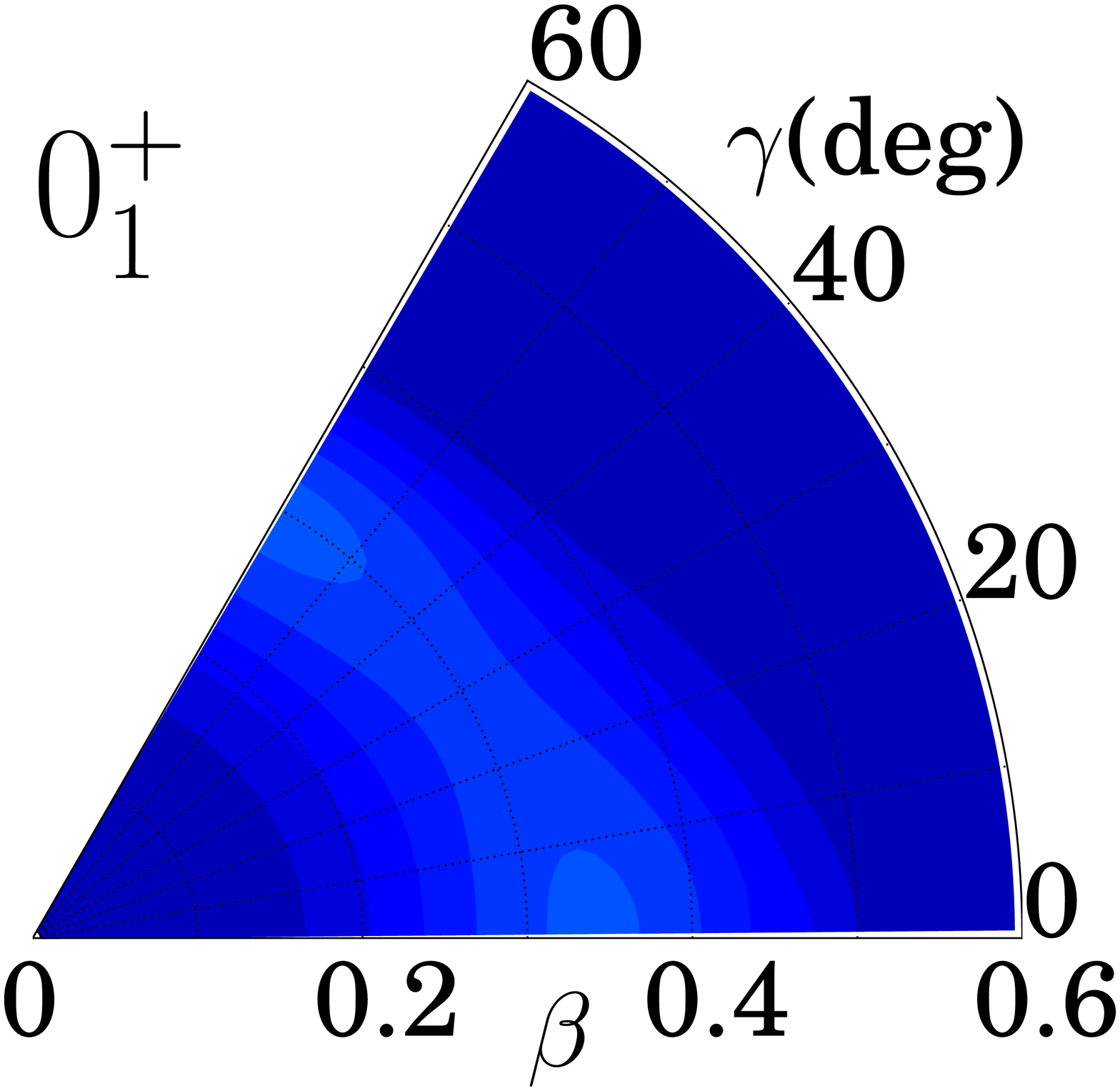} &
 \includegraphics[width=30mm]{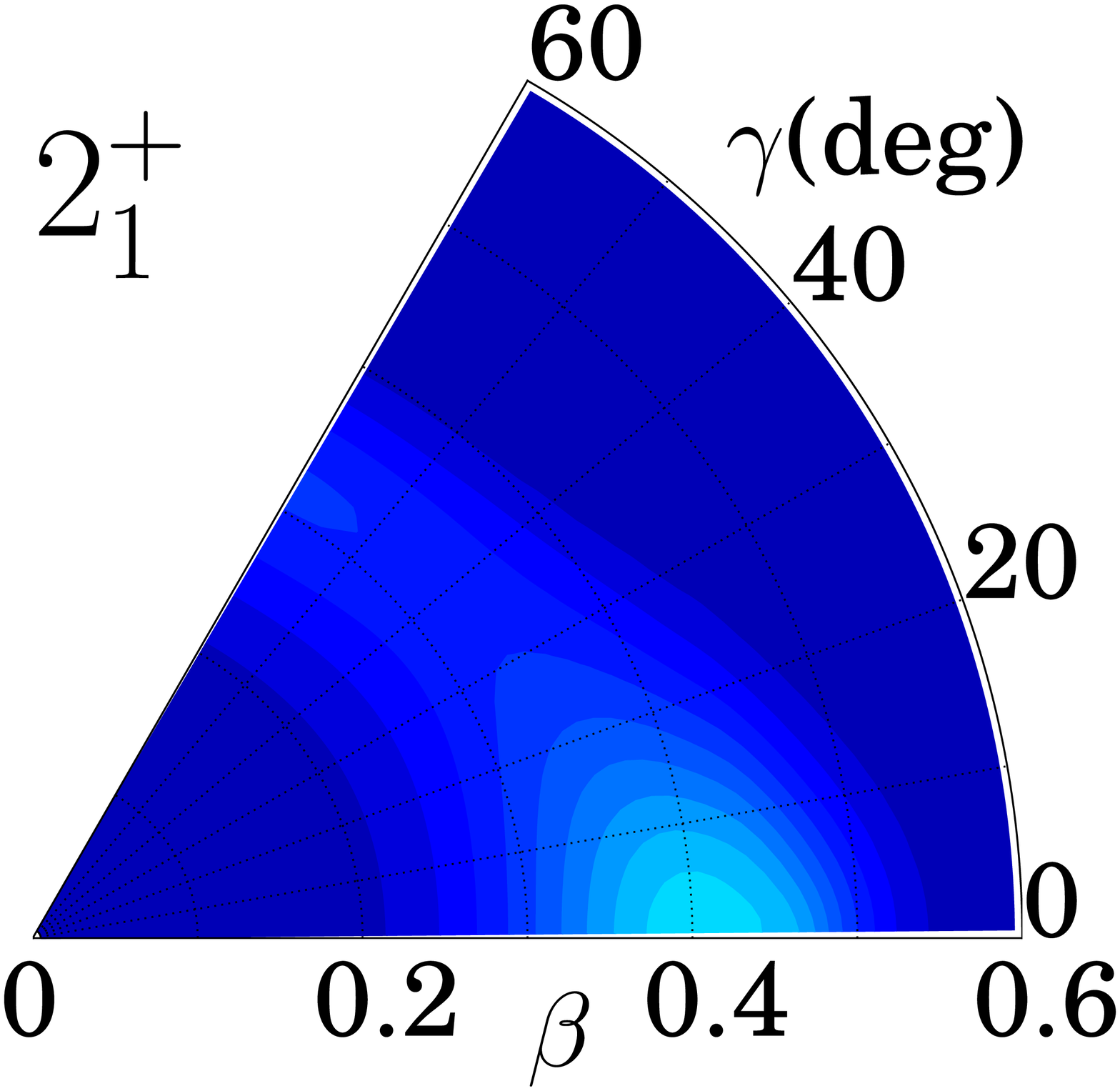} &
 \includegraphics[width=30mm]{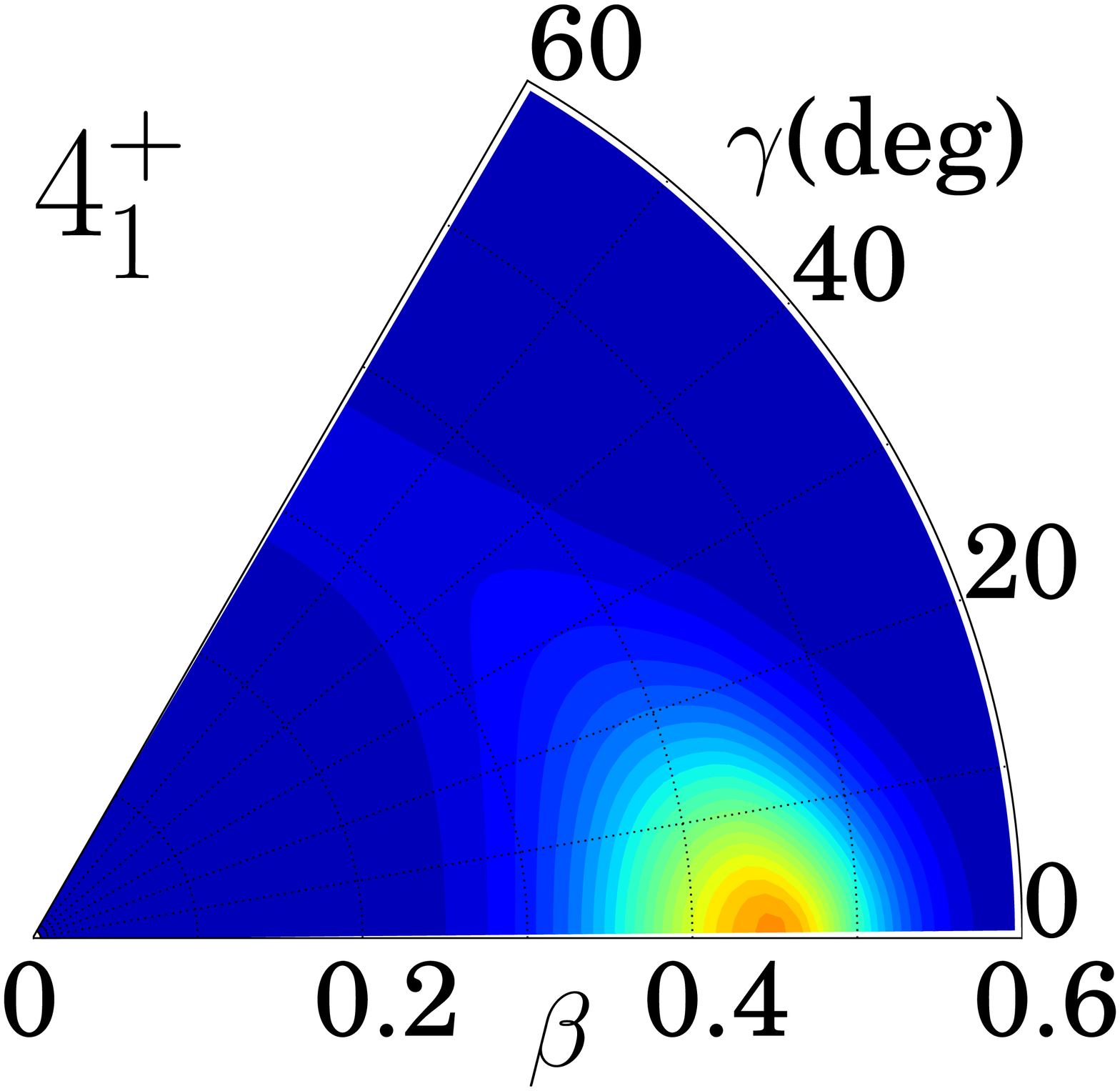} &
 \includegraphics[width=30mm]{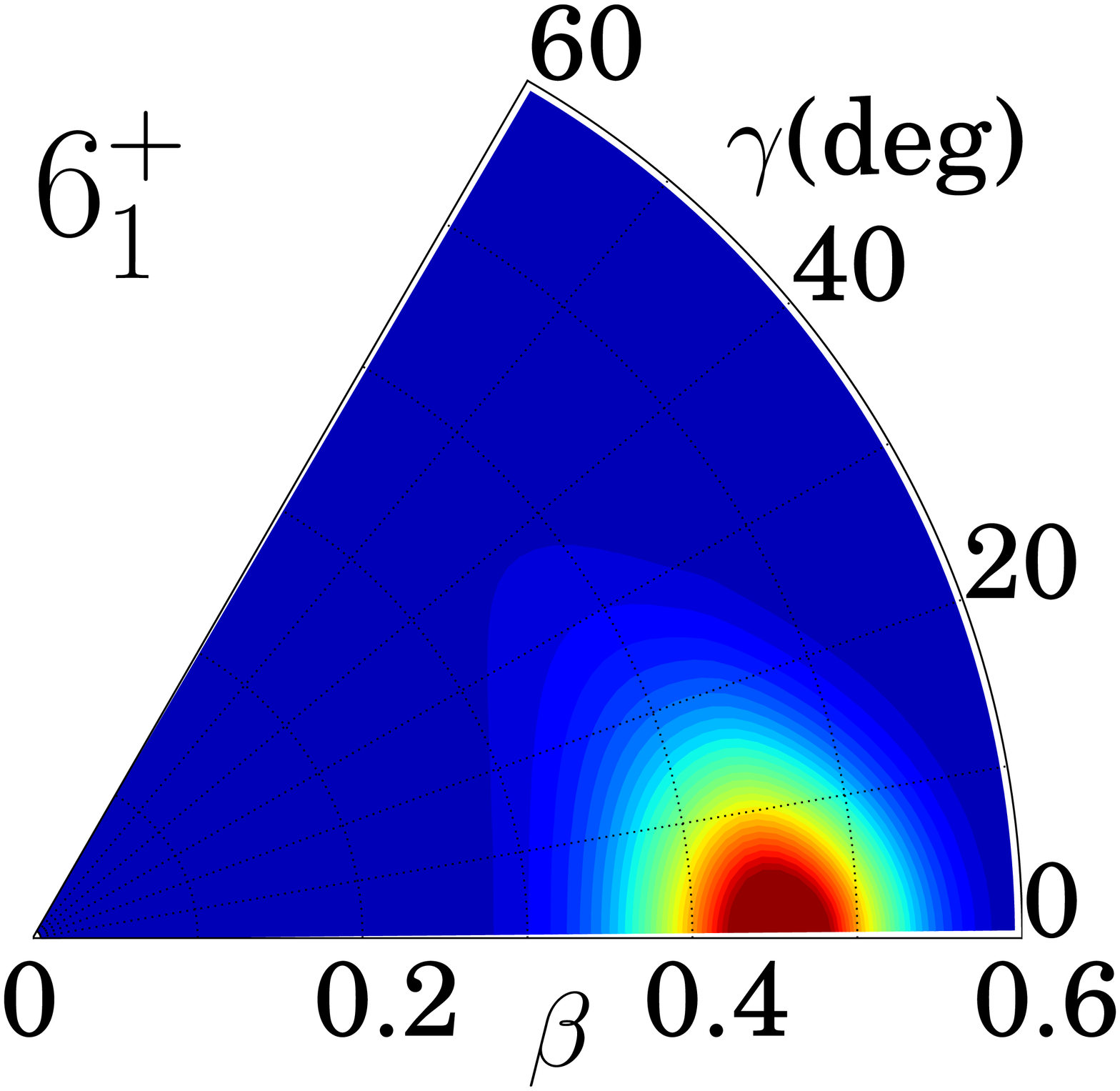} & \\
 \includegraphics[width=30mm]{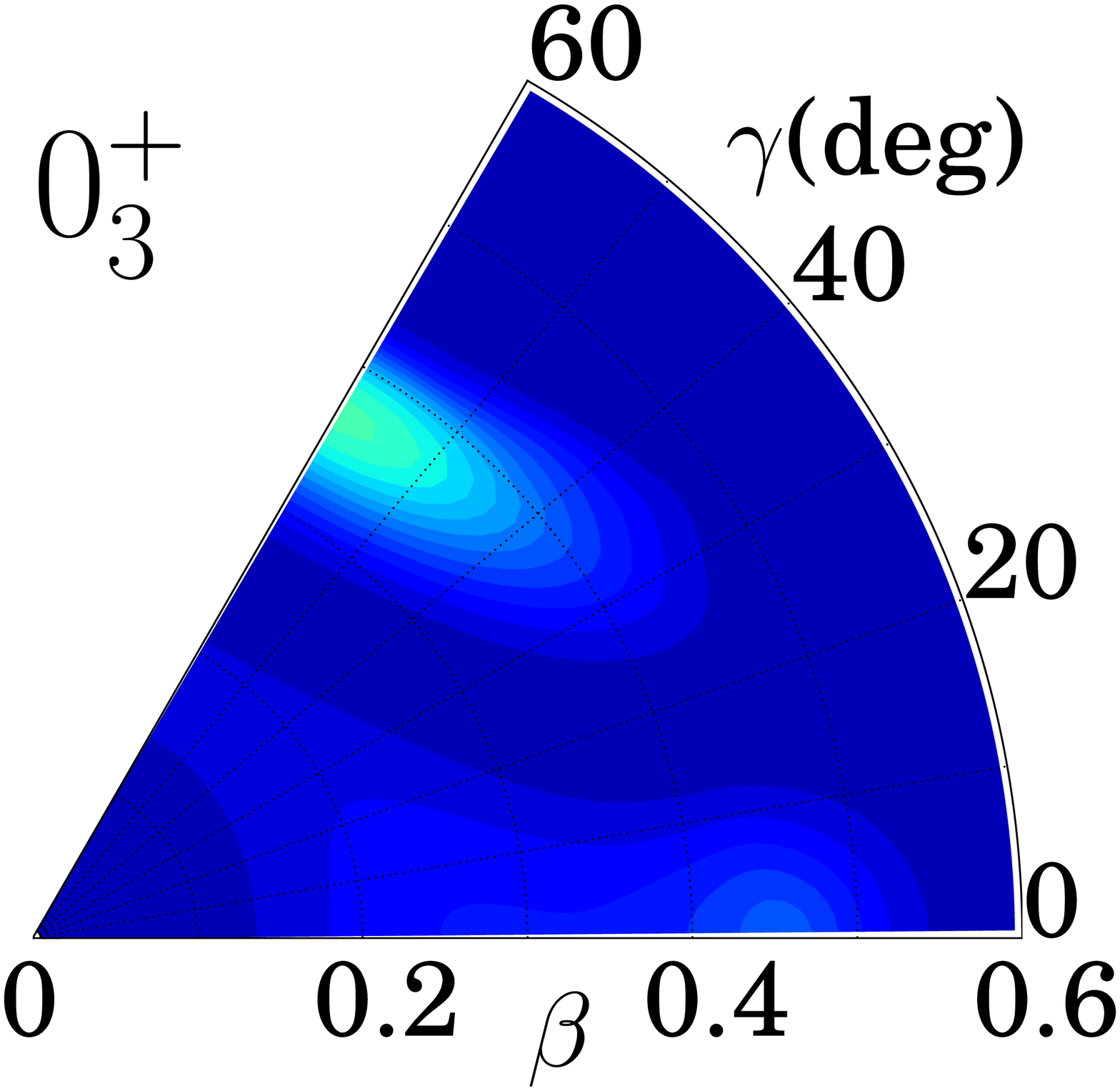} &
 \includegraphics[width=30mm]{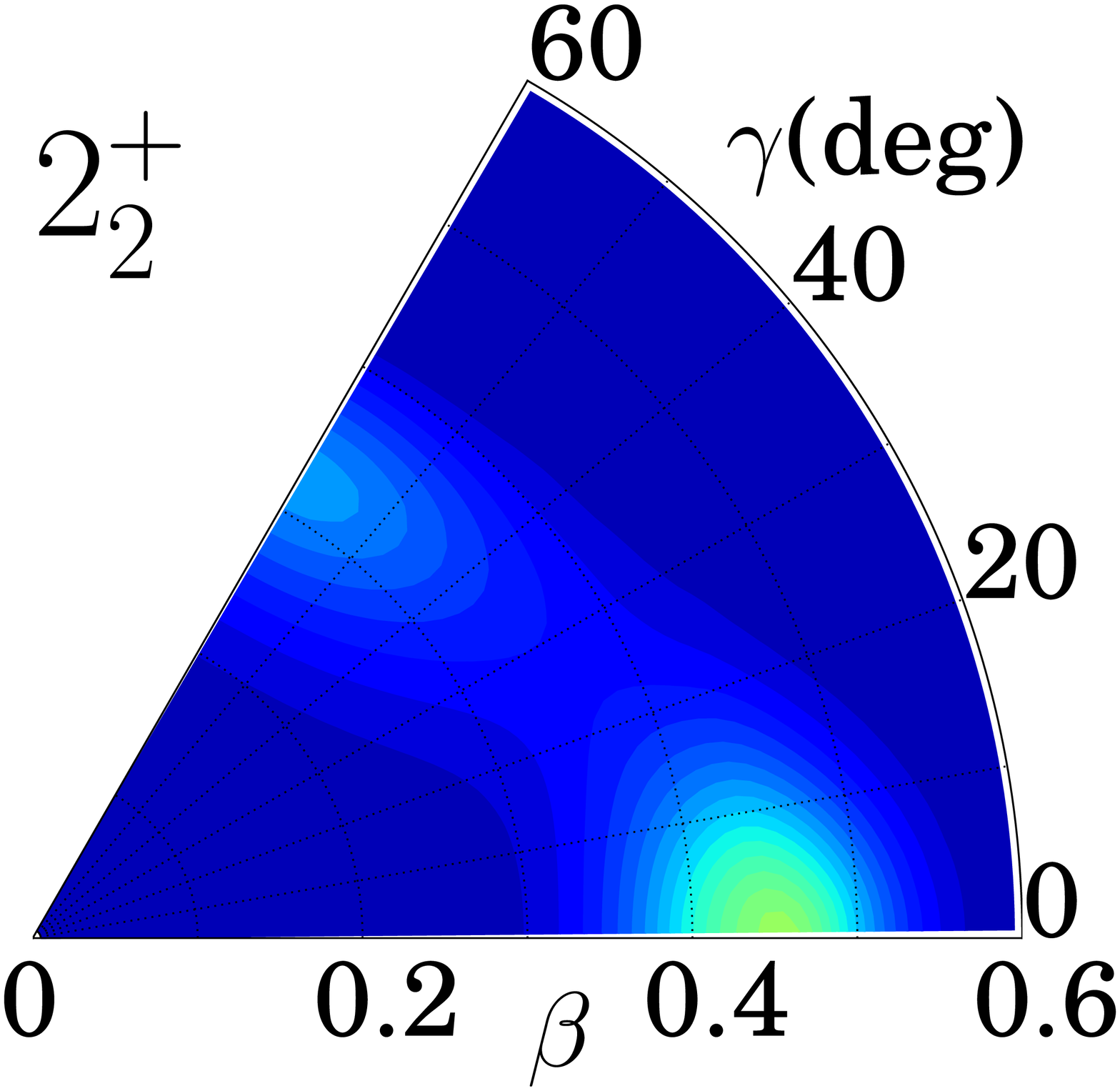} &
 \includegraphics[width=30mm]{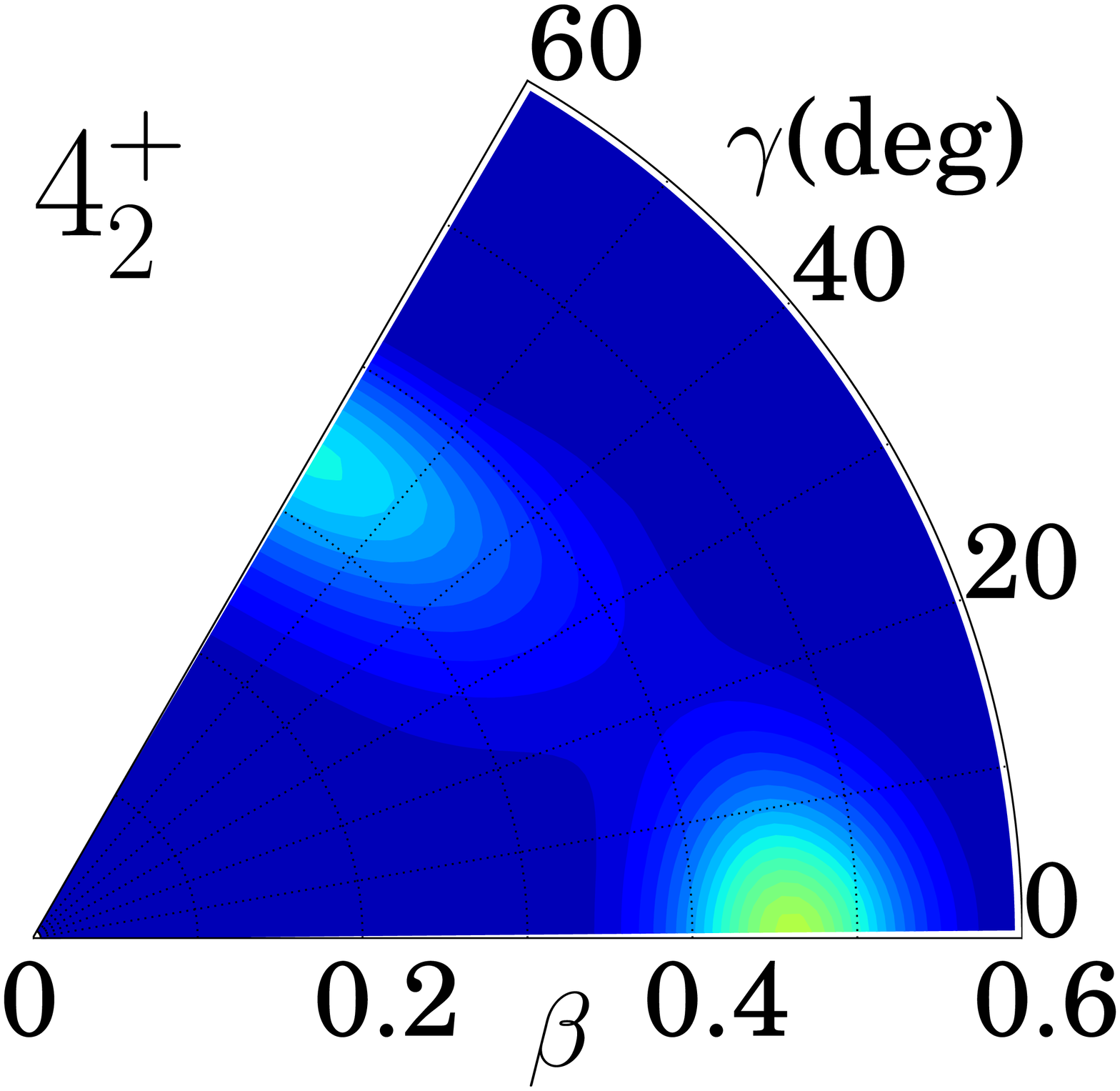} &
 \includegraphics[width=30mm]{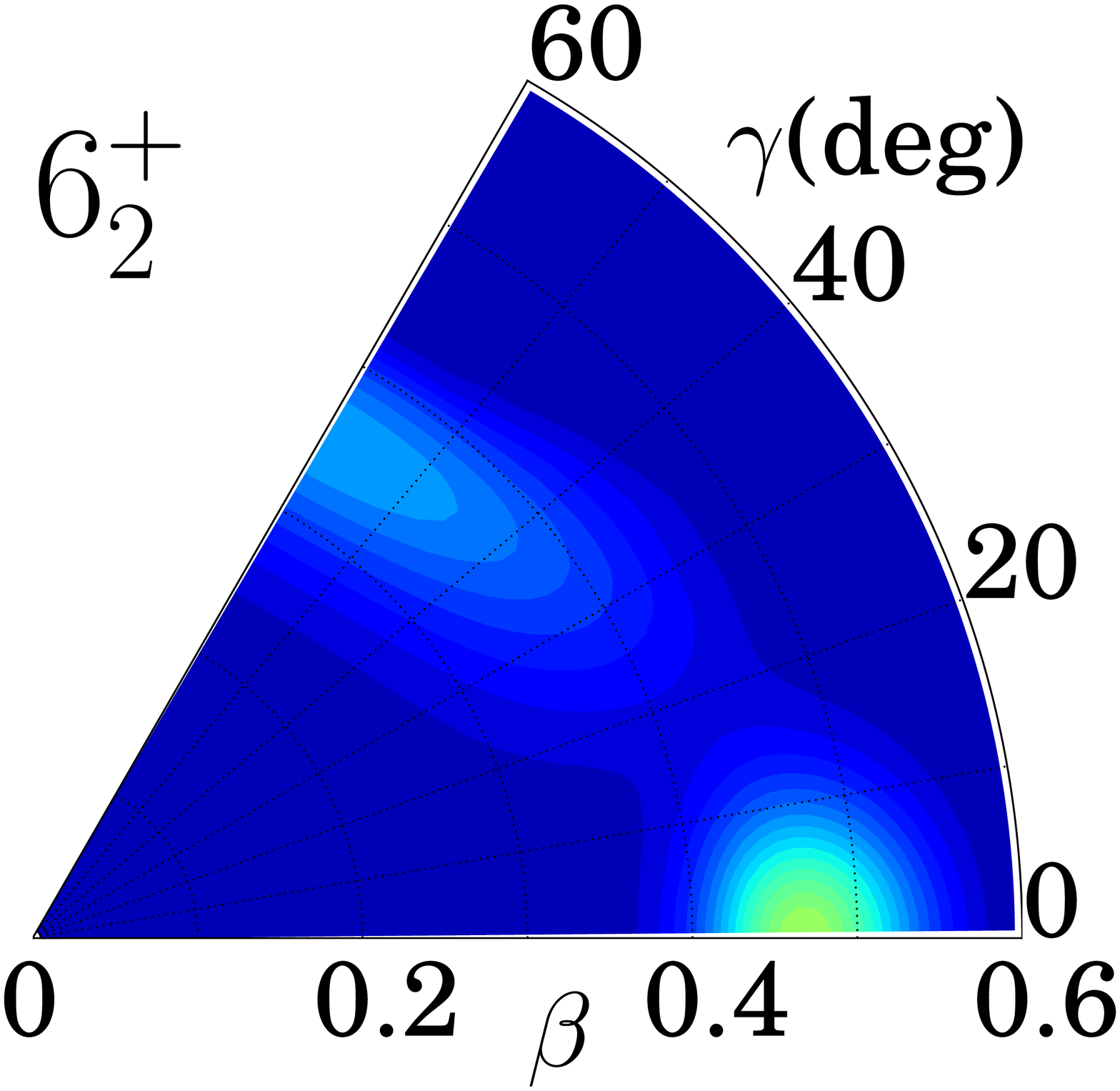} &
 \includegraphics[height=25mm]{colorbar.eps} \\
 \includegraphics[width=30mm]{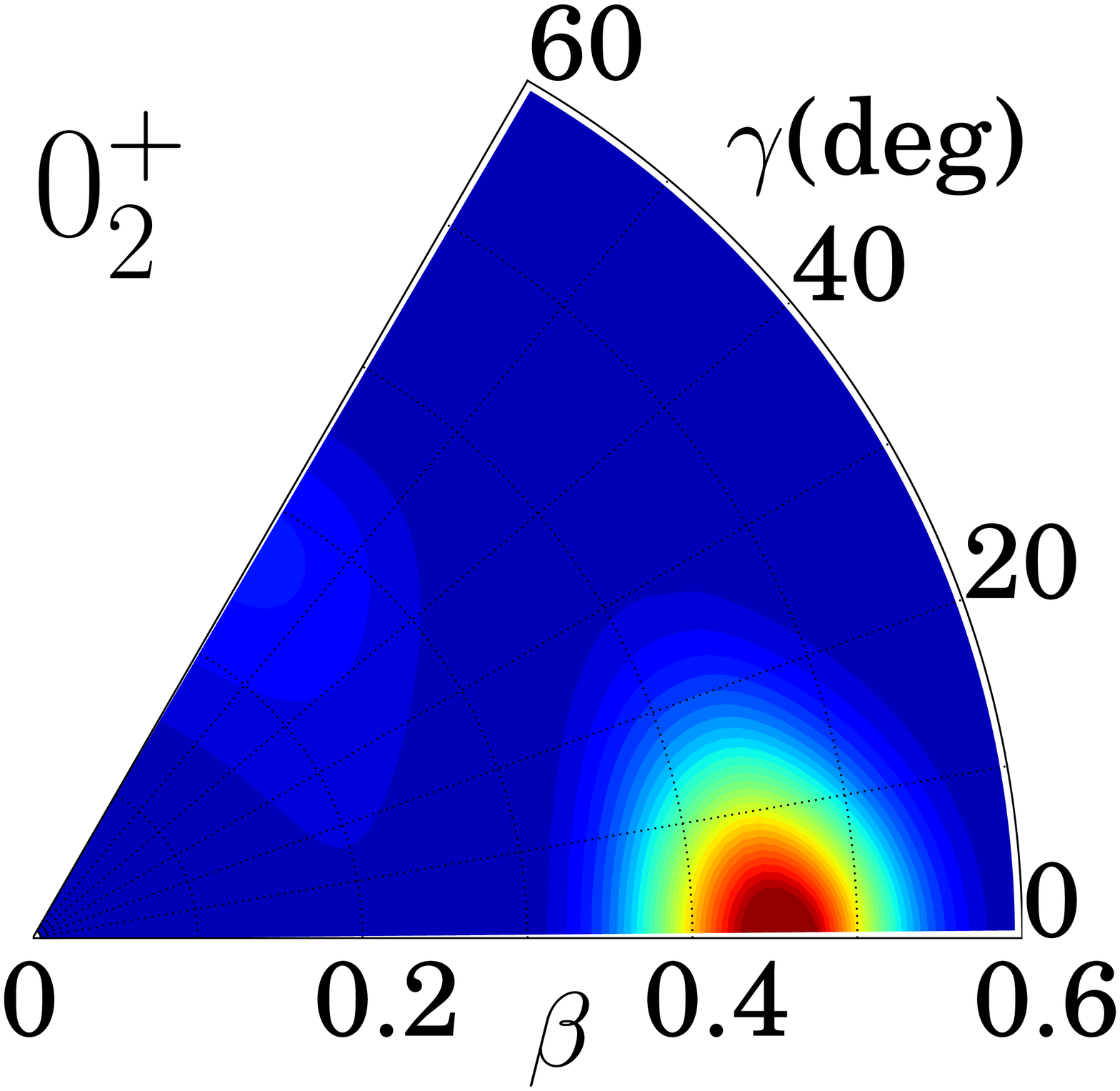} &
 \includegraphics[width=30mm]{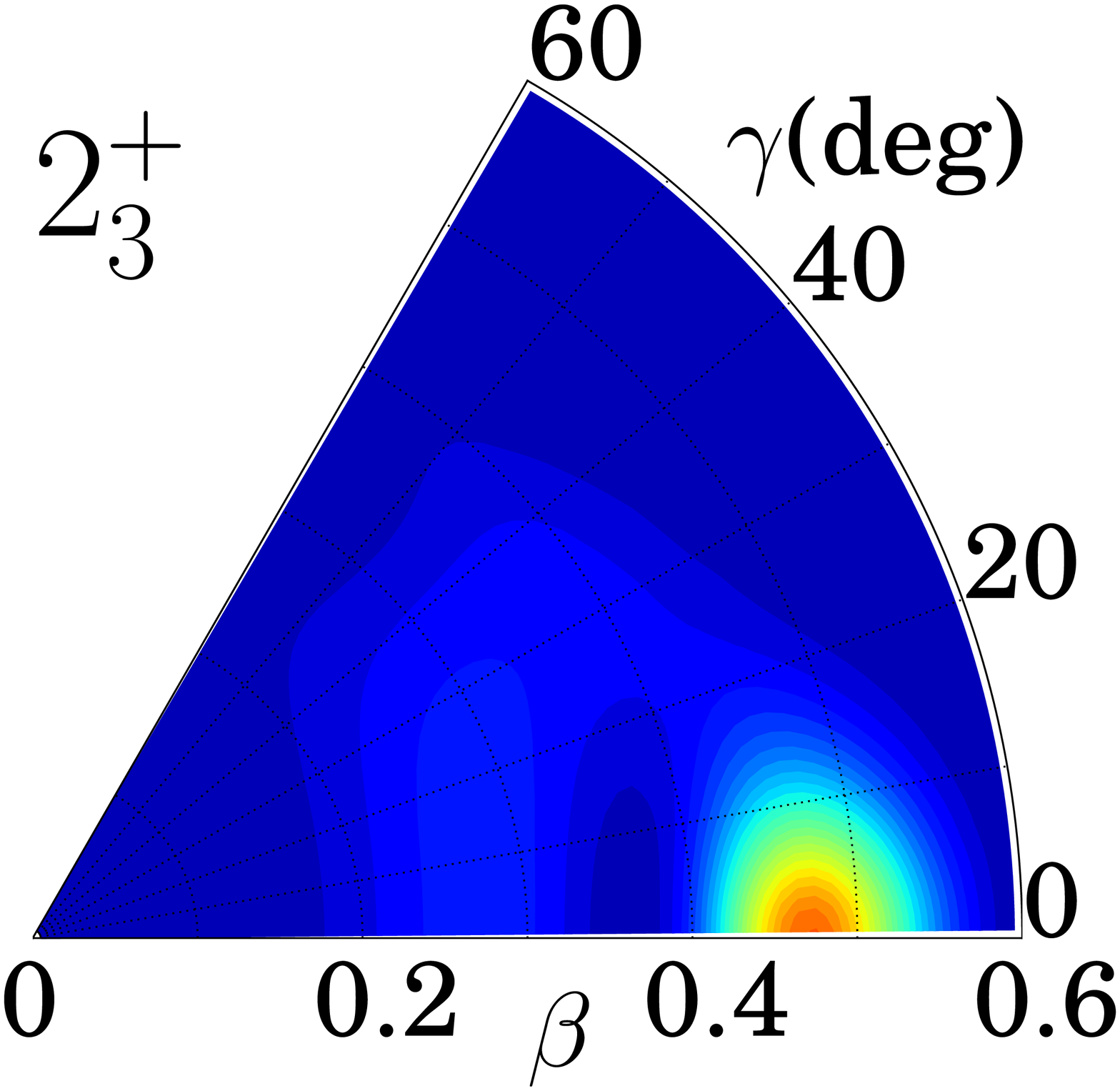} & & &
\end{tabular}
\caption{\label{fig:72Se-wave}(Color online) 
Same as Fig.~\ref{fig:68Se-wave} but for $^{72}$Se.}
\end{figure*}

We have calculated collective wave functions
solving the collective Schr\"odinger equation  (\ref{eq:Schroedinger}) 
and evaluated excitation spectra, quadrupole transition probabilities, 
and spectroscopic quadrupole moments. 
The results for low-lying states in $^{68,70,72}$Se are presented in
Figs.~\ref{fig:68Se_energy}--\ref{fig:72Se-wave}. 


\noindent

\begin{figure}
\begin{center}
 \includegraphics[width=80mm]{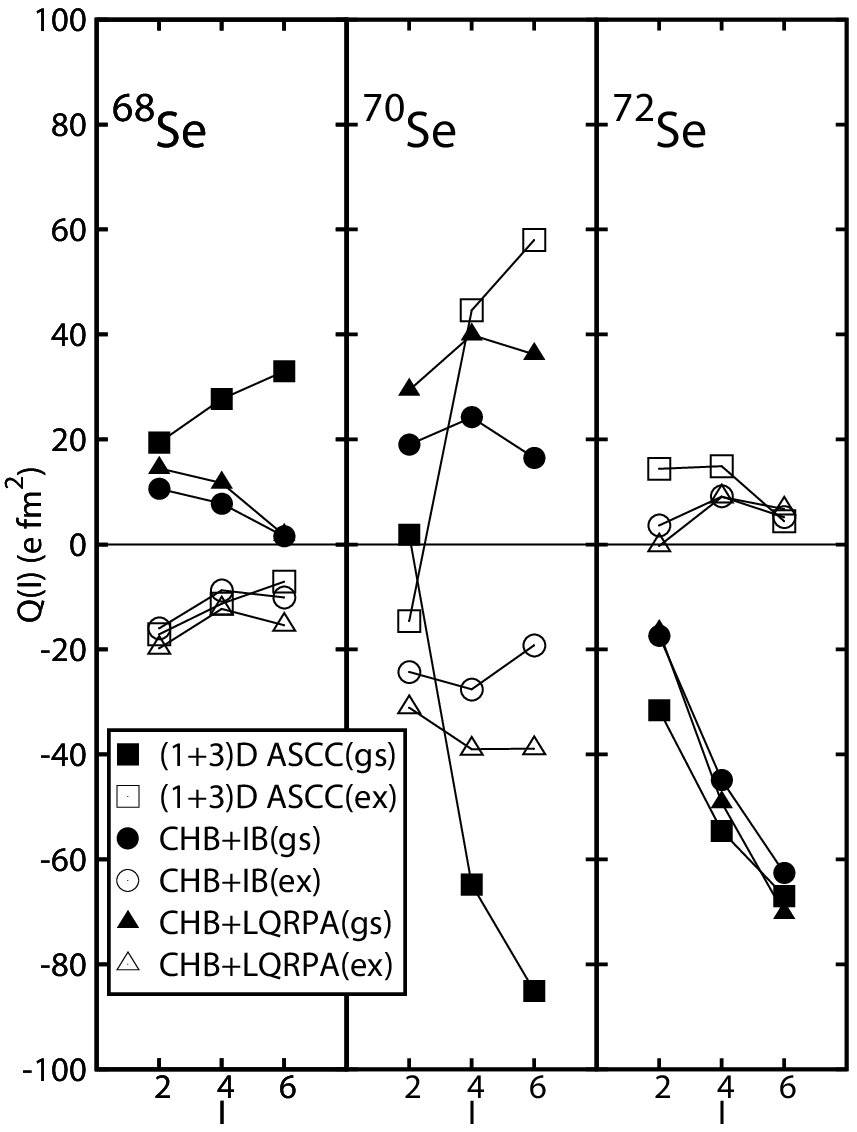}
\end{center}
\caption{\label{fig:qmoment}
Spectroscopic quadrupole moments for $^{68,70,72}$Se.
Values calculated with the LQRPA collective masses are shown with the triangles. 
For comparison, values calculated with the IB collective masses and 
those obtained with the (1+3)D version of the ASCC method are also shown 
with the squares and the circles, respectively.  
The filled symbols show the values for the yrast states, 
while the open symbols those for the yrare states.}
\end{figure}


In Figs.~\ref{fig:68Se_energy}, \ref{fig:70Se_energy}, and \ref{fig:72Se_energy}, 
excitation spectra and $B(E2)$ values for $^{68}$Se, $^{70}$Se, and $^{72}$Se, 
calculated with the CHB+LQRPA method,  
are displayed together with experimental data. 
The eigenstates are labeled with $I^\pi = 0^+,   2^+, 4^+$, and $6^+$.
In these figures, results obtained using the IB cranking masses    
are also shown for the sake of comparison. 
Furthermore, the results calculated with the (1+3)D version of the ASCC method   
reported in our previous paper \cite{hinohara:014305} are 
shown also for comparison with the 5D calculations. 
We use the abbreviation (1+3)D to indicate 
that a single collective coordinate along the collective path 
describing large-amplitude vibration and three rotational angles 
associated with the rotational motion are taken into account in these calculations.
The classification of the calculated low-lying states into families of 
two or three rotational bands is made according to the properties of their 
vibrational wave functions. 
These vibrational wave functions are displayed in Figs.~\ref{fig:68Se-wave}, \ref{fig:70Se-wave}, and \ref{fig:72Se-wave}.  
In these figures,  only the $\beta^4$ factor in the volume element (\ref{eq:metric}) 
are multiplied to the vibrational wave functions  squared
leaving the $\sin 3\gamma$ factor aside. 
This is because all vibrational wave functions look like triaxial and the probability 
at the oblate and prolate shapes vanish  if  
the $\sin 3\gamma$ factor is multiplied by them. 

Let us first summarize the results of the CHB+LQRPA calculation.  
The most conspicuous feature of the low-lying states in these proton-rich Se isotopes 
is the dominance of the large-amplitude vibrational motion 
in the triaxial shape degree of freedom. 
In general, the vibrational wave function extends over the triaxial region 
between the oblate $(\gamma=60^\circ$)  and the prolate ($\gamma=0^\circ$) shapes. 
In particular, this is the case for the $0^+$ states causing their peculiar behaviors;  
for instance,  
we obtain two excited $0^+$ states located slightly below or above the $2_2^+$ state. 
Relative positions between these excited states are quite sensitive to the interplay 
of large-amplitude $\gamma$-vibrational modes and the $\beta$-vibrational modes. 
This result of calculation is consistent with the available experimental data 
where the excited $0^+$ state has not yet been found, 
but more experimental data are needed to examine the validity of the theoretical prediction. 
Below, let us examine characteristic features of the theoretical spectra 
more closely for individual nuclei.  

For $^{68}$Se, we obtain the third band in low energy.   
The $0_2^+$ and $2^+_3$ states belonging to this band are also 
shown in Fig.~\ref{fig:68Se_energy}. 
Their vibrational wave functions exhibit nodes in the $\beta$ direction 
(see Fig.~\ref{fig:68Se-wave}) indicating that a $\beta$-vibrational mode is excited on top of 
the large-amplitude $\gamma$ vibrations. 
As a matter of course,  this kind of state is outside of the scope of the 
(1+3)D calculation.
The vibrational wave functions of the yrast $2_1^+$ and $4_1^+$ states 
exhibit localization in a region around the oblate shape, 
while the yrare $2_2^+, 4_2^+$, and $6_2^+$ states  localize 
around the prolate shape. 
It is apparent, however, that all the wave functions significantly extend 
from $\gamma=0^\circ$  to $60^\circ$ over the triaxial region, 
indicating $\gamma$-soft character of these states. 
In particular, the yrare $4_2^+$ and $6_2^+$ wave functions 
exhibit two-peak structure consisting of the prolate and oblate peaks. 
The peaks of the vibrational wave function gradually shift toward 
a region of larger $\beta$ as the angular momentum increases.  
This is a centrifugal effect decreasing the rotational energy by increasing 
the moment of inertia.
In the (1+3)D calculation, this effect is absent  
because the collective path is fixed at the ground state. 
Thus, the 5D calculation yields, for example, a much larger value for   
$B(E2;6_1^+\rightarrow 4_1^+)$ in comparison with the (1+3)D calculation. 
Actually, in the 5D CHB+LQRPA calculation, 
the wave function of the yrast $6_1^+$ state localizes in the triaxial region 
(see Fig.~\ref{fig:68Se-wave}) 
where the moment of inertia takes a maximum value.   
This leads to a small value for the spectroscopic quadrupole moment (see Fig.~\ref{fig:qmoment}) 
because of the cancellation between the contributions 
from the oblate-like and prolate-like regions.
This cancellation mechanism due to the large-amplitude $\gamma$ fluctuation 
is effective also in other states; 
although the spectroscopic quadrupole moments of the yrast  
$2_1^+$ and $4_1^+$ (yrare $2_2^+, 4_2^+$, and $6_2^+$) states 
are positive (negative) indicating their oblate-like (prolate-like) character,  
their absolute magnitudes are rather small. 
 
The $E2$-transition probabilities exhibit a pattern reminiscent of 
the $\gamma$-unstable situation; for instance, 
$B(E2; 6_2^+ \to 6_1^+)$, $B(E2; 4_2^+ \to 4_1^+)$,
and $B(E2; 2_2^+ \to 2_1^+)$ are much larger than   
$B(E2; 6_2^+ \to 4_1^+)$, $B(E2; 4_2^+ \to 2_1^+)$,
and $B(E2; 2_2^+ \to 0_1^+)$; see Fig.~\ref{fig:68Se_energy}. 
Thus, the low-lying states in $^{68}$Se may be characterized as 
an intermediate situation between the oblate-prolate shape coexistence and 
the Wilets-Jean $\gamma$-unstable model \cite{PhysRev.102.788}. 
Using the phenomenological Bohr-Mottelson collective Hamiltonian,  
we have shown in Ref.~\cite{PTP.123.129} that it is possible to describe  
the oblate-prolate shape coexistence and  
the $\gamma$-unstable situation in a unified way   
varying a few parameters controlling the degree of oblate-prolate asymmetry 
in the collective potential and the collective masses. 
The two-peak structure seen in the $4_2^+$ and $6_2^+$ states 
may be considered as one of the characteristics of the intermediate situation.   
It thus appears that the excitation spectrum for $^{68}$Se (Fig.~\ref{fig:68Se_energy})  
serves as a typical example of the transitional phenomena from 
the $\gamma$-unstable to the oblate-prolate shape coexistence situations. 

Let us make a comparison between the spectra in Fig.~\ref{fig:68Se_energy}  
obtained with the LQRPA collective masses and that with the IB cranking masses. 
It is obvious that the excitation energies are appreciably overestimated in the latter. 
This result is as expected from the too low values of the IB cranking masses.  
The result of our calculation is in qualitative agreement with the 
HFB-based configuration-mixing calculation reported by Ljungvall et al.
\cite{ljungvall:102502} in that 
both calculations indicate the oblate (prolate) dominance for the 
yrast (yrare) band in $^{68}$Se. 
Quite recently, the $B(E2; 2_1^+ \to 0_1^+)$ value has been measured 
in experiment \cite{obertelli:031304}. 
The calculated value (492 $e^2$fm$^4$) is in fair agreement with 
the experimental data (432 $e^2$fm$^4$).

The result of calculation for $^{70}$Se (Figs.~\ref{fig:70Se_energy} 
and \ref{fig:70Se-wave}) is similar 
to that for  $^{68}$Se. 
The vibrational wave functions of the yrast $2_1^+, 4_1^+$, and $6_1^+$ states 
localize in a region around the oblate shape, exhibiting, at the same time,    
long tails in the triaxial direction. 
We note here that, differently from the $^{68}$Se case, 
the $6_1^+$ wave function keeps the oblate-like structure. 
On the other hand, the yrare $2_2^+, 4_2^+$, and $6_2^+$ states 
localize around the prolate shape, exhibiting, at the same time,    
small secondary bumps around the oblate shape. 
For the yrare $2_2^+$ state, we obtain a strong  
oblate-prolate shape mixing in the (1+3)D calculation \cite{hinohara:014305}. 
This mixing becomes weaker in the present 5D calculation,  
resulting in the reduction of the $B(E2; 4_1^+ \to 2_2^+)$ value.  
Similarly to $^{68}$Se, we obtain two excited $0^+$ states in low energy. 
We see considerable oblate-prolate shape mixings in their vibrational wave 
functions, but, somewhat differently from those in $^{68}$Se, 
the second and third $0^+$ states in $^{70}$Se 
exhibit clear peaks at the oblate and prolate shapes, respectively,  
Their energy ordering is quite sensitive to the interplay of 
the large-amplitude $\gamma$ vibration and  the $\beta$ vibrational modes.
The calculated spectrum for $^{70}$Se is in fair agreement with  
the recent experimental data \cite{0954-3899-28-10-307} , 
although the $B(E2)$ values between the yrast states are overestimated.  

The result of calculation for $^{72}$Se (Figs.~\ref{fig:72Se_energy} 
and \ref{fig:72Se-wave}) presents  
a feature somewhat different from those for $^{68}$Se and $^{70}$Se; 
that is, the yrast $2_1^+, 4_1^+$, and $6_1^+$ states 
localize around the prolate shape instead of the oblate shape. 
The localization develops with increasing angular momentum.   
On the other hand, similarly to the $^{68,70}$Se cases, 
the yrare $2_2^+, 4_2^+$, and $6_2^+$ states exhibit the two-peak structure. 
The spectroscopic quadrupole moments of the 
$2_1^+, 4_1^+$, and $6_1^+$ states are negative, and  
their absolute magnitude increases with increasing angular momentum 
(see Fig.~\ref{fig:qmoment}) reflecting the developing prolate character in the yrast band, 
while those of the yrare states are small because of the two-peak structure 
of their vibrational wave functions, that is, due to the  
cancellation of the contributions from the prolate-like and oblate-like regions.   
Also for $^{72}$Se, we obtain two excited $0^+$ states in low energy, 
but they show features somewhat different from the corresponding 
excited $0^+$ states in $^{68,70}$Se.  
Specifically, the vibrational wave functions of 
the second and third $0^+$ states 
exhibit peaks at the prolate and oblate shape, respectively. 
As seen in Fig.~\ref{fig:72Se_energy}, 
our results of calculation for the excitation energies and $B(E2)$ values are 
in good agreement with the recent experimental data \cite{ljungvall:102502}  
for the yrast  $2_1^+, 4_1^+$, and $6_1^+$ states in  $^{72}$Se. 
Experimental $E2$-transition data are awaited 
for understanding the nature of the observed excited band. 

\section{\label{sec:conclusion}Conclusions}

On the basis of the ASCC method, 
we have developed a practical microscopic approach, called CHFB+LQRPA,  
of deriving the 5D quadrupole collective Hamiltonian 
and confirmed its efficiency by applying it to the oblate-prolate shape 
coexistence/mixing phenomena in proton-rich $^{68,70,72}$Se.   
The results of numerical calculation for the excitation energies and $B(E2)$ values 
are in good agreement with the recent experimental data 
\cite{obertelli:031304,ljungvall:102502}  
for the yrast  $2_1^+, 4_1^+$, and $6_1^+$ states in these nuclei.
It is shown that the time-odd components of the moving mean-field 
significantly increase the vibrational and rotational collective masses and 
make the theoretical spectra in much better agreement with the experimental data 
than calculations using the IB cranking masses. 
Our analysis clearly indicates that 
low-lying states in these nuclei possess a transitional character   
between the oblate-prolate shape coexistence 
and the so-called $\gamma$ unstable situation 
where large-amplitude triaxial-shape fluctuations play a dominant role.  

Finally, we would like to list a few issues for the future 
that seems particularly interesting.
First, fully self-consistent solution of the ASCC equations for 
determining the two-dimensional collective hypersurface  
and examination of the validity of the approximations 
adopted in this paper in the derivation of the CHFB+LQRPA scheme.  
Second, application to various kind of collective spectra 
associated with large-amplitude collective motions near the yrast 
lines (as listed in Ref.~\cite{0954-3899-37-6-064018}). 
Third, possible extension of the quadrupole collective Hamiltonian
by explicitly treating the pairing vibrational degrees of freedom as 
additional collective coordinates.  
Fourth, use of the Skyrme energy functionals + density-dependent 
contact pairing interaction in place of the P+Q force,  
and then modern density functionals currently under active development. 
Fifth, application of the CHFB+LQRPA scheme to fission dynamics. 
The LQRPA approach enables us to evaluate, without need of numerical 
derivatives, the collective inertia masses including the time-odd mean-field 
effects.

\begin{acknowledgments}
Two of the authors (K. S. and N. H.) are supported by the Junior Research Associate
Program and the Special Postdoctoral Researcher Program of RIKEN, respectively. 
The numerical calculations were carried out on Altix3700 BX2 at Yukawa 
Institute for Theoretical Physics in Kyoto University
and RIKEN Cluster of Clusters (RICC) facility.
This work is supported by Grants-in-Aid for Scientific Research
(Nos. 20105003, 20540259, and 21340073) from the Japan Society for the 
Promotion of Science and the JSPS Core-to-Core
Program ``International Research Network for Exotic Femto Systems.''
\end{acknowledgments}

\bibliography{../../../Bibtex/paper}

\begin{thebibliography}{56}
\expandafter\ifx\csname natexlab\endcsname\relax\def\natexlab#1{#1}\fi
\expandafter\ifx\csname bibnamefont\endcsname\relax
  \def\bibnamefont#1{#1}\fi
\expandafter\ifx\csname bibfnamefont\endcsname\relax
  \def\bibfnamefont#1{#1}\fi
\expandafter\ifx\csname citenamefont\endcsname\relax
  \def\citenamefont#1{#1}\fi
\expandafter\ifx\csname url\endcsname\relax
  \def\url#1{\texttt{#1}}\fi
\expandafter\ifx\csname urlprefix\endcsname\relax\def\urlprefix{URL }\fi
\providecommand{\bibinfo}[2]{#2}
\providecommand{\eprint}[2][]{\url{#2}}

\bibitem[{\citenamefont{Bohr and Mottelson}(1998)}]{BMvol2}
\bibinfo{author}{\bibfnamefont{A.}~\bibnamefont{Bohr}} \bibnamefont{and}
  \bibinfo{author}{\bibfnamefont{B.~R.} \bibnamefont{Mottelson}},
  \emph{\bibinfo{title}{Nuclear Structure}}, vol.~\bibinfo{volume}{II}
  (\bibinfo{publisher}{W-A. Benjamin Inc., 1975; World Scientific},
  \bibinfo{year}{1998}).

\bibitem[{\citenamefont{Belyaev}(1965)}]{Belyaev196517}
\bibinfo{author}{\bibfnamefont{S.~T.} \bibnamefont{Belyaev}},
  \bibinfo{journal}{Nucl. Phys.} \textbf{\bibinfo{volume}{64}},
  \bibinfo{pages}{17 } (\bibinfo{year}{1965}).

\bibitem[{\citenamefont{Kumar and Baranger}(1967)}]{Kumar1967608}
\bibinfo{author}{\bibfnamefont{K.}~\bibnamefont{Kumar}} \bibnamefont{and}
  \bibinfo{author}{\bibfnamefont{M.}~\bibnamefont{Baranger}},
  \bibinfo{journal}{Nucl. Phys. A} \textbf{\bibinfo{volume}{92}},
  \bibinfo{pages}{608 } (\bibinfo{year}{1967}).

\bibitem[{\citenamefont{Pr\'{o}chniak and
  Rohozi\'{n}ski}(2009)}]{0954-3899-36-12-123101}
\bibinfo{author}{\bibfnamefont{L.}~\bibnamefont{Pr\'{o}chniak}}
  \bibnamefont{and} \bibinfo{author}{\bibfnamefont{S.~G.}
  \bibnamefont{Rohozi\'{n}ski}}, \bibinfo{journal}{J. Phys. G}
  \textbf{\bibinfo{volume}{36}}, \bibinfo{pages}{123101}
  (\bibinfo{year}{2009}).

\bibitem[{\citenamefont{Fischer et~al.}(2003)\citenamefont{Fischer, Lister, and
  Balamuth}}]{PhysRevC.67.064318}
\bibinfo{author}{\bibfnamefont{S.~M.} \bibnamefont{Fischer}},
  \bibinfo{author}{\bibfnamefont{C.~J.} \bibnamefont{Lister}},
  \bibnamefont{and} \bibinfo{author}{\bibfnamefont{D.~P.}
  \bibnamefont{Balamuth}}, \bibinfo{journal}{Phys. Rev. C}
  \textbf{\bibinfo{volume}{67}}, \bibinfo{pages}{064318}
  (\bibinfo{year}{2003}).

\bibitem[{\citenamefont{Fischer et~al.}(2000)\citenamefont{Fischer, Balamuth,
  Hausladen, Lister, Carpenter, Seweryniak, and
  Schwartz}}]{PhysRevLett.84.4064}
\bibinfo{author}{\bibfnamefont{S.~M.} \bibnamefont{Fischer}},
  \bibinfo{author}{\bibfnamefont{D.~P.} \bibnamefont{Balamuth}},
  \bibinfo{author}{\bibfnamefont{P.~A.} \bibnamefont{Hausladen}},
  \bibinfo{author}{\bibfnamefont{C.~J.} \bibnamefont{Lister}},
  \bibinfo{author}{\bibfnamefont{M.~P.} \bibnamefont{Carpenter}},
  \bibinfo{author}{\bibfnamefont{D.}~\bibnamefont{Seweryniak}},
  \bibnamefont{and} \bibinfo{author}{\bibfnamefont{J.}~\bibnamefont{Schwartz}},
  \bibinfo{journal}{Phys. Rev. Lett.} \textbf{\bibinfo{volume}{84}},
  \bibinfo{pages}{4064} (\bibinfo{year}{2000}).

\bibitem[{\citenamefont{Obertelli et~al.}(2009)\citenamefont{Obertelli,
  Baugher, Bazin, Delaroche, Flavigny, Gade, Girod, Glasmacher, G\"{o}rgen,
  Grinyer et~al.}}]{obertelli:031304}
\bibinfo{author}{\bibfnamefont{A.}~\bibnamefont{Obertelli}},
  \bibinfo{author}{\bibfnamefont{T.}~\bibnamefont{Baugher}},
  \bibinfo{author}{\bibfnamefont{D.}~\bibnamefont{Bazin}},
  \bibinfo{author}{\bibfnamefont{J.~P.} \bibnamefont{Delaroche}},
  \bibinfo{author}{\bibfnamefont{F.}~\bibnamefont{Flavigny}},
  \bibinfo{author}{\bibfnamefont{A.}~\bibnamefont{Gade}},
  \bibinfo{author}{\bibfnamefont{M.}~\bibnamefont{Girod}},
  \bibinfo{author}{\bibfnamefont{T.}~\bibnamefont{Glasmacher}},
  \bibinfo{author}{\bibfnamefont{A.}~\bibnamefont{G\"{o}rgen}},
  \bibinfo{author}{\bibfnamefont{G.~F.} \bibnamefont{Grinyer}},
  \bibnamefont{et~al.}, \bibinfo{journal}{Phys. Rev. C}
  \textbf{\bibinfo{volume}{80}}, \bibinfo{eid}{031304} (\bibinfo{year}{2009}).

\bibitem[{\citenamefont{Ljungvall et~al.}(2008)\citenamefont{Ljungvall,
  G\"orgen, Girod, Delaroche, Dewald, Dossat, Farnea, Korten, Melon, Menegazzo
  et~al.}}]{ljungvall:102502}
\bibinfo{author}{\bibfnamefont{J.}~\bibnamefont{Ljungvall}},
  \bibinfo{author}{\bibfnamefont{A.}~\bibnamefont{G\"orgen}},
  \bibinfo{author}{\bibfnamefont{M.}~\bibnamefont{Girod}},
  \bibinfo{author}{\bibfnamefont{J.-P.} \bibnamefont{Delaroche}},
  \bibinfo{author}{\bibfnamefont{A.}~\bibnamefont{Dewald}},
  \bibinfo{author}{\bibfnamefont{C.}~\bibnamefont{Dossat}},
  \bibinfo{author}{\bibfnamefont{E.}~\bibnamefont{Farnea}},
  \bibinfo{author}{\bibfnamefont{W.}~\bibnamefont{Korten}},
  \bibinfo{author}{\bibfnamefont{B.}~\bibnamefont{Melon}},
  \bibinfo{author}{\bibfnamefont{R.}~\bibnamefont{Menegazzo}},
  \bibnamefont{et~al.}, \bibinfo{journal}{Phys. Rev. Lett.}
  \textbf{\bibinfo{volume}{100}}, \bibinfo{eid}{102502} (\bibinfo{year}{2008}).

\bibitem[{\citenamefont{Inglis}(1954)}]{PhysRev.96.1059}
\bibinfo{author}{\bibfnamefont{D.~R.} \bibnamefont{Inglis}},
  \bibinfo{journal}{Phys. Rev.} \textbf{\bibinfo{volume}{96}},
  \bibinfo{pages}{1059} (\bibinfo{year}{1954}).

\bibitem[{\citenamefont{Beliaev}(1961)}]{Beliaev1961322}
\bibinfo{author}{\bibfnamefont{S.~T.} \bibnamefont{Beliaev}},
  \bibinfo{journal}{Nucl. Phys.} \textbf{\bibinfo{volume}{24}},
  \bibinfo{pages}{322 } (\bibinfo{year}{1961}).

\bibitem[{\citenamefont{Kumar}(1974)}]{Kumar1974189}
\bibinfo{author}{\bibfnamefont{K.}~\bibnamefont{Kumar}},
  \bibinfo{journal}{Nucl. Phys. A} \textbf{\bibinfo{volume}{231}},
  \bibinfo{pages}{189 } (\bibinfo{year}{1974}).

\bibitem[{\citenamefont{Pomorski et~al.}(1977)\citenamefont{Pomorski,
  Kaniowska, Sobiczewski, and Rohozi\'{n}ski}}]{Pomorski1977394}
\bibinfo{author}{\bibfnamefont{K.}~\bibnamefont{Pomorski}},
  \bibinfo{author}{\bibfnamefont{T.}~\bibnamefont{Kaniowska}},
  \bibinfo{author}{\bibfnamefont{A.}~\bibnamefont{Sobiczewski}},
  \bibnamefont{and} \bibinfo{author}{\bibfnamefont{S.~G.}
  \bibnamefont{Rohozi\'{n}ski}}, \bibinfo{journal}{Nucl. Phys. A}
  \textbf{\bibinfo{volume}{283}}, \bibinfo{pages}{394 } (\bibinfo{year}{1977}).

\bibitem[{\citenamefont{Rohozi\'{n}ski
  et~al.}(1977)\citenamefont{Rohozi\'{n}ski, Dobaczewski, Nerlo-Pomorska,
  Pomorski, and Srebrny}}]{Rohozinski197766}
\bibinfo{author}{\bibfnamefont{S.~G.} \bibnamefont{Rohozi\'{n}ski}},
  \bibinfo{author}{\bibfnamefont{J.}~\bibnamefont{Dobaczewski}},
  \bibinfo{author}{\bibfnamefont{B.}~\bibnamefont{Nerlo-Pomorska}},
  \bibinfo{author}{\bibfnamefont{K.}~\bibnamefont{Pomorski}}, \bibnamefont{and}
  \bibinfo{author}{\bibfnamefont{J.}~\bibnamefont{Srebrny}},
  \bibinfo{journal}{Nucl. Phys. A} \textbf{\bibinfo{volume}{292}},
  \bibinfo{pages}{66 } (\bibinfo{year}{1977}).

\bibitem[{\citenamefont{Dudek et~al.}(1980)\citenamefont{Dudek, Dudek,
  Ruchowska, and Skalski}}]{Z.Phys.A294.341Dudek}
\bibinfo{author}{\bibfnamefont{J.}~\bibnamefont{Dudek}},
  \bibinfo{author}{\bibfnamefont{W.}~\bibnamefont{Dudek}},
  \bibinfo{author}{\bibfnamefont{E.}~\bibnamefont{Ruchowska}},
  \bibnamefont{and} \bibinfo{author}{\bibfnamefont{J.}~\bibnamefont{Skalski}},
  \bibinfo{journal}{Z. Phys. A} \textbf{\bibinfo{volume}{294}},
  \bibinfo{pages}{341 } (\bibinfo{year}{1980}).

\bibitem[{\citenamefont{Baranger and V\'en\'eroni}(1978)}]{Baranger1978123}
\bibinfo{author}{\bibfnamefont{M.}~\bibnamefont{Baranger}} \bibnamefont{and}
  \bibinfo{author}{\bibfnamefont{M.}~\bibnamefont{V\'en\'eroni}},
  \bibinfo{journal}{Ann. Phys.} \textbf{\bibinfo{volume}{114}},
  \bibinfo{pages}{123 } (\bibinfo{year}{1978}).

\bibitem[{\citenamefont{Dobaczewski and Skalski}(1981)}]{Dobaczewski1981123}
\bibinfo{author}{\bibfnamefont{J.}~\bibnamefont{Dobaczewski}} \bibnamefont{and}
  \bibinfo{author}{\bibfnamefont{J.}~\bibnamefont{Skalski}},
  \bibinfo{journal}{Nucl. Phys. A} \textbf{\bibinfo{volume}{369}},
  \bibinfo{pages}{123 } (\bibinfo{year}{1981}).

\bibitem[{\citenamefont{Villars}(1977)}]{Villars1977269}
\bibinfo{author}{\bibfnamefont{F.}~\bibnamefont{Villars}},
  \bibinfo{journal}{Nucl. Phys. A} \textbf{\bibinfo{volume}{285}},
  \bibinfo{pages}{269 } (\bibinfo{year}{1977}).

\bibitem[{\citenamefont{Goeke and Reinhard}(1978)}]{Goeke1978328}
\bibinfo{author}{\bibfnamefont{K.}~\bibnamefont{Goeke}} \bibnamefont{and}
  \bibinfo{author}{\bibfnamefont{P.-G.} \bibnamefont{Reinhard}},
  \bibinfo{journal}{Ann. Phys.} \textbf{\bibinfo{volume}{112}},
  \bibinfo{pages}{328 } (\bibinfo{year}{1978}).

\bibitem[{\citenamefont{Rowe and Bassermann}(1976)}]{Can.J.Phys.54.1941Rowe}
\bibinfo{author}{\bibfnamefont{D.~J.} \bibnamefont{Rowe}} \bibnamefont{and}
  \bibinfo{author}{\bibfnamefont{R.}~\bibnamefont{Bassermann}},
  \bibinfo{journal}{Can. J. Phys.} \textbf{\bibinfo{volume}{54}},
  \bibinfo{pages}{1941 } (\bibinfo{year}{1976}).

\bibitem[{\citenamefont{Marumori}(1977)}]{PTP.57.112}
\bibinfo{author}{\bibfnamefont{T.}~\bibnamefont{Marumori}},
  \bibinfo{journal}{Prog. Theor. Phys.} \textbf{\bibinfo{volume}{57}},
  \bibinfo{pages}{112} (\bibinfo{year}{1977}).

\bibitem[{\citenamefont{Libert et~al.}(1999)\citenamefont{Libert, Girod, and
  Delaroche}}]{PhysRevC.60.054301}
\bibinfo{author}{\bibfnamefont{J.}~\bibnamefont{Libert}},
  \bibinfo{author}{\bibfnamefont{M.}~\bibnamefont{Girod}}, \bibnamefont{and}
  \bibinfo{author}{\bibfnamefont{J.-P.} \bibnamefont{Delaroche}},
  \bibinfo{journal}{Phys. Rev. C} \textbf{\bibinfo{volume}{60}},
  \bibinfo{pages}{054301} (\bibinfo{year}{1999}).

\bibitem[{\citenamefont{Pal et~al.}(1975)\citenamefont{Pal, Zawischa, and
  Speth}}]{ZPhysA272_387}
\bibinfo{author}{\bibfnamefont{M.~K.} \bibnamefont{Pal}},
  \bibinfo{author}{\bibfnamefont{D.}~\bibnamefont{Zawischa}}, \bibnamefont{and}
  \bibinfo{author}{\bibfnamefont{J.}~\bibnamefont{Speth}}, \bibinfo{journal}{Z.
  Phys. A} \textbf{\bibinfo{volume}{272}}, \bibinfo{pages}{387 }
  (\bibinfo{year}{1975}).

\bibitem[{\citenamefont{Walet et~al.}(1991)\citenamefont{Walet, Do~Dang, and
  Klein}}]{PhysRevC.43.2254}
\bibinfo{author}{\bibfnamefont{N.~R.} \bibnamefont{Walet}},
  \bibinfo{author}{\bibfnamefont{G.}~\bibnamefont{Do~Dang}}, \bibnamefont{and}
  \bibinfo{author}{\bibfnamefont{A.}~\bibnamefont{Klein}},
  \bibinfo{journal}{Phys. Rev. C} \textbf{\bibinfo{volume}{43}},
  \bibinfo{pages}{2254} (\bibinfo{year}{1991}).

\bibitem[{\citenamefont{Almehed and Walet}(2004)}]{Almehed2004163}
\bibinfo{author}{\bibfnamefont{D.}~\bibnamefont{Almehed}} \bibnamefont{and}
  \bibinfo{author}{\bibfnamefont{N.~R.} \bibnamefont{Walet}},
  \bibinfo{journal}{Phys. Lett. B} \textbf{\bibinfo{volume}{604}},
  \bibinfo{pages}{163 } (\bibinfo{year}{2004}).

\bibitem[{\citenamefont{Marumori et~al.}(1980)\citenamefont{Marumori, Maskawa,
  Sakata, and Kuriyama}}]{PTP.64.1294}
\bibinfo{author}{\bibfnamefont{T.}~\bibnamefont{Marumori}},
  \bibinfo{author}{\bibfnamefont{T.}~\bibnamefont{Maskawa}},
  \bibinfo{author}{\bibfnamefont{F.}~\bibnamefont{Sakata}}, \bibnamefont{and}
  \bibinfo{author}{\bibfnamefont{A.}~\bibnamefont{Kuriyama}},
  \bibinfo{journal}{Prog. Theor. Phys.} \textbf{\bibinfo{volume}{64}},
  \bibinfo{pages}{1294} (\bibinfo{year}{1980}).

\bibitem[{\citenamefont{Giannoni and Quentin}(1980)}]{PhysRevC.21.2060}
\bibinfo{author}{\bibfnamefont{M.~J.} \bibnamefont{Giannoni}} \bibnamefont{and}
  \bibinfo{author}{\bibfnamefont{P.}~\bibnamefont{Quentin}},
  \bibinfo{journal}{Phys. Rev. C} \textbf{\bibinfo{volume}{21}},
  \bibinfo{pages}{2060} (\bibinfo{year}{1980}).

\bibitem[{\citenamefont{Dang et~al.}(2000)\citenamefont{Dang, Klein, and
  Walet}}]{Dang200093}
\bibinfo{author}{\bibfnamefont{G.~D.} \bibnamefont{Dang}},
  \bibinfo{author}{\bibfnamefont{A.}~\bibnamefont{Klein}}, \bibnamefont{and}
  \bibinfo{author}{\bibfnamefont{N.~R.} \bibnamefont{Walet}},
  \bibinfo{journal}{Phys. Rep.} \textbf{\bibinfo{volume}{335}},
  \bibinfo{pages}{93 } (\bibinfo{year}{2000}).

\bibitem[{\citenamefont{Matsuyanagi et~al.}(2010)\citenamefont{Matsuyanagi,
  Matsuo, Nakatsukasa, Hinohara, and Sato}}]{0954-3899-37-6-064018}
\bibinfo{author}{\bibfnamefont{K.}~\bibnamefont{Matsuyanagi}},
  \bibinfo{author}{\bibfnamefont{M.}~\bibnamefont{Matsuo}},
  \bibinfo{author}{\bibfnamefont{T.}~\bibnamefont{Nakatsukasa}},
  \bibinfo{author}{\bibfnamefont{N.}~\bibnamefont{Hinohara}}, \bibnamefont{and}
  \bibinfo{author}{\bibfnamefont{K.}~\bibnamefont{Sato}}, \bibinfo{journal}{J.
  Phys. G} \textbf{\bibinfo{volume}{37}}, \bibinfo{pages}{064018}
  (\bibinfo{year}{2010}).

\bibitem[{\citenamefont{Nik\v{s}i\'{c}
  et~al.}(2007)\citenamefont{Nik\v{s}i\'{c}, Vretenar, Lalazissis, and
  Ring}}]{niksic:092502}
\bibinfo{author}{\bibfnamefont{T.}~\bibnamefont{Nik\v{s}i\'{c}}},
  \bibinfo{author}{\bibfnamefont{D.}~\bibnamefont{Vretenar}},
  \bibinfo{author}{\bibfnamefont{G.~A.} \bibnamefont{Lalazissis}},
  \bibnamefont{and} \bibinfo{author}{\bibfnamefont{P.}~\bibnamefont{Ring}},
  \bibinfo{journal}{Phys. Rev. Lett.} \textbf{\bibinfo{volume}{99}},
  \bibinfo{eid}{092502} (\bibinfo{year}{2007}).

\bibitem[{\citenamefont{Nik\v{s}i\'{c}
  et~al.}(2009)\citenamefont{Nik\v{s}i\'{c}, Li, Vretenar, Pr\'{o}chniak, Meng,
  and Ring}}]{niksic:034303}
\bibinfo{author}{\bibfnamefont{T.}~\bibnamefont{Nik\v{s}i\'{c}}},
  \bibinfo{author}{\bibfnamefont{Z.~P.} \bibnamefont{Li}},
  \bibinfo{author}{\bibfnamefont{D.}~\bibnamefont{Vretenar}},
  \bibinfo{author}{\bibfnamefont{L.}~\bibnamefont{Pr\'{o}chniak}},
  \bibinfo{author}{\bibfnamefont{J.}~\bibnamefont{Meng}}, \bibnamefont{and}
  \bibinfo{author}{\bibfnamefont{P.}~\bibnamefont{Ring}},
  \bibinfo{journal}{Phys. Rev. C} \textbf{\bibinfo{volume}{79}},
  \bibinfo{eid}{034303} (\bibinfo{year}{2009}).

\bibitem[{\citenamefont{Li et~al.}(2009)\citenamefont{Li, Nik\v{s}i\'{c},
  Vretenar, Meng, Lalazissis, and Ring}}]{li:054301}
\bibinfo{author}{\bibfnamefont{Z.~P.} \bibnamefont{Li}},
  \bibinfo{author}{\bibfnamefont{T.}~\bibnamefont{Nik\v{s}i\'{c}}},
  \bibinfo{author}{\bibfnamefont{D.}~\bibnamefont{Vretenar}},
  \bibinfo{author}{\bibfnamefont{J.}~\bibnamefont{Meng}},
  \bibinfo{author}{\bibfnamefont{G.~A.} \bibnamefont{Lalazissis}},
  \bibnamefont{and} \bibinfo{author}{\bibfnamefont{P.}~\bibnamefont{Ring}},
  \bibinfo{journal}{Phys. Rev. C} \textbf{\bibinfo{volume}{79}},
  \bibinfo{eid}{054301} (\bibinfo{year}{2009}).

\bibitem[{\citenamefont{Li et~al.}(2010)\citenamefont{Li, Nik\v{s}i\'{c},
  Vretenar, and Meng}}]{PhysRevC.81.034316}
\bibinfo{author}{\bibfnamefont{Z.~P.} \bibnamefont{Li}},
  \bibinfo{author}{\bibfnamefont{T.}~\bibnamefont{Nik\v{s}i\'{c}}},
  \bibinfo{author}{\bibfnamefont{D.}~\bibnamefont{Vretenar}}, \bibnamefont{and}
  \bibinfo{author}{\bibfnamefont{J.}~\bibnamefont{Meng}},
  \bibinfo{journal}{Phys. Rev. C} \textbf{\bibinfo{volume}{81}},
  \bibinfo{pages}{034316} (\bibinfo{year}{2010}).

\bibitem[{\citenamefont{Girod et~al.}(2009)\citenamefont{Girod, Delaroche,
  G\"{o}rgen, and Obertelli}}]{Girod200939}
\bibinfo{author}{\bibfnamefont{M.}~\bibnamefont{Girod}},
  \bibinfo{author}{\bibfnamefont{J.-P.} \bibnamefont{Delaroche}},
  \bibinfo{author}{\bibfnamefont{A.}~\bibnamefont{G\"{o}rgen}},
  \bibnamefont{and}
  \bibinfo{author}{\bibfnamefont{A.}~\bibnamefont{Obertelli}},
  \bibinfo{journal}{Phys. Lett. B} \textbf{\bibinfo{volume}{676}},
  \bibinfo{pages}{39 } (\bibinfo{year}{2009}).

\bibitem[{\citenamefont{Delaroche et~al.}(2010)\citenamefont{Delaroche, Girod,
  Libert, Goutte, Hilaire, P\'eru, Pillet, and Bertsch}}]{PhysRevC.81.014303}
\bibinfo{author}{\bibfnamefont{J.~P.} \bibnamefont{Delaroche}},
  \bibinfo{author}{\bibfnamefont{M.}~\bibnamefont{Girod}},
  \bibinfo{author}{\bibfnamefont{J.}~\bibnamefont{Libert}},
  \bibinfo{author}{\bibfnamefont{H.}~\bibnamefont{Goutte}},
  \bibinfo{author}{\bibfnamefont{S.}~\bibnamefont{Hilaire}},
  \bibinfo{author}{\bibfnamefont{S.}~\bibnamefont{P\'eru}},
  \bibinfo{author}{\bibfnamefont{N.}~\bibnamefont{Pillet}}, \bibnamefont{and}
  \bibinfo{author}{\bibfnamefont{G.~F.} \bibnamefont{Bertsch}},
  \bibinfo{journal}{Phys. Rev. C} \textbf{\bibinfo{volume}{81}},
  \bibinfo{pages}{014303} (\bibinfo{year}{2010}).

\bibitem[{\citenamefont{Bender et~al.}(2003)\citenamefont{Bender, Heenen, and
  Reinhard}}]{RevModPhys.75.121}
\bibinfo{author}{\bibfnamefont{M.}~\bibnamefont{Bender}},
  \bibinfo{author}{\bibfnamefont{P.-H.} \bibnamefont{Heenen}},
  \bibnamefont{and} \bibinfo{author}{\bibfnamefont{P.-G.}
  \bibnamefont{Reinhard}}, \bibinfo{journal}{Rev. Mod. Phys.}
  \textbf{\bibinfo{volume}{75}}, \bibinfo{pages}{121} (\bibinfo{year}{2003}).

\bibitem[{\citenamefont{Matsuo et~al.}(2000)\citenamefont{Matsuo, Nakatsukasa,
  and Matsuyanagi}}]{PTP.103.959}
\bibinfo{author}{\bibfnamefont{M.}~\bibnamefont{Matsuo}},
  \bibinfo{author}{\bibfnamefont{T.}~\bibnamefont{Nakatsukasa}},
  \bibnamefont{and}
  \bibinfo{author}{\bibfnamefont{K.}~\bibnamefont{Matsuyanagi}},
  \bibinfo{journal}{Prog. Theor. Phys.} \textbf{\bibinfo{volume}{103}},
  \bibinfo{pages}{959} (\bibinfo{year}{2000}).

\bibitem[{\citenamefont{Ring and Schuck}(1980)}]{Ring-Schuck}
\bibinfo{author}{\bibfnamefont{P.}~\bibnamefont{Ring}} \bibnamefont{and}
  \bibinfo{author}{\bibfnamefont{P.}~\bibnamefont{Schuck}},
  \emph{\bibinfo{title}{The Nuclear Many-Body Problem}}
  (\bibinfo{publisher}{Springer-Verlag}, \bibinfo{year}{1980}).

\bibitem[{\citenamefont{Bes and Sorensen}(1969)}]{Bes-Sorensen}
\bibinfo{author}{\bibfnamefont{D.~R.} \bibnamefont{Bes}} \bibnamefont{and}
  \bibinfo{author}{\bibfnamefont{R.~A.} \bibnamefont{Sorensen}},
  \emph{\bibinfo{title}{Advances in Nuclear Physics}}, vol.~\bibinfo{volume}{2}
  (\bibinfo{publisher}{Prenum Press}, \bibinfo{year}{1969}).

\bibitem[{\citenamefont{Baranger and
  Kumar}(1968{\natexlab{a}})}]{Baranger1968490}
\bibinfo{author}{\bibfnamefont{M.}~\bibnamefont{Baranger}} \bibnamefont{and}
  \bibinfo{author}{\bibfnamefont{K.}~\bibnamefont{Kumar}},
  \bibinfo{journal}{Nucl. Phys. A} \textbf{\bibinfo{volume}{110}},
  \bibinfo{pages}{490 } (\bibinfo{year}{1968}{\natexlab{a}}).

\bibitem[{\citenamefont{Hinohara et~al.}(2006)\citenamefont{Hinohara,
  Nakatsukasa, Matsuo, and Matsuyanagi}}]{PTP.115.567}
\bibinfo{author}{\bibfnamefont{N.}~\bibnamefont{Hinohara}},
  \bibinfo{author}{\bibfnamefont{T.}~\bibnamefont{Nakatsukasa}},
  \bibinfo{author}{\bibfnamefont{M.}~\bibnamefont{Matsuo}}, \bibnamefont{and}
  \bibinfo{author}{\bibfnamefont{K.}~\bibnamefont{Matsuyanagi}},
  \bibinfo{journal}{Prog. Theor. Phys.} \textbf{\bibinfo{volume}{115}},
  \bibinfo{pages}{567} (\bibinfo{year}{2006}).

\bibitem[{\citenamefont{Rainovski et~al.}(2002)\citenamefont{Rainovski,
  Schnare, Schwengner, Plettner, K\"aubler, D\"onau, Ragnarsson, Eberth,
  Steinhardt, Thelen et~al.}}]{0954-3899-28-10-307}
\bibinfo{author}{\bibfnamefont{G.}~\bibnamefont{Rainovski}},
  \bibinfo{author}{\bibfnamefont{H.}~\bibnamefont{Schnare}},
  \bibinfo{author}{\bibfnamefont{R.}~\bibnamefont{Schwengner}},
  \bibinfo{author}{\bibfnamefont{C.}~\bibnamefont{Plettner}},
  \bibinfo{author}{\bibfnamefont{L.}~\bibnamefont{K\"aubler}},
  \bibinfo{author}{\bibfnamefont{F.}~\bibnamefont{D\"onau}},
  \bibinfo{author}{\bibfnamefont{I.}~\bibnamefont{Ragnarsson}},
  \bibinfo{author}{\bibfnamefont{J.}~\bibnamefont{Eberth}},
  \bibinfo{author}{\bibfnamefont{T.}~\bibnamefont{Steinhardt}},
  \bibinfo{author}{\bibfnamefont{O.}~\bibnamefont{Thelen}},
  \bibnamefont{et~al.}, \bibinfo{journal}{J. Phys. G}
  \textbf{\bibinfo{volume}{28}}, \bibinfo{pages}{2617} (\bibinfo{year}{2002}).

\bibitem[{\citenamefont{Palit et~al.}(2001)\citenamefont{Palit, Jain, Joshi,
  Sheikh, and Sun}}]{PhysRevC.63.024313}
\bibinfo{author}{\bibfnamefont{R.}~\bibnamefont{Palit}},
  \bibinfo{author}{\bibfnamefont{H.~C.} \bibnamefont{Jain}},
  \bibinfo{author}{\bibfnamefont{P.~K.} \bibnamefont{Joshi}},
  \bibinfo{author}{\bibfnamefont{J.~A.} \bibnamefont{Sheikh}},
  \bibnamefont{and} \bibinfo{author}{\bibfnamefont{Y.}~\bibnamefont{Sun}},
  \bibinfo{journal}{Phys. Rev. C} \textbf{\bibinfo{volume}{63}},
  \bibinfo{pages}{024313} (\bibinfo{year}{2001}).

\bibitem[{\citenamefont{Baranger and
  Kumar}(1968{\natexlab{b}})}]{Baranger1968241}
\bibinfo{author}{\bibfnamefont{M.}~\bibnamefont{Baranger}} \bibnamefont{and}
  \bibinfo{author}{\bibfnamefont{K.}~\bibnamefont{Kumar}},
  \bibinfo{journal}{Nucl. Phys. A} \textbf{\bibinfo{volume}{122}},
  \bibinfo{pages}{241 } (\bibinfo{year}{1968}{\natexlab{b}}).

\bibitem[{\citenamefont{Kumar and Baranger}(1968)}]{Kumar1968273}
\bibinfo{author}{\bibfnamefont{K.}~\bibnamefont{Kumar}} \bibnamefont{and}
  \bibinfo{author}{\bibfnamefont{M.}~\bibnamefont{Baranger}},
  \bibinfo{journal}{Nucl. Phys. A} \textbf{\bibinfo{volume}{122}},
  \bibinfo{pages}{273 } (\bibinfo{year}{1968}).

\bibitem[{\citenamefont{Hinohara et~al.}(2007)\citenamefont{Hinohara,
  Nakatsukasa, Matsuo, and Matsuyanagi}}]{PTP.117.451}
\bibinfo{author}{\bibfnamefont{N.}~\bibnamefont{Hinohara}},
  \bibinfo{author}{\bibfnamefont{T.}~\bibnamefont{Nakatsukasa}},
  \bibinfo{author}{\bibfnamefont{M.}~\bibnamefont{Matsuo}}, \bibnamefont{and}
  \bibinfo{author}{\bibfnamefont{K.}~\bibnamefont{Matsuyanagi}},
  \bibinfo{journal}{Prog. Theor. Phys.} \textbf{\bibinfo{volume}{117}},
  \bibinfo{pages}{451} (\bibinfo{year}{2007}).

\bibitem[{\citenamefont{Hinohara et~al.}(2008)\citenamefont{Hinohara,
  Nakatsukasa, Matsuo, and Matsuyanagi}}]{PTP.119.59}
\bibinfo{author}{\bibfnamefont{N.}~\bibnamefont{Hinohara}},
  \bibinfo{author}{\bibfnamefont{T.}~\bibnamefont{Nakatsukasa}},
  \bibinfo{author}{\bibfnamefont{M.}~\bibnamefont{Matsuo}}, \bibnamefont{and}
  \bibinfo{author}{\bibfnamefont{K.}~\bibnamefont{Matsuyanagi}},
  \bibinfo{journal}{Prog. Theor. Phys.} \textbf{\bibinfo{volume}{119}},
  \bibinfo{pages}{59} (\bibinfo{year}{2008}).

\bibitem[{\citenamefont{Hinohara et~al.}(2009)\citenamefont{Hinohara,
  Nakatsukasa, Matsuo, and Matsuyanagi}}]{hinohara:014305}
\bibinfo{author}{\bibfnamefont{N.}~\bibnamefont{Hinohara}},
  \bibinfo{author}{\bibfnamefont{T.}~\bibnamefont{Nakatsukasa}},
  \bibinfo{author}{\bibfnamefont{M.}~\bibnamefont{Matsuo}}, \bibnamefont{and}
  \bibinfo{author}{\bibfnamefont{K.}~\bibnamefont{Matsuyanagi}},
  \bibinfo{journal}{Phys. Rev. C} \textbf{\bibinfo{volume}{80}},
  \bibinfo{eid}{014305} (\bibinfo{year}{2009}).

\bibitem[{\citenamefont{Thouless and Valatin}(1962)}]{Thouless1962211}
\bibinfo{author}{\bibfnamefont{D.~J.} \bibnamefont{Thouless}} \bibnamefont{and}
  \bibinfo{author}{\bibfnamefont{J.~G.} \bibnamefont{Valatin}},
  \bibinfo{journal}{Nucl. Phys.} \textbf{\bibinfo{volume}{31}},
  \bibinfo{pages}{211 } (\bibinfo{year}{1962}).

\bibitem[{\citenamefont{Bengtsson and Ragnarsson}(1985)}]{Bengtsson198514}
\bibinfo{author}{\bibfnamefont{T.}~\bibnamefont{Bengtsson}} \bibnamefont{and}
  \bibinfo{author}{\bibfnamefont{I.}~\bibnamefont{Ragnarsson}},
  \bibinfo{journal}{Nucl. Phys. A} \textbf{\bibinfo{volume}{436}},
  \bibinfo{pages}{14 } (\bibinfo{year}{1985}).

\bibitem[{\citenamefont{Nilsson and Ragnarsson}(1995)}]{Nilsson-Ragnarsson}
\bibinfo{author}{\bibfnamefont{S.~G.} \bibnamefont{Nilsson}} \bibnamefont{and}
  \bibinfo{author}{\bibfnamefont{I.}~\bibnamefont{Ragnarsson}},
  \emph{\bibinfo{title}{Shapes and Shells in Nuclear Structure}}
  (\bibinfo{publisher}{Cambridge University Press}, \bibinfo{year}{1995}).

\bibitem[{\citenamefont{Yamagami et~al.}(2001)\citenamefont{Yamagami,
  Matsuyanagi, and Matsuo}}]{Yamagami2001579}
\bibinfo{author}{\bibfnamefont{M.}~\bibnamefont{Yamagami}},
  \bibinfo{author}{\bibfnamefont{K.}~\bibnamefont{Matsuyanagi}},
  \bibnamefont{and} \bibinfo{author}{\bibfnamefont{M.}~\bibnamefont{Matsuo}},
  \bibinfo{journal}{Nucl. Phys. A} \textbf{\bibinfo{volume}{693}},
  \bibinfo{pages}{579 } (\bibinfo{year}{2001}).

\bibitem[{\citenamefont{Sakamoto and Kishimoto}(1990)}]{Sakamoto1990321}
\bibinfo{author}{\bibfnamefont{H.}~\bibnamefont{Sakamoto}} \bibnamefont{and}
  \bibinfo{author}{\bibfnamefont{T.}~\bibnamefont{Kishimoto}},
  \bibinfo{journal}{Phys. Lett. B} \textbf{\bibinfo{volume}{245}},
  \bibinfo{pages}{321 } (\bibinfo{year}{1990}).

\bibitem[{\citenamefont{Baranger and Kumar}(1965)}]{Baranger1965113}
\bibinfo{author}{\bibfnamefont{M.}~\bibnamefont{Baranger}} \bibnamefont{and}
  \bibinfo{author}{\bibfnamefont{K.}~\bibnamefont{Kumar}},
  \bibinfo{journal}{Nucl. Phys.} \textbf{\bibinfo{volume}{62}},
  \bibinfo{pages}{113 } (\bibinfo{year}{1965}).

\bibitem[{\citenamefont{Hamamoto and Nazarewicz}(1994)}]{PhysRevC.49.2489}
\bibinfo{author}{\bibfnamefont{I.}~\bibnamefont{Hamamoto}} \bibnamefont{and}
  \bibinfo{author}{\bibfnamefont{W.}~\bibnamefont{Nazarewicz}},
  \bibinfo{journal}{Phys. Rev. C} \textbf{\bibinfo{volume}{49}},
  \bibinfo{pages}{2489} (\bibinfo{year}{1994}).

\bibitem[{\citenamefont{Wilets and Jean}(1956)}]{PhysRev.102.788}
\bibinfo{author}{\bibfnamefont{L.}~\bibnamefont{Wilets}} \bibnamefont{and}
  \bibinfo{author}{\bibfnamefont{M.}~\bibnamefont{Jean}},
  \bibinfo{journal}{Phys. Rev.} \textbf{\bibinfo{volume}{102}},
  \bibinfo{pages}{788} (\bibinfo{year}{1956}).

\bibitem[{\citenamefont{Sato et~al.}(2010)\citenamefont{Sato, Hinohara,
  Nakatsukasa, Matsuo, and Matsuyanagi}}]{PTP.123.129}
\bibinfo{author}{\bibfnamefont{K.}~\bibnamefont{Sato}},
  \bibinfo{author}{\bibfnamefont{N.}~\bibnamefont{Hinohara}},
  \bibinfo{author}{\bibfnamefont{T.}~\bibnamefont{Nakatsukasa}},
  \bibinfo{author}{\bibfnamefont{M.}~\bibnamefont{Matsuo}}, \bibnamefont{and}
  \bibinfo{author}{\bibfnamefont{K.}~\bibnamefont{Matsuyanagi}},
  \bibinfo{journal}{Prog. Theor. Phys.} \textbf{\bibinfo{volume}{123}},
  \bibinfo{pages}{129} (\bibinfo{year}{2010}).

\end{thebibliography}

\end{document}